\documentclass[preprint,12pt,authoryear]{elsarticle}
\usepackage[margin=2.5cm]{geometry}%

\usepackage{amssymb}
\usepackage{amsmath}
\usepackage{multirow,array}
\usepackage{enumitem}
\usepackage{caption}
\usepackage{tikz}
\usepackage{tabularx}
\usepackage{float}
\usepackage{booktabs}
\usetikzlibrary{trees}
\usepackage{url,hyperref,lineno,microtype,subcaption}
\usepackage[noend,ruled,vlined,linesnumbered,boxed]{algorithm2e}
\DeclareMathOperator*{\argmax}{argmax}
\DeclareMathOperator{\sign}{sign}

\journal{arXiv}

\begin{document}

\begin{frontmatter}



\title{On the Computational Complexity of Ethics: Moral Tractability for Minds and Machines}


\author[inst1,inst2]{Jakob Stenseke}

\affiliation[inst1]{organization={Department of Philosophy, Lund University},
            addressline={Helgonavagen 3, Lund 221 00}, 
            country={Sweden}}

\affiliation[inst2]{country={jakob.stenseke@fil.lu.se, 0000-0001-8579-3975}}

\begin{abstract}
Why should moral philosophers, moral psychologists, and machine ethicists care about computational complexity? Debates on whether artificial intelligence (AI) can or should be used to solve problems in ethical domains have mainly been driven by what AI can or cannot do in terms of human capacities. In this paper, we tackle the problem from the other end by exploring what kind of moral machines are possible based on what computational systems can or cannot do. To do so, we analyze normative ethics through the lens of computational complexity. First, we introduce computational complexity for the uninitiated reader and discuss how the complexity of ethical problems can be framed within Marr’s three levels of analysis. We then study a range of ethical problems based on consequentialism, deontology, and virtue ethics, with the aim of elucidating the complexity associated with the problems themselves (e.g., due to combinatorics, uncertainty, strategic dynamics), the computational methods employed (e.g., probability, logic, learning), and the available resources (e.g., time, knowledge, learning). The results indicate that most problems the normative frameworks pose lead to tractability issues in every category analyzed. Our investigation also provides several insights about the computational nature of normative ethics, including the differences between rule- and outcome-based moral strategies, and the implementation-variance with regard to moral resources. We then discuss the consequences complexity results have for the prospect of moral machines in virtue of the trade-off between optimality and efficiency. Finally, we elucidate how computational complexity can be used to inform both philosophical and cognitive-psychological research on human morality by advancing the Moral Tractability Thesis (MTT).
\end{abstract}




\end{frontmatter}


\tableofcontents

\section{Introduction}
\label{sec:Introd}


Computational systems of hardware and software continue to enter and transform a growing number of human domains. As autonomous vehicles, virtual teachers, and carebots augment or even take over traditional human roles of drivers, educators, and caretakers, it becomes hard to ignore the need for systems that align with the norms and moral standards associated by such roles.\footnote{Unless specified, terms such as ``AI system'', ``machine'', and ``computer'' will be used interchangeably to denote computational systems of hardware and software.} These concerns have spawned the interdisciplinary field of \emph{machine ethics}, which broadly explores the prospects of implementing ethics into machines \citep{Wallach2008, Anderson2011}. Lying in the intersection of computer science and moral philosophy, machine ethics encompasses a spectrum of more or less interconnected research aims, including work that addresses the challenges of value alignment \citep{gabriel2020artificial}, explainability \citep{gunning2019xai}, and safety \citep{Amodei2016} of existing AI methods, the development of systems tackling various ethical dilemmas \citep{Cervantes2020,Tolmeijer2020}, and theoretical debates on whether and to what extent artificial moral agents are feasible or desirable \citep{floridi2004morality,Behdadi2020}.

The feasibility debate has, in turn, mainly been driven by what AI systems can or cannot do in terms of human capacities; whether artificial agents could be autonomous or have free will \citep{hellstrom2013moral}, be equipped with human-like rationality \citep{purves2015autonomous}, or capable of conscious experience \citep{himma2009artificial}. However, by centering on capacities that remain elusive and conceptually opaque from a computational perspective, debates on artificial morality fails to engage with the technical dimensions of AI, and as a result, they become practically otiose for the design and development of ethical machines \citep{mabaso2021computationally,Behdadi2020,Stenseke2022IDR}. Another issue that obscures the feasibility of moral machines is the absence of systemic evaluation tools \citep{Tolmeijer2020}. In machine ethics, there are at present no domain-specific nor general benchmarks that can be used to evaluate the performance of different ethical systems. Consequentially, since evaluations of systems are limited to the experimental conditions of their particular implementation, the scalability of solutions and generalizability of results are severely restricted.

In this paper, we address these issues by exploring what kind of moral machines are possible based on the ethical problems computational systems can or cannot solve effectively. To do so, we analyze normative ethics through the lens of computational complexity theory, which classifies problems in terms of the resources (e.g., time and space) a computer requires to solve them. While previous work have discussed computational limitations for moral machines more informally \citep{brundage2014limitations,StensekeBalkenius2022}, and provided embryonic complexity analyses of ethical actions \citep{reynolds2005computational} and plans \citep{Lindner2020}, the computational complexity of ethics and its potential relevance for machine ethics remains largely unexplored. For instance, if artificial systems were to operate in ethical domains where time is of the essence (e.g., a self-driving ambulance), it is crucial that such systems can make efficient as well as competent ethical decisions. Furthermore, if human moral cognition is constrained by tractability \citep{van2008tractable}, the analysis might also serve moral psychology and normative theory by constraining the space of problems an agent following a certain normative theory can be reasonably expected to solve.

In the rest of the paper, concepts and theories from both moral philosophy and computer science are introduced and explained in a way that is friendly for readers with a limited background in one or both areas. It is structured as follows. First, we give an introduction to computational complexity and tractability with the aim of explaining their relevance for the uninitiated reader (section \ref{sec:salad}). In section \ref{sec:complexity}, we survey previous implementations in machine ethics and discuss various interpretations of the complexity of ethics using Marr's three levels of analysis (\ref{sec:Marr3}), which motivates the analysis of problems posed by normative theory (computational level, \ref{sec:level1computational}) that are solved through a variety of computational methods (algorithmic level, \ref{sec:level2algorithm}) by a deterministic Turing Machine (implementation level, \ref{sec:level3implementation}). We then explore the complexity of various ethical problems based on consequentialism (\ref{sec:consequentialism}), deontology (\ref{sec:deontology}), and virtue ethics (\ref{sec:learning}). The main aim is to elucidate the complexity associated with the problems themselves (e.g., due to uncertainty, combinatorics, strategic dynamics, and generality), the available resources (e.g., time, cognition, and domain knowledge), and the computational methods employed to tackle the problems (e.g., probability, logic, and learning). The results indicate that most problems the normative theories pose lead to intractability issues (a succinct summary is given in \ref{table:summary}), and especially if the prescriptive ideal should be optimally satisfied. In particular, based on the intractability (and undecidability) stemming from combinatorics of action plans (\ref{sec:combinatorics}), probabilistic causal inference (\ref{sec:causalinference}), dynamic and partially observable environments (\ref{sec:dynamicenvironments}), general rules (\ref{sec:deontologyGenerality}), strategic dynamics (\ref{sec:deoGameTheory}), logic (\ref{sec:deontologySemantics}), semantics (\ref{sec:deoProblemOfSemantics}) and learning (\ref{sec:ComplexityOfLearning}), we firmly conclude that perfect moral machines are impossible. Our investigation also provides additional insights regarding the computational nature of the normative theories, including (i) the differences between action- and outcome-based strategies, (ii) the benefits of moral hybrids (\ref{sec:deontologyconshybrids}), and (iii) the extreme implementation-variance with regard to moral resources. In section \ref{sec:discussion}, we discuss the consequences the results have for the prospects of moral machines by focusing on the trade-off between optimality and efficiency, the equivocal role of normative theory, and the intimate relationship between different moral resources. Finally, we  demonstrate how computational tractability can be used to inform both philosophical and psychological research on human morality by advancing the Moral Tractability Thesis.



\section{The Complexity of Making a Salad}
\label{sec:salad}
Let us begin with an illustrative example.\footnote{The reader who is already familiar with computational complexity is advised to skip to section \ref{sec:complexity}.} There is a high chance that you have stumbled upon a salad bar where you can choose ingredients to your own liking.\footnote{The example is inspired by the excellent introduction to complexity analysis given in \cite{van2019cognition}.} The question is, what ingredients do you pick in order to create the best tasting salad? Let us assume that you can immediately assess the tastiness of each ingredient in isolation and give them a ``taste value'' ($v$) on a scale ranging from the most off-putting ($-10$) to the most delicious ($+10$). With these values, you find that one efficient way of putting together a decent salad is to exclusively pick ingredients ($I$) with a positive $v$ ($v(I) > 0$), or a $v$ that is higher than a certain threshold (e.g., $v(I) > 5$). Let us name this strategy $\Psi$. In fact, as a queue is lining up behind the salad bar, you appreciate the speed $\Psi$ allows you to make a salad: you only have to visit each ingredient once and check whether they are sufficiently tasty to be included in your mix. Furthermore, you realize that the performance of $\Psi$ grows, in the worst-case, linearly with the number of salad ingredients. This means that, regardless of how many ingredients there could be, $\Psi$ will always be efficient: for any input — in this case, $n$ number of ingredients — the time it takes to make a salad will closely mirror the size of the input (i.e., 1000 ingredients equals 1000 visits to distinct ingredients).

But upon further reflection, you realize that something is odd with $\Psi$. It asks you to put sun-dried tomatoes on top of pineapple. You imagine how the saltiness of sun-dried tomatoes mixes with the sweet-sourness of pineapple as they traverse the taste buds of your tongue. Your immediate disgust of the image reveals a fatal flaw of $\Psi$: even if these two ingredients were given some of the highest taste values ($v > 9$), their combination yields a taste value that is terribly off-putting ($v = -10$). You realize that $\Psi$ violates a fundamental principle of gastronomy, namely, that combinations of ingredients yield taste values that do not necessarily correspond with the tastiness of its individual ingredients. We can call this principle the Combinatorial Principle of Gastronomy (CPG).

Luckily, you have a perfect gustatory imagination and can immediately assess the taste value of any given combination of ingredients. How do you find the optimal combination of ingredients in a way that maximizes taste value and adheres to CPG? We can formally describe this as the following computational problem:

\begin{center}
    \textsc{Optimal salad following CPG}
\end{center}
\begin{quote}
\textbf{Input:} A salad bar as a set $SB = \{I_{1}, I_{2}, ..., I_{n}\}$ of $n$ ingredients and a value function $v$ that assigns a taste value to every subset (or salad) $S \subseteq SB$.

\textbf{Output:} A salad $S \subseteq SB$ such that $v(S)$ is maximized over all possible salads in the salad bar ($S \subseteq SB$).
\end{quote}

You realize that there is a straight-forward strategy, you call it $\Phi$, that is guaranteed to produce an optimal salad while satisfying CPG: simply imagine the taste value of \emph{each} possible subset $S \subseteq SB$ and pick the salad with the highest $v(S)$. But you have a feeling that there must be a catch with $\Phi$. You do some basic combinatorics: if there was only one ingredient, e.g., $\{cucumber\}$, there would be one possible salad (made entirely of cucumber); two ingredients yield three distinct salads, e.g., $\{cucumber\},\{onion\},\{cucumber,onion\}$; three ingredients make seven; four make fifteen; etc. You determine that the number of possible salads grows \emph{exponentially} with the number of ingredients, so that $n$ ingredients produce $2^{n}-1$ possible salads. The salad bar you are currently facing has 30 ingredients, which presents $2^{30}-1 = 1,073,741,823$ distinct salads. Since your otherwise extraordinary gustatory system can only assess the taste of one salad per second, $\Phi$ asks you to imagine salads for roughly 34 years, that is, if you were to optimally satisfy the combinatorial principle of gastronomy. Unfortunately, you have already wasted more than enough time, and the people in the queue behind you are very upset.

The example serves to draw five important lessons about computational complexity:

(1) The first is that many decision problems that we encounter in everyday life can be formulated in similar ways, from planning an itinerary, packing a bag for a trip, or inviting a selection of friends to a birthday party in your small apartment. And as we will see throughout this paper, ethical problems are no exception. You might wonder why it matters so much to find the optimal salad; at worst, you end up with a poor-tasting salad, which is far from a disastrous consequence. But would you be so quick to disregard optimal results if the problem was a matter of life and death? And even if you do not care much about the combinatorial principle of gastronomy, there might be moral principles that are fundamental to your ethical life.

(2) The second lesson is that the complexity of a problem can be expressed in terms of the resources an agent or algorithm requires to solve it. For computational systems, the two most interesting resources are \emph{time} and \emph{space}. The latter conventionally denotes the size of computer memory (e.g., bits), whereas the former refers to the number of machine operations (or synonymously used terms such as ``computations'', ``calculations'', ``steps'', or ``state transitions''). Why measure time in terms of machine operations and not in seconds or minutes? The reason is that, while the real-time speed of computers solving a problem by running some algorithm $A$ can vary greatly, the amount of machine operations they need to execute $A$ remain unchanged. A 21st century computer and one from the 1960's both have to consider $2^{n}-1$ salads if they were to produce the best tasting salad following $\Phi$, even if the modern computer could potentially do so a million times faster. Importantly, this forms the basis of the Invariance Thesis,\footnote{More formally, the thesis states that given two machines $M_{1}$ and $M_{2}$, and a given computational problem $\Theta$, the complexity of $\Theta$ executed by $M_{1}$ and $M_{2}$ will differ at most by a polynomial amount. That is, if $M_{1}$ is able to compute $\Theta$ in time $t$, $M_{2}$ can compute $\Theta$ in $t^{c}$, where $c$ is a constant. The thesis is widely accepted among computer scientists provided that $M_{1}$ and $M_{2}$ are any type of Turing machine or any other reasonable model of computation (e.g., cellular automata, neural networks) and the input is reasonably encoded (e.g., it does not involve irrelevant information). See \cite{garey1979computers}.} which allows us to analyze and compare the worst-case complexity that is inherent to computational problems independent of specific machines.

(3) This leads to the third lesson, which is the simple observation that some problems are more complex than others. If a problem is undecidable, it means that it can be proven that no algorithm can be constructed to solve the problem.\footnote{The halting problem is an example of an undecidable problem: in 1936, Alan Turing proved that there is no algorithm that can determine whether an arbitrary program eventually halts. Decidability will be further addressed in section \ref{sec:DeoDescriptive}.} Among the decidable problems, the most important distinction is between problems that are \emph{tractable} and \emph{intractable}. Crudely put, a problem is tractable if it can be solved using a `realistic' amount of resources. For most computational theorists, however, tractable is synonymous with ``computable in polynomial time''. This means that the runtime (number of machine operations) of an algorithm is upperbounded by a polynomial expression in its input, i.e., of the type $n^{c}$ (where $c$ is some positive constant). This includes functions that show logarithmic ($\log n)$, linear ($n$), quadratic ($n^{2}$), or cubic ($n^{3}$) growth in time as the input $n$ increases. The class of decision problems that can be solved in polynomial time by a deterministic Turing machine is called P, capturing the notion of decision problems with ``effective'' decision procedures \citep{cobham1965intrinsic}. Conversely, problems that cannot be solved in polynomial time are called intractable as their runtime grows exponentially ($c^{n}$), by a factorial ($n!)$, or super-exponentially ($n^{n}$). This notion of tractability is illustrated in the difference between decision procedure $\Psi$ and $\Phi$. For $\Psi$, salad-making time will never grow more than linearly in relation to the number of ingredients. Using Big O notation, which expresses an asymptomtotic upperbound\footnote{``asymptotic'' means that we can ignore lower order polynomials and constants when we describe a function. For instance, the function $f(n) = 4n + n^{3}$ is written as $O(n^{3})$, since $4n$ becomes insignificant compared to $n^{3}$ as $n$ increases.} of a function (in this case, a mapping between input size $n$ and time), the time complexity of $\Psi$ is $O(n)$. By contrast, performing an exhaustive search over all possible salads of the salad bar leads $\Phi$ to the exponential $O(2^{n})$. Even if your gustatory imagination could utilize the speed of the parallelized neural computation of your brain, which allowed you to imagine one billion salads per second, a salad bar of 50 ingredients would still take you 13 days to master, and a bar of 60 takes you roughly 37 billion years. Of course, no sane person would spend that much time imagining the taste of different salads. But the problem with problems remains: if we want to solve them effectively, we  might need to give up our requirement of optimality. Instead of ``best imaginable'', we need compromises that are ``good enough'' given the available resources. As such, intractable problems can present an uncomfortable trade-off between ideal and feasible. And it is precisely how this uncomfortable trade-off affects ethical decisions for computational agents that will be the topic of this paper.

Part of the reason why there is no effective way to make an optimal salad following CPG is captured in the widely believed conjecture P $\neq$ NP. It states that decision problems that have solutions which can be \emph{checked} (or verified) effectively cannot necessarily be solved effectively. To be more precise, it states that the complexity class P does not equal NP: the class of decision problems solvable in polynomial time by a \emph{non-deterministic} TM, or equivalently, decision problems where solutions can be verified in polynomial time. $\Phi$ exemplifies such a case. Even if you can check the taste of any combination of salad ingredients quickly (polynomial time), there is no deterministic procedure that allows you to find the optimal; you still have to check the entire space of combinations to ensure that you have the optimal subset. In fact, finding the optimal salad following CPG is NP-hard, which means that it is \emph{at least} as hard as the hardest problem in NP. More formally, a problem $X$ is NP-hard when \emph{every} problem in NP can be \emph{reduced} in polynomial time to $X$. This means that if we assume that a solution for $X$ takes one unit of time, the solution can be used to solve every problem in NP in polynomial time. A closely related property is the notion of completeness. An NP-complete problem is both NP-hard \emph{and} belongs to NP. Note that, while P and NP are classes of decision problems — which can be framed as a yes/no-type question — NP-hard problems are not restricted to decision problems as such; they are simply at least as hard as the hardest decision versions of the same problem. For instance, while decision variants of NP-hard problems might be NP-complete — e.g., the Boolean satisfiability problem (SAT) or subset sum problem (SSP) — other variants of the same problem, e.g., framed as optimization or search problems, are not (they are not decision problems). Again, this is illustrated in our example: salad making following CPG is NP-hard since it is an optimization version of the subset sum problem (SSP), which is NP-complete.

Furthermore, note also that, while NP-hardness only denotes a general lower bound, it does not say anything about an upper bound, which might be more informative for understanding exactly \emph{how} hard a problem is.\footnote{For instance, the halting problem is NP-hard but undecidable (it is not decidable in a finite amount of operations); the true quantified Boolean formula language (QBF) is NP-hard but decidable in polynomial space (PSPACE-complete) \citep{garey1979computers}.} In computational complexity theory, classes of computational problems are instead defined by the upper bound (or constraints) on the amount of resources they require in the worst-case (formalized using Big O notation). In turn, this allows us to describe general hierarchies of how complex problems are. For instance, problems solvable in polynomial time by a deterministic TM are also solvable by a non-deterministic TM, which implies that P is a subset ($\subseteq$) of NP. Similarly, it is widely believed that P $\subseteq$ NP $\subseteq$ PSPACE $\subseteq$ EXPTIME $\subseteq$ EXPSPACE (see Appendix (\ref{sec:AppendixA}) for a summary of the complexity classes used in this article). The worst-case analysis can be motivated by the fact that an algorithm needs to consider all possible inputs of a problem, which includes the worst-case input. But in the analysis of algorithms, there are several other essential tools to study complexity. For instance, if the lower and upper bound coincide, we have a \emph{tight bound}. Again, there is such a tight bound on the time complexity of making an optimal salad following $\Phi$: we have to imagine the taste of at least \emph{and} at most $2^{n}-1$ salads to ensure optimality. Alternatively, we could imagine that salad bars were arranged in ways that allowed for exploitation, e.g., sorted in rows of pre-made combinations. If so, we could measure the time complexity of an algorithm in terms of how many operations it required to make a salad over a number of different salad bars (inputs), and see how it performed in the best-case, average-case, and worst-case.\footnote{However, note that such performance measures would not work for finding the optimal salad following $\Phi$, since it does not matter in which way the ingredients are arranged.} In short, computational complexity provides a smorgasbord of analytical tools to understand the difficulty of problems and their algorithmic solutions.

(4) The fourth lesson, and a corollary of the third, is that the way an agent solves a problem ultimately depends on its \emph{resources}, broadly construed. Besides time and memory-size, these resources include heuristics (efficient strategies), cognition (capacities for perceiving and acting in the world),\footnote{Throughout this paper, the term ``cognition'' will be broadly used to denote all sorts of information-processing that enables capacities such as perception, action, reasoning, and learning. As such, it differs from cognitivism in meta-ethics (the view that moral language can express propositions that can be true or false) and conceptions of cognition that emphasize prefrontal activity (e.g., thinking, memory, judgement) in contrast to 'back of the brain' sensory processing \citep{block2019wrong}.}, knowledge, and learning. In reality, you might mix aspects of $\Psi$ and $\Phi$. You might select a few key ingredients as a basis that you already know yields a reasonably tasty salad, and imagine whether this basis could benefit from further additions. Drawing from your vast experience of cooking — combining previous trial-and-error, general rules of thumb, and educated guess-work — you are able to quickly put together an almost perfect salad while still adhering to the CPG (albeit not optimally). In fact, your stomach might already know what kind of salad it craves before you even see what the bar offers; you only have to pick up the ingredients. In such cases, a low input-size (e.g., 10 ingredients) could be a curse rather than a blessing, since you find that a critical ingredient is missing. The main point is that, although problems might be intractable regarding some specific resource (e.g., time), or due to the choice of strategy (e.g., $\Phi$), it is hard to tell in a given situation whether an effective solution could be obtained via other means (e.g., using some different strategy or given more of a certain resource). Importantly, this leads to a distinction between the \emph{problem itself} (e.g., put together a tasty salad), and \emph{how} the problem is solved (e.g., follow $\Phi$). And while the distinction between problem and solution might be relatively clear in computational contexts (e.g., between problem and algorithmic solution), we will dedicate much effort in this paper to elucidate their difference in moral contexts.


\section{Computational Complexity of Ethics}
\label{sec:complexity}
What is the computational complexity of ethics? First, we should note that ``ethics'' is a multifaceted and equivocal concept that permeates many levels of analysis across different disciplines. Throughout the ages, moral philosophers have in more or less systematic ways tried to resolve questions regarding what is morally ``good'' and ``bad''. In modern times, Anglophone analytical ethics is conventionally divided into (i) \emph{applied ethics} (determining what is ``good'' and ``bad'' in particular instances), (ii) \emph{normative ethics} (advancing standards and principles of what is ``good'' and ``bad''), and (iii) \emph{meta-ethics} (determining the meaning and nature of morality). But the landscape of ethics stretches far beyond these divisions. From a biological point of view, it includes the evolutionary foundations of cooperation (as extensively studied in game theory \citep{axelrod1981,nowak2006five}), where morality can be viewed as an adaptive solution to the problem of competition among self-interested organisms,\footnote{Or alternatively put, the function of morality is to alleviate the failures of rationality \citep{ullmann2015emergence}.} from individual cells \citep{hummert2014evolutionary} to human beings \citep{Leben2018}. The landscape gets further complicated if we also consider the social, psychological, and cognitive dimensions, e.g., how ethical behavior is intertwined with the empathy, emotions, and reasoning of embodied agents, and carried out by highly distributed and parallel cognitive systems \citep{newen2018oxford,Feldman2015}. Far from being `fixed', moral behavior is something which is developed and actively refined through experience.\footnote{From the pioneering work of \cite{kohlberg1977moral}, through refinements by \cite{rest1999dit2}, moral psychology has grown into a mature paradigm that investigates the link between morality and cognitive development.} Beyond individuals, ethics is also manifested at the level of societies and culture; maintained and transformed through practices and institutions, mediated through the language of ideology and religion, and with justifications that ranges from divine authority (e.g., word of God), maintaining political order \citep{hobbes1651}, to the promotion of liberty \citep{mill1859} or justice \citep{Rawls1971}.


Hence, to delimit our investigation, we will focus on the complexity of ethical problems as they have been framed within the field of \emph{machine ethics}. The majority of technical work in machine ethics has been focusing on normative ethics, or more specifically, how certain tenets or aspects of a normative theory can be implemented so that an artificial agent acts in accordance with the theory \citep{Cervantes2020,Tolmeijer2020}. As such, it can be viewed as a form of \emph{applied} normative ethics, since it primarily centers on the practical implementation of a certain theory as opposed to discussions about what theory that should be. In their exhaustive survey of implementations, \cite{Tolmeijer2020} has suggested that approaches to moral machines can be characterized along three broad dimensions: ethical theory, implementation, and technology. The first dimension denotes the ethical theory used, which includes normative frameworks such as deontology \citep{Anderson2008ETHELTA,Malle2017,Shim2017}, consequentialism \citep{Abel2016,Armstrong2015MotivatedVS,Cloos2005}, virtue ethics \citep{Stenseke2021,Govindarajulu2019, Howard2017}, and hybrids \citep{Dehghani2008, Thornton2016}. The second dimension, following a division proposed by \cite{allen2005artificial}, considers \emph{how} ethics is implemented in the system, e.g., whether it is through a `bottom-up' learning process, carried out via `top-down' principles, or in a combination of both top-down and bottom-up processing. The technical dimension, in turn, considers the computational techniques used to realize the implementation, which include methods from various AI paradigms such as logical reasoning (e.g., inductive, deductive, and abductive logic), machine learning (e.g., neural networks, reinforcement learning, evolutionary computing), and probability (e.g., Bayesian and Markov models).

\subsection{The Complexity of Ethics Following Marr's Three-Level Analysis}
\label{sec:Marr3}

Based on these considerations, how can we frame the computational complexity of ethics for machines? Recalling the final lesson in the previous section, we first need to find some way of distinguishing \emph{problems} as such from \emph{how} these problems are solved. This distinction is reflected in the influential scheme proposed by \cite{marr1981vision}. Marr suggested that the information processing of a cognitive system can be explained on three distinct yet complementary levels of analysis: (i) \emph{Computational level}, (ii) \emph{Algorithmic level}, and (iii) \emph{Implementation level}. The computational level describes the problem itself (e.g., an input-output mapping), the algorithmic level specifies the algorithmic process (e.g., strategy or heuristic) that is performed to tackle the problem, and the implementation level specifies how the algorithmic process is realized by the physical hardware of the system (e.g., neurons or circuits). These levels can be illustrated using the salad bar example: (i) the computational level specifies the number of ingredients, value functions (e.g., tastiness of individual ingredients or combinations of ingredients), and desired output (maximally tasty salad); (ii) the algorithmic level describes the problem-solving process (such as $\Phi$); (iii) the implementation level describes the way a brain or machine implements the problem-solving process physically. Each of these levels of a system can be analyzed independently. For instance, since one and the same computational problem can be solved by a range of different algorithmic procedures, we can describe a cognitive system at the computational level independently of the algorithmic level, and thus have a computational-level theory of the computational system. Likewise, since an algorithm can be physically realized in a range of different systems — e.g., silicon or carbon — we might have an algorithmic-level theory of a cognitive system that does not require us to explain how it is physically implemented. Nevertheless, Marr argued that it is easier to elucidate the workings of a cognitive system through the top-down lens, i.e., by starting from the problem it solves as opposed to the precise mechanisms it uses to solve it \citep{marr1977artificial,marr1981vision}.\footnote{To clarify, this particular notion of ``top-down'', from computation (top), to algorithm, to implementation (bottom), is distinct from the common use in cognitive psychology, where ``bottom-up'' processing starts from the sensory input, and ``top-down'' processes centers around interpreting the incoming information based on knowledge, experience, and expectations.} The reason is that higher-level explanations make commitments about the lower-levels, which in turn forms a hierarchy of underdetermination. For instance, if we conjecture that a cognitive system solves problem $P$ at the computational level, we might be uncertain or agnostic with regards to the specific algorithm it employs to compute $P$. However, if our conjecture should carry any explanatory value beyond the computational level, we must commit to the idea that at least \emph{some} algorithm can compute $P$. If it can be proven that no such algorithm exists, then our problem is undecidable. Similarly, if we believe that a system solves $P$ using algorithm $A$, we commit to the idea that \emph{some} physical system can realize $A$.

\subsubsection{Level 1: Computational Problem}
\label{sec:level1computational}

How do we fit ethical problems into Marr's scheme? More precisely, what is the algorithmic level and what is the computational level of ethical problems posed by normative theory (NT)? First, we note that normative ethics blurs the line between Marr's first two levels. In particular, its prescriptive component is intimately linked with its action-guidance, i.e., by answering what \emph{is} good (e.g., adherence to moral duties), it tells you how to \emph{do} good (e.g., only perform actions that adhere to moral duties).\footnote{Normative theories that put less emphasis on actions might present an interesting exception; for instance, versions of virtue ethics that emphasize \emph{being} rather than \emph{doing}. However, rather than resolving the distinction, it only pushes it to the blur between \emph{flourishing} and the character traits that enable an agent to flourish. Furthermore, the idea that virtue ethics cannot offer action-guidance have also been criticized; see e.g., \cite{Hursthouse1999} for a virtue theoretic take on action-guidance.} In turn, this opens up a range of possible interpretations, and we will address three:

(1) \emph{NT as algorithmic-level solution to generalized morality} — In the most general sense, if the computational-level problem is phrased as ``do what is moral'', we might interpret an NT as an algorithmic-level solution to the computational problem ``how to be moral in general''. This interpretation would capture the generality ambition of NTs in human contexts (or at least in philosophical discourse on NT); that an NT should provide general answers or standards regarding right and wrong that are applicable to a range of particular instances. An agent that is committed to $NT_{1}$ would only be moral insofar as it adheres to $NT_{1}$ in its general behavior.\footnote{Or at least, the agent uses $NT_{1}$ as its main criterion to evaluate whether an action is moral.} Nevertheless, it is hard to see how one could feasibly frame such a broad interpretation in the formalism required by a computational complexity analysis; it would entail some form of general-purpose algorithm — e.g., in terms of a value, principle, or maxim — that provides solutions to all possible moral dilemmas.\footnote{The Golden Rule or Kant's categorical imperative \citep{kant1785} might be paradigmatic examples of such general-purpose algorithms, which we will discuss in section \ref{sec:deontology}.}

(2) \emph{NT as algorithmic-level solutions to specific moral problems} — A similar but more narrow interpretation is that NTs provide algorithmic-level strategies that can be used to solve \emph{specific} moral problems. This interpretation seems to, at least prima facie, capture everyday usage of the term ``moral dilemma'', i.e., a decision problem that arises as a conflict between two or more NTs (where ``NTs'' might as well be replaced with values, duties, virtues, or norms). We could, for instance, specify the computational-level problem as the trolley problem in order to draw attention to the conflict between action- and outcome-based NTs: is it morally right to save 5 people even if it involves actions that are intrinsically bad (e.g., murder)? Note, however, that the moral complexity (or undecidability) of such a problem does not reside in the computational-level problem itself, but rather in how the conflict between algorithmic-level solutions should be resolved (e.g., through the doctrine of double effect \citep{foot1967problem}). Regardless, the interpretation is still consistent with the view that different NTs could be employed to solve different problems, depending on the nature of the problem and the available resources. This seems to resonate with experimental studies that shows that humans are flexible with regard to the moral strategies they employ in different contexts \citep{capraro2018right,conway2013deontological,greene2008cognitive}. Intuitively, facing some ethical problem $E_{1}$, you might be reluctant to perform a certain action because you find the act immoral in itself (according to $NT_{1}$), while facing some other ethical problem $E_{2}$, no action seem immoral in itself, yet some actions lead to outcomes that seem more preferable than others (according to another theory, $NT_{2}$). That is, if no conflict arises between $NT_{1}$ and $NT_{2}$, you simply pick the one that is best suited for the computational-level problem at hand. Under this interpretation it would be possible to, at least in principle, assess whether some NT is more computationally efficient than another with regard to the same computational-level problem.\footnote{Similarly, even if we believed that only one NT is correct, that does not necessarily mean that we cannot find an alternative theory useful due to its computational efficiency (even if we generally dislike the theory from a moral standpoint).} However, could we ask whether it is more successful, morally speaking? It seems unlikely that an answer can be provided without resolving further meta-theoretical issues.\footnote{For instance, it is plausible that, while a solution provided by $NT_{1}$ could be the most computationally efficient, it could also violate some principle from $NT_{2}$, yet, no efficient solution exists for $NT_{2}$ (e.g., requires exponential time). The issue is thus: what is the most successful NT if we assume that the agent believes that $NT_{2}$ is morally superior to $NT_{1}$?} Perhaps more problematically, the interpretation seems to posit that ethical problems are, in some meaningful way, distinctly invariant from the ways they could be solved. To the contrary, the algorithmic solution (NT) seems to depend on the nature of the computational problem itself, and how it affords an algorithmic solution via some NT; affordances that are already embedded at the computational level. If a specific problem is only decidable or tractable for a particular NT, it thus seems more fair to treat it as a computational-level problem in its own right. For instance, we could imagine an ethical problem space which only contains information about obligations, and no information regarding outcomes; it is thus decidable for obligation-based NTs while being undecidable for outcome-based NTs. This naturally leads to an even more narrow interpretation, and the one we will primarily focus on in our analysis:

(3) \emph{Specific moral problems posed by NT as computational-level problems} — Instead of placing NTs at the algorithmic level, we could define specific computational-level problems as they are framed by a \emph{specific} NT. In turn, this allows us to be agnostic about the precise procedure that is carried out at the algorithmic-level: we only have to assume that such a procedure exists. As such, (3) provides a number of conveniences for machine ethicists, including (i) answers regarding what \emph{is} moral, or what is morally good to \emph{do} (as prescribed by the modeled theory), (ii) blueprints for action-guidance that can assist algorithmic design and choice of computational method(s), and (iii) means of evaluating performance (e.g., an apt deontological agent successfully adheres to moral duties and rules). Importantly, the narrowness of (3) allows one to ignore theoretical issues that plague (1) and (2): in contrast to (3), it does not have any generality ambition (and could thus be adopted to specific contexts or domains); in contrast to (2), the relevant action-guiding aspects of the modeled NT are already embedded at the computational-level problem description. I.e., while (3) accommodates the fact that different ethical problems — e.g., the information provided in a certain environment — give rise to different affordances with regard to ethical behavior, (2) does not. Perhaps most importantly, (3) allows us to analyze the algorithmic level of ethical problems, while (2) treats the normative theory as the algorithm itself, which potentially obscures the analysis of how such a procedure is actually carried out.

\begin{figure}
\centering
\includegraphics[width=0.99\textwidth]{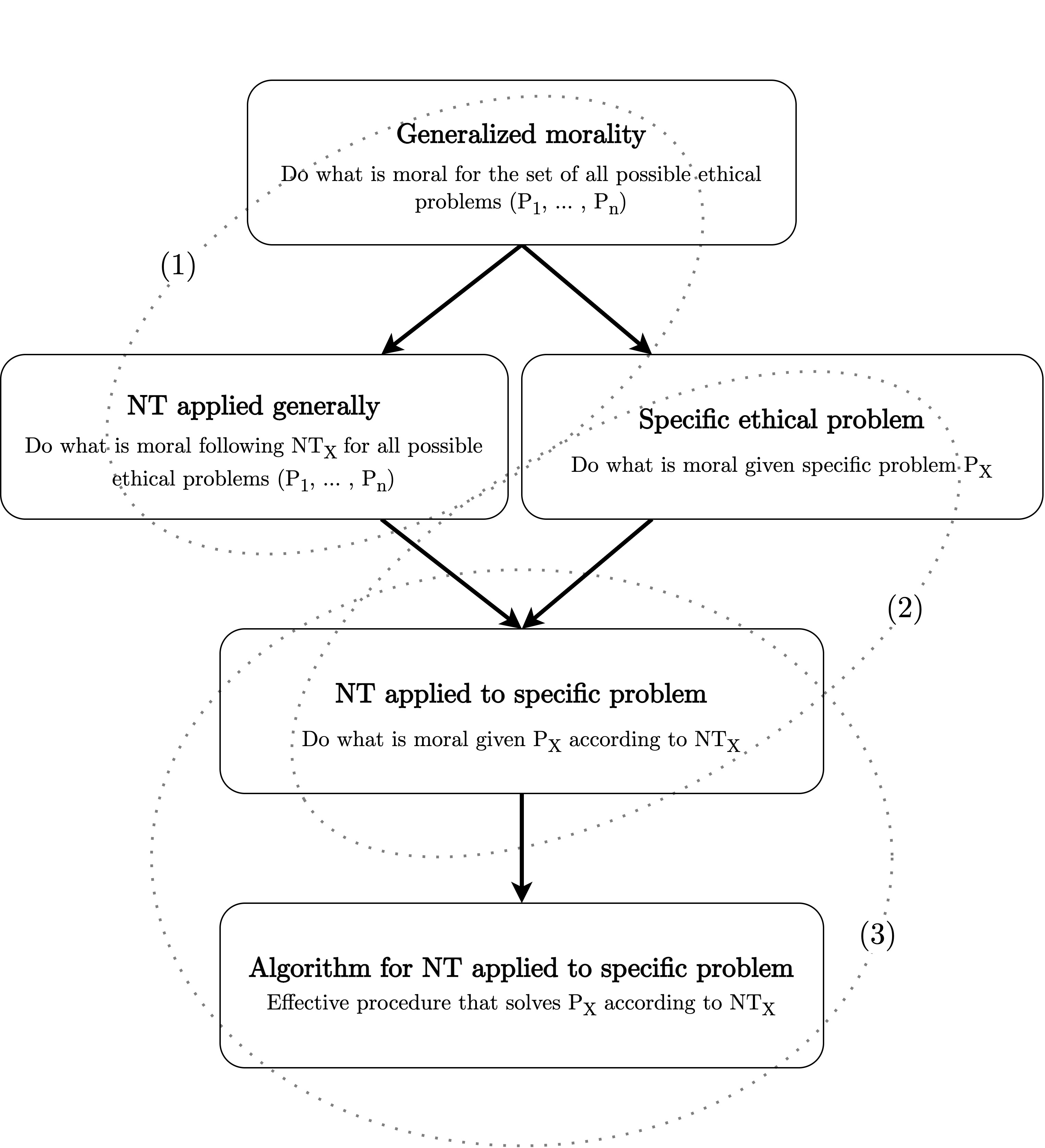}
\caption{Hierarchy of ethical computations. Arrows indicate dependency (i.e., A $\rightarrow$ B means that solutions to A depends on solutions to B). The dotted ellipses capture the computational (top) and algorithmic (bottom) level of the three interpretations (1)-(3).}
\label{fig:hierarchyofethics}
\end{figure}

Another principal strength with (3), with respect to a complexity analysis, is that it constitutes an \emph{essential} yet the \emph{least complex} aspect of ethical computing, in the sense that both (1) and (2) presuppose that an agent can perform computations of type (3). In other words, since interpretation (1) is a generalization of (2), which, in turn, depends on specific instantiations of (3), they form a hierarchy of ethical computations (illustrated in Figure \ref{fig:hierarchyofethics}). That is, to solve the generalized moral problem (interpretation 1) following some normative theory, e.g., $NT_{1}$, it requires that an agent can also apply $NT_{1}$ to specific moral problems (interpretation 2); to apply $NT_{1}$ in the particular (interpretation 2), it requires that an agent can apply $NT_{1}$ in the very way it is framed by $NT_{1}$ (interpretation 3). Thus, if some specific (3)-type computation is undecidable, it would follow that it is undecidable for type (1) and (2) computations of the same problem; it is undecidable for the NT in the specific case (2) and thus in the general case (1). The difference between the interpretations is further illustrated in Table \ref{table:1}.

\begin{table*}
\centering
    \begin{tabular}{  p{3.4cm} |  p{2.6cm} | p{2.6cm} | p{2.6cm}}
& \textbf{Computation} & \textbf{Algorithm} & \textbf{Implementation} \\
\textbf{Interpretation} & \small What is the problem?\par & \small What is the solution?\par & \small How is it implemented?\par
\\\hline
\small(1) NT as algorithmic solution to generalized morality\par
& \small Do what is moral (general behavior)\par
& \small NT$_X$ (applied generally)\par
& \small Mind/machine\par\\\hline
\small(2) NT or NTs as algorithmic solution(s) to specific moral problems\par
& \small Specific moral problem $P$\par     
& \small NT$_X$ (applied to $P$)\par
& \small Mind/machine \par
\\\hline
\small(3) Specific moral problem $P$ posed by specific NT$_X$\par
& \small $P$ as framed by NT$_X$\par
& \small Computational methods\par
& \small Mind/machine\par
\\\hline
\end{tabular}
\caption{Three interpretations on how ethical problems under normative theory can be framed and analyzed within Marr's three levels.\label{table:1}}
\end{table*}

\subsubsection{Level 2: Algorithm}
\label{sec:level2algorithm}

Thus, we believe that a natural way to analyze the computational complexity of ethics is to focus on problems posed by normative theory (computational level), that are solved through a variety of computational methods (algorithmic level), by a deterministic Turing Machine (implementation level). Of course, this still leaves a rather vast interpretative leeway regarding what goes on at the algorithmic and implementation level. To find the most effective algorithmic solution to a well-defined problem is often an empirical question, and answers are continuously revised in light of new advancements in programming techniques (e.g., breaking down a problem into simpler sub-problems through dynamic programming) or heuristics (e.g., exploiting regularities in the problem). More importantly, it also depends on what we accept as a solution. If we believe that the NT strictly dictates that the system should find the optimal solution to a problem, it entails that the algorithmic level should follow some procedure that is guaranteed to produce an optimal solution; a so-called \emph{exact} algorithm.\footnote{For instance, this is analogous to the way $\Phi$ produces an optimal (albeit intractable) solution to the salad problem following the CPG.} A less strict interpretation is to accept solutions that are ``close enough'' to the optimal; so-called \emph{approximate} algorithms. Although approximate algorithms are not guaranteed to find an optimal solution, they guarantee that the solution is within some fixed distance to the optimal one (i.e., there is a provable bound on the ratio between the optimal and approximated solution). The difference between approximate and exact makes all the difference with regard to tractability, since many problems that have intractable exact solutions can be approximated in polynomial time \citep{williamson2011design}.

We will mainly focus on exact solutions for two interrelated reasons: (i) it is prescribed by the normative ideal (following the strict interpretation), and (ii) it allows us to focus on problems as opposed to algorithms. The first reason can be supported by the following consideration: if approximate solutions are acceptable, how can we motivate that a solution is within an acceptable distance to the optimal? Note that, although an approximation yields a provable guarantee of the distance, this distance can still be arbitrarily large.\footnote{We can, for instance, imagine a dilemma where the optimal solution has a moral value of 100, but the best approximation only yields a value of 50.} It seems as if we then need to also define what an acceptable distance is, which might vary greatly from case to case. Furthermore, many real-world problems exhibit no identifiable structure that can be exploited, and as such, they yield no efficient approximation algorithms \citep{Nievergelt1995}. This naturally leads to the second reason, which is simply that it is easier to compare exact as opposed to approximate solutions, as we do not need to define the conditions under which an approximation is sufficiently close.

Of course, moral theorists might rightfully point out that we should not be interested in exact or optimal solutions to moral problems, but rather, we should understand them in terms of what is ``permissible'' or ``impermissible''. For instance, an action might be permissible even if it is suboptimal, and morality does not require us to do anything more than what is permissible, as long as we avoid what is impermissible. This line of reasoning might, in turn, serve to justify the use of suboptimal approximations. However, this would obscure the difference between optimality as a mathematical concept and as a moral concept. Moral permissibility could, for instance, be construed \emph{as} the mathematical optimal; i.e., some fixed point or metric to evaluate behavior against. Alternatively, moral permissibility could be construed as a mathematical approximation of some fixed notion of moral optimality. But then, again, we are led back to the same dilemma we wanted to avoid: in each case, we need to justify how a given approximation is acceptable given a certain threshold of moral permissibility. It is important to note, though, that this does not exclude the possibility that such approximations can be justified in relation to permissibility in particular contexts, but rather that such an analysis is beyond the scope of this paper.

\subsubsection{Level 3: Implementation}
\label{sec:level3implementation}

On the level of implementation, we will adopt the most widely used model of computation: the Turing Machine (TM) \citep{turing1936computable}. More specifically, since a TM is a mathematical model of computation, it denotes any physical system that can realize a TM (i.e., it is Turing complete). Turing claimed that every function that can be computed by an algorithm can be computed by a TM. The thesis gained credence when Turing showed how his notion of computability was equivalent to the independently suggested proposal by \cite{church1936note}. This forms the basis for the Church-Turing thesis, which in turn has been shown to be equivalent to many other forms of computation \citep{herken1995universal}. Simply put, it means that any general-purpose system (e.g., computer or computer language) can simulate the computational aspects of any other general-purpose system. We will also assume that P $\neq$ NP (discussed in section \ref{sec:salad}). Like the Church-Turing thesis, it is another widely accepted conjecture among computer scientists, even if it remains to be proven.\footnote{Note that the Church-Turing thesis is not a conjecture in the mathematical sense, but rather a hypothesis about the nature of computation; it cannot be proven since its notion of effective calculability is defined informally.}

Importantly, if we can show that a problem is NP-hard, it means that we cannot expect to find an efficient solution to it, where ``efficient'' means ``solvable in polynomial time for a deterministic TM'' (P-tractability).\footnote{P $\neq$ NP entails that we cannot expect to find effective solutions to NP-complete problems, and NP-hard problems which can be translated to NP-complete decision variants. Note that many NP-hard problems would still remain intractable even if P $=$ NP, e.g., if they are complete for complexity classes that are believed to encompass NP (e.g., PSPACE or EXPTIME).} Therefore, if ethical problems solved by computational methods are NP-hard, we cannot expect computational systems to solve them efficiently, and as such, it would yield direct consequences for the feasibility of moral machines. However, even if P $\neq$ NP and the Church-Turing thesis have near-universal acceptance, it is crucial to address a few caveats regarding the limitations and relevance for the notion of P-tractability. For instance, P-intractability is of no major concern if it is guaranteed that the input size remains sufficiently small (e.g., a salad bar with 5 ingredients only yields 32 possible combinations). Importantly, simply because a problem is P-intractable, it does not mean that it cannot be solved effectively under other reasonable conceptions of tractability. In fact, many NP-hard problems can be solved by algorithms whose runtime is superpolynomial in only \emph{some} part of its input (input parameter), while the runtime is polynomial in the overall input size.\footnote{It is this very observation that has motivated the development of parameterized complexity \citep{downey2012parameterized}, and the class of fixed-parameter tractable problems (FPT). See also \cite{fellows2002parameterized} and \cite{niedermeier2006invitation}.} Conversely, large constants in polynomial functions, e.g., $n^{100}$, are P-tractable even if they might fail to capture any intuitive notion of ``effective''. Furthermore, time consumption might also be significantly reduced with alternative models of computation, e.g., utilizing parallelization, random access memory, or quantum computing. While the Invariance Thesis — along with the closely related extended Church-Turing thesis \citep{kaye2006introduction,bernstein1997quantum} — states that no machine can be super-polynomially faster than a deterministic TM,\footnote{Note that it is generally believed that the Invariance Thesis applies to both parallel and serial models of computation, see, e.g., \cite{frixione2001tractable,parberry1994circuit,tsotsos1990analyzing}.} it remains to be seen whether and to what extent it can be falsified in light of future advancements in computing.\footnote{E.g., it is not unlikely that \emph{some} future computer could at least solve \emph{some} problems in polynomial time that are currently intractable.} The main point is that, although P-tractability constitutes an indispensable tool for the formal study of effective computing in theory and practice, it should not be interpreted as drawing a definitive line, across the board, between what is tractable and what is not. And while P-tractability has direct consequences for moral machines, a related yet even more convoluted question is whether it could provide any relevant insight into the moral cognition of humans (a question we will return to in section \ref{sec:discussion}).

To divide our problem space, we will focus on three types of moral machines: causal engines (section \ref{sec:consequentialism}), rule-followers (section \ref{sec:deontology}), and moral learners (section \ref{sec:learning}). The main reason is that nearly all implementations in machine ethics take one of these approaches \citep{Tolmeijer2020}. Another reason is that these types each correspond to a prominent normative framework: consequentialism is about predicting future events (causal engines), deontology is about adhering to rules or duties, and virtue ethics emphasizes learning.\footnote{Of course, as we will see later on, in many cases the line between these theories and types become blurry.} In order to be subject to a complexity analysis, we will also assume that ethical problems can be cast as well-defined computational problems, which means that they have clearly defined initial conditions and goals (e.g., in terms of specific input and output conditions) which can be represented through conventional data types (e.g., booleans, integers, floating points) and structures (e.g., arrays, lists, graphs, trees). While these simplifying conditions might do little justice to the vastly rich and potentially ill-defined problems agents might face in the real world, it can be motivated by the fact that real-world ethical problems, given that they are decidable at all, are at least as rich in information as their simplified computational counterpart. In technical terms, we assume that well-defined computational problems represent a reasonable lower-bound on the information-theoretic nature of ethical problems in real-world environments. Finally, we will mainly focus on time rather than space complexity for the simple reason that accessing and storing memory consumes time, which means that memory consumption is often upperbounded by time consumption \citep{garey1979computers}.

\section{Consequentialism and Causal Engines}
\label{sec:consequentialism}
Consequentialism is a family of normative theories that puts outcomes at the center of moral evaluation. While all consequentialists agree on the moral importance of outcomes, they might disagree on what a good outcome is, or alternatively, what \emph{makes} an outcome good. For instance, utilitarianism — arguably the most influential branch of consequentialist theories — prescribes actions that maximize utility, where utility can be understood as the overall well-being of the individuals affected \citep{bentham1789,mill1861utilitarianism}, satisfaction of their preferences \citep{singer2011}, reduction of their suffering \citep{smart1956extreme}, or the well-fare of their state \citep{sen1979utilitarianism}. There are also many nuances regarding the way outcomes are morally important, e.g., whether intended consequences matter (as opposed to only actual consequences), whether they depend on the perspective of the acting agent (i.e., agent-relative as opposed to agent-neutral), whether indirect consequences matter (as opposed to the direct consequences of the act itself), for whom they matter (e.g., a limited set of individuals or all sentient beings on earth), and for how long (e.g., only immediate outcomes or for all eternity) \citep{sep-consequentialism}. Nevertheless, what is common to all forms is the commitment to the moral value of future events. Therefore, any agent — artificial or biological — committed to consequentialism must be able to make predictions about the future, insofar as they are committed to carrying out the prescriptions of the theory in practice. This is why successful consequentialist agents rely on so called ``causal engines'', a term we use to broadly refer to the information processing that supports causal cognition.

Note that, in some way or another, most biological organisms care about the consequences of their actions, as it greatly increases their chance of survival. Intuitively, causal cognition appears to be critical for many essential capabilities such as avoiding harm, problem-solving, and planning. Experimental results indicate that human children, as young as eight months, can make inferences based on cause and effect \citep{sobel2006blickets}. This might suggest that some form of pre-reflective capacity for causal inference could be deeply engraved in our very biological being, reflecting the predictive processing that many believe to be \emph{the} central function of nervous systems \citep{friston2010free,hohwy2013predictive,keller2018predictive}. However, unlike biological organisms, machines did not develop causal engines through an evolutionary process. Instead, an artificial system's ability to follow consequentialism relies on computational techniques, often stemming from the families of statistical, Bayesian, and Markovian modeling \citep{casella2021statistical}. It is also common to view machine learning methods as a form of ``predictive analytics'' in the sense that algorithms learn to make better predictions based on experience; e.g., in supervised learning via human-generated data, in reinforcement learning through an interactive process of trial-and-error. But consequentialism is not solely about making predictions about the future. It is also about evaluating, from the set of possible outcomes, what outcomes are morally preferable over others. That is, even if a consequentialist agent could predict the outcomes of all possible actions with godlike accuracy and speed, it does not necessarily mean that it can easily decide, with the same speed, which the optimal outcome is.

In light of these considerations, this section will explore the computational complexity of three general types of consequentialist problems: combinatorics of determining the optimal outcome (\ref{sec:combinatorics}), causal inference (\ref{sec:causalinference}), and decisions in dynamic and partially observable environments under different time horizons (\ref{sec:dynamicenvironments}). The section is written so as to incrementally introduce uninitiated readers to time complexity analysis, probability theory (Bayesian Networks), and stochastic methods (Markov Decision Processes).

\subsection{The Combinatorics of Outcomes}
\label{sec:combinatorics}

In the most simplified case, we could think of the problem a consequentialist face when they compare the moral value of different outcomes, given that the agent can already determine what these outcomes are. In this way, we can ignore the complexity of the causal inference itself so as to isolate the problem of optimal outcome evaluation. In complexity theoretical terms, we assume that the agent has access to a so-called \emph{oracle machine}, which is able to provide answers regarding causal events in a single operation. For instance, if the agent asks ``what happens if I perform action $a$?'', the oracle gives an answer of the type ``action $a$ yields an outcome with a moral value of $v$''.\footnote{To encompass many versions of utilitarianism, we will remain agnostic about the exact nature of the utility that ought to be maximized; the only important thing is that it can be represented as a numerical value.} The most trivial computational problem of this kind can be formalized in the following way:

\begin{center}
    \textsc{C1 — Optimal outcome following consequentialism}
\end{center}
\begin{quote}
\textbf{Input:} An environment as a set $E = \{a_{1}, a_{2}, ..., a_{n}\}$ of $n$ possible actions and a value function $v$ that assigns an outcome value to each action $a \in E$.

\textbf{Output:} An action $a \in E$ such that $v(a)$ is maximized over all possible actions in $E$.
\end{quote}

An optimal solution can be guaranteed by the following generic exhaustive-search algorithm:

\begin{algorithm}[H]
$h \gets 0$ // outcome value (0 set as default)\\
$j \gets 0$ // index of action (0 set as default)\\
 \For{$a_{i} = a_{1}$ \textbf{\upshape{to}} $a_{n} \in E$}{
 $h_{i} \gets v(a_{i})$ // call oracle\\
  \uIf{$h_{i} > h$}{
   $h \gets h_{i}$ // update highest value\\
   $j \gets i$ // update index of highest value\\
   }
 \textbf{end if}
   }
 \textbf{end for}\\
  \textbf{return} $j$ // return index of action with highest outcome value
 \caption{Exhaustive search with causal oracle}
\end{algorithm}

In short, the algorithm initializes default values for outcomes (step 1) and the index of actions (step 2). It then loops through each action in the environment (step 3), calls the oracle (step 4), checks if the outcome of that action is higher than the current highest (step 5), and if so, updates the highest outcome value (step 6) and its index (step 7). Finally, it halts after returning the index of the highest outcome (step 10). If we assume that each instruction requires an equal amount of time ($1$) to be executed, we can count the precise number of machine operations the algorithm needs to solve \textsc{C1} in the following way: lines 1, 2, and 10 needs to be executed just once (3), lines 3-7 needs to be executed $n$ times each ($5n$), and 8 and 9 can be ignored (as they are flow control statements), which yields a total of $3 + 5n$. In Big O, this collapses into $O(n)$. In other words, the time complexity grows linearly ($O(n)$) to the size of the input. Importantly, regardless of how fast a machine can execute the other instructions, to ensure optimality, it must ask a number of questions to the oracle which is at least equal to the number of possible actions. I.e., if there are 10 actions, the agents must make, at minimum, 10 calls to the oracle.

What happens if we allow for multiple values? For instance, we could assume that the agent has a set of two or more outcome values that needs to be checked for each action-outcome (e.g., pleasure, fairness, trust, etc.). This yields the following problem:

\begin{center}
    \textsc{C2 — Optimal combination of values}
\end{center}
\begin{quote}
\textbf{Input:} Same as \textsc{C1} with the addition of a set of outcome value functions $V = \{v_{1}, v_{2}, ..., v_{i}\}$ assigned to each $a \in E$.

\textbf{Output:} An action $a$ such that $v(a)$ is maximized over all $v \in V$ and $a \in E$.
\end{quote}

If we posit that the values interact trivially, in the sense that values can be summarized $v_{1}(a) + v_{2}(a) + ... + v_{i}(a)$ to yield a single total value $V(a)$ (i.e., obeying the law of additivity), the optimal action $a^*$ can be formally expressed as:
\begin{equation}
a^* := \argmax_{a \in E} V(a) := \{a \in E : \sum_{m=1}^{i} v_{m}(\acute{a}) \le \sum_{m=1}^{i} v_{m}(a) \: \text{for all} \: \acute{a} \in E \}
\end{equation}
If the agent needs to make distinct calls to the oracle for each value, the time complexity is the product of $n$ (number of actions) and $i$ (number of values), yielding $O(ni)$. If $i$ is equal to the number of actions, the runtime grows quadratically in relation to $n$, which still yields the polynomial $O(n^{2})$.\footnote{To show this result in an algorithmic procedure, we can simply extend the exhaustive search (Algorithm 1) to iterate $n$ actions over $i$ values, e.g., by adding one additional \textbf{for}-loop for each $i$, or as a nested loop over values 1 to $i$ within the loop over actions 1 to $n$.}

We have so far only been focusing on the moral evaluation of a single action. But in ethical decision problems of the real world, it is possible to perform multiple actions. However, as illustrated in the salad example, the possibility of combining actions can present tractability issues that are inherent to permutations of combinatorial structures. To show how this affects the computational complexity of consequentialism,\footnote{See \cite{Lindner2020} for a similar analysis of action plans based on the SAS$^+$ formalism.} we define an action plan $\varphi = \{a_{1}, a_{2}, ..., a_{n}\}$ as a distinct non-empty set of $n$ actions presented by the environment such that $\varphi \subseteq E$. We then augment \textsc{C1} to describe the following problem:


\begin{center}
    \textsc{C3 — Optimal plan of up to two distinct actions}
\end{center}
\begin{quote}
\textbf{Input:} An environment as a set $E = \{a_{1}, a_{2}, ..., a_{n}\}$ of $n$ possible actions and a function $v$ that assigns an outcome value to each action plan $\varphi \subseteq E$.

\textbf{Output:} An action plan $\varphi$ such that $v(\varphi)$ is maximized over all $\varphi \subseteq E$, no $a \in \varphi$ is identical to itself (i.e., the same action cannot be performed more than once), and $\lvert \varphi \rvert \le 2$.
\end{quote}

The only way to solve \textsc{C3} is to make a number of calls to the oracle which is equal to the number of possible action plans (with a maximum of two actions). This number will grow triangularly ($\frac{n(n+1)}{2}$) — i.e., half of a square — with the number of actions.\footnote{As the famous story goes, Carl Friedrich Gauss quickly identified the formula for this series at a young age when asked by his teacher to add all the numbers between 1 and 100.} This tractable procedure would satisfy the Combinatorial Principle of Actions (CPO), i.e., that action plans yield outcome values that does not necessarily correspond to the sum of its individual actions if performed in isolation. It would, however, violate a fundamental principle of causality: that the resulting outcome of two causal events, action $x$ and action $y$, depends on the order in which $x$ and $y$ occurs. We can call this the Principle of Causal Order (PCO). In order to satisfy PCO when solving \textsc{C3}, the consequentialist must make an additional triangle of calls to the oracle, which completes the quadratic growth of $n(n-1)$. In asymptotic Big O, however, solving \textsc{C3} in either way results in a time complexity of $O(n^{2})$, which is still comfortably within tractable bounds.

The computational complexity of action-outcomes becomes an issue for the consequentialist when we generalize problems of type \textsc{C3}, e.g., to account for $n$ number of actions:

\begin{center}
    \textsc{C4 — Optimal plan of up to $n$ distinct actions}
\end{center}
\begin{quote}
\textbf{Input:} Same as \textsc{C3}.
\textbf{Output:} An action plan $\varphi$ such that $v(\varphi)$ is maximized over all $\varphi \subseteq E$, no $a \in \varphi$ is identical to itself, and $\lvert \varphi \rvert \le n$.
\end{quote}

The time complexity of an exact algorithm that solves \textsc{C4} while adhering to the CPO is $O(2^{n})$. In other words, there is no polynomial-time tractable procedure for consequentialists who try to solve problems of type \textsc{C4}.\footnote{This is analogous to the salad problem following $\Phi$ (where the CPO is exchanged for the equivalent CPG).} Worse still, if the consequentialist should also adhere to the PCO, an exact algorithm would yield the factorial growth of $O(n!)$ \citep{oeisFACTORIAL}.\footnote{This series is called ``the number of permutations of nonempty subsets of $\{1,...,n\}$'' \citep{oeisFACTORIAL}, and can be expressed as the floor function of $en! - 1$, where $e$ denotes Euler's number.} More broadly, it is well-known that many planning tasks are PSPACE-complete \citep{bylander1991complexity,bylander1994computational,littman1998computational}.\footnote{See also \cite{backstrom1995complexity} for some tractable results for SAS+ planning.}

Note that, while this intractability might not constitute a detrimental issue in practice — e.g., for small inputs, say, four possible actions, solving \textsc{C4} following CPO requires 15 calls, whereas following CPO and PCO requires 64 — \textsc{C4} still presupposes a large set of other non-trivial assumptions that might not hold in real-world situations. For instance, it assumes that agents cannot perform the same action more than once, and that the problem space remains static while the agent computes the solution. By contrast, real-world environments might present a potentially infinite set of possible actions in a state space which is only partially observable and changes in continuous time (which we will return to in sec. \ref{sec:dynamicenvironments}). Above all, the agent cannot make any calls to a causal oracle but needs to rely on its own causal engine; which leads us to the complexity of causal inference.

\subsection{Causal Inference}
\label{sec:causalinference}

As soon as we enter the realm of uncertainty, we cannot guarantee that the performance of any system will be optimal. Instead, the best we can hope for is optimal according to our ``best guesses'', i.e., in virtue of what we believe or know (Bayesian optimality). This is part of the reason why many versions of utilitarianism are revised so as to stress the maximization of ``expected'' as opposed to actual utility \citep{broome1987utilitarianism}. It also forms the basis for the expected utility hypothesis \citep{von1947theory}, which is widely used in decision theory and economics to model rational choice, preferences, and risk appetite (i.e., openness and aversion to risk) when payoffs are unknown.\footnote{Specifically, the hypothesis states that an agent chooses between alternatives by comparing expected utility values, which is commonly calculated as the weighted sum of utility values $U$ multiplied by their probabilities $P$, in the sense that $\textstyle \sum U(x_{i})P_{i}$.} Note, however, that different ways to model probability leaves room for interpretations that carry moral weight, in the sense that different normative principles can guide how decisions under uncertainty should be tackled.

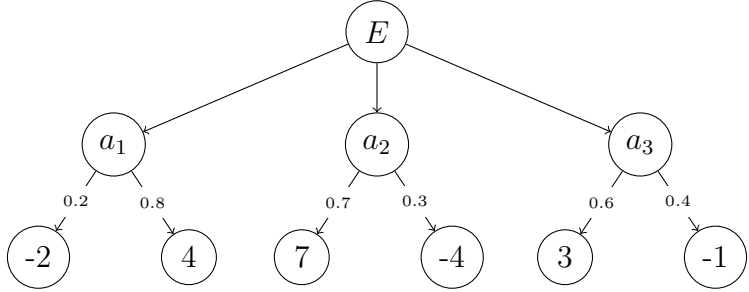
\begin{figure}
\begin{center}
    \textsc{C5 — Optimal action under uncertainty}
\begin{tikzpicture}[level distance=1.5cm,
level 1/.style={sibling distance=3.5cm},
->,
level 2/.style={sibling distance=2cm},
tree node/.style={circle,draw},
every child node/.style={tree node}]
\node[tree node] (Root) {$E$}
    child {
    node {$a_{1}$} 
    child { node {-2} edge from parent node[fill=white] {\tiny{$0.2$}} }
    child { node {4} edge from parent node[fill=white] {\tiny{$0.8$}} }
}
child {
    node {$a_{2}$}
    child { node {7} edge from parent node[fill=white] {\tiny{$0.7$}} }
    child { node {-4} edge from parent node[fill=white] {\tiny{$0.3$}} }
    }
    child {
    node {$a_{3}$}
    child { node {3} edge from parent node[fill=white] {\tiny{$0.6$}} }
    child { node {-1} edge from parent node[fill=white] {\tiny{$0.4$}} }
};
\end{tikzpicture}
\end{center}
    \caption{A directed acyclic graph (DAG), representing a decision problem under uncertainty.}
    \label{fig:DAG1}
\end{figure}

This is illustrated in the following problem, represented as a directed acyclic graph (Figure \ref{fig:DAG1}). The graph shows an environment with three actions, each with a probability of yielding one out of two possible outcome values. If we simply want to maximize expected utility regardless of risk, we can simply add the product of each outcome value with their respective probability — e.g., $0.2(-2) + 0.8(4)$ for $a_{1}$ — and select the action with the highest expected utility. Alternatively, a more risk averse option would be to select the action with the \emph{best} worst-case outcome (a decision rule called ``minmax'', i.e., maximizing the minimum gain). While these two decision-procedures make little difference with regards to runtime — like our solution to \textsc{C2}, both take $O(no)$ time, where $o$ refers to the number of outcomes for each action — they make a significant moral difference.

However, like \textsc{C1-C4}, \textsc{C5} still assumes some sort of Bayesian oracle, which is able to infer the exact posterior probabilities that certain events (outcomes) will occur given certain causes (actions). More broadly, causal inference can be understood as the ability to identify \emph{what causes what}, e.g., ``what is the cause (or causes) of phenomenon $X$?'', ``what is the effect (or effects) of $Y$?'', and ``what is the causal relationship between $X$ and $Y$?''. None of these questions are trivial; indeed, scientific endeavors are to a large extent driven by answering causal question through a combination of carefully collected data, a vast set of statistical modeling techniques, and causal reasoning capacities such as deductive (deducing from given premises), inductive (inferring from observations), and abductive reasoning (inference to the best explanation).

One essential aspect of causal inference is to determine posterior probabilities based on prior knowledge, i.e., to the determine the likelyhood of $A$ given evidence or belief $B$. In statistical modeling, the Bayesian interpretation of probability offers a popular response to this challenge. Bayesian methods — e.g., Bayesian inference, networks, and statistics — are all based on Thomas Bayes' theorem, which states that the probability of $A$ given $B$ is provided by the equation $\Pr(A|B)=\frac{\Pr(B|A)\Pr(A)}{\Pr(B)}$.\footnote{Applying Bayes' rule to determine the likelyhood of some causal event $A$ given $B$ is a trivial procedure given that we have (i) some prior probability that $A$ (before $B$) $P(A)$, (ii) some estimated evidence for the probability that $P(B)$ without $A$ (and $P(B) \neq 0$), and (iii) an estimate of the converse likelihood that $B$ happens given that $A$.} Bayesian modeling have been used to address and model a vast range of cognitive phenomena, such as motor control \citep{kording2006bayesian}, symbolic reasoning \citep{oaksford2001probabilistic}, animal learning \citep{courville2006bayesian}, causal learning and inference \citep{steyvers2003inferring, griffiths2005structure}, inductive learning \citep{tenenbaum2006theory}, goal inference \citep{baker2007goal}, and consciousness \citep{lau2007higher}.

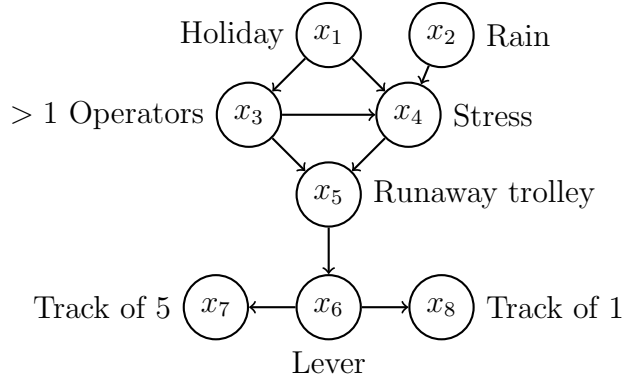
\begin{figure}
\begin{center}
\begin{tikzpicture}[node distance={15mm}, thick, main/.style = {draw, circle}]
\node[main,label=left:{Holiday}] (1) {$x_1$}; 
\node[main,label=right:{Rain}] (2) [right of=1] {$x_2$};
\node[main,label=left:{$>1$ Operators}] (3) [below left of=1] {$x_3$};
\node[main,label=right:{Stress}] (4) [below right of=1] {$x_4$};
\node[main,label=right:{Runaway trolley}] (5) [below left of=4] {$x_5$};
\node[main,label=below:{Lever}] (6) [below of=5] {$x_6$};
\node[main,label=left:{Track of 5}] (7) [left of=6] {$x_7$};
\node[main,label=right:{Track of 1}] (8) [right of=6] {$x_8$};
\draw[->] (1) -- (3);
\draw[->] (1) -- (4);
\draw[->] (2) -- (4);
\draw[->] (3) -- (5);
\draw[->] (3) -- (4);
\draw[->] (4) -- (5);
\draw[->] (5) -- (6);
\draw[->] (6) -- (7);
\draw[->] (6) -- (8);
\end{tikzpicture}
\end{center}
    \caption{A Bayesian network representing the causal relationships of eight Boolean variables for the Bayesian Trolley Problem (\textsc{C6}). Since their introduction in the 1980's, Bayesian networks have facilitated evidence-based prediction in complex domains such as medical diagnosis \citep{lucas2000probabilistic} and weather forecasting \citep{cofino2002bayesian}.}
    \label{fig:bayesianNet}
\end{figure}

Among the most powerful and widely used extensions of Bayes theorem is the construction of graphical models, called Bayesian networks (BNs) \citep{pearl1985bayesian}, which can succinctly represent a large set of variables and their conditional dependencies as a single DAG (Figure \ref{fig:bayesianNet}). BNs have been particularly useful in addressing the learning of causal relationships in humans \citep{l2008bayesian}. While the nodes of a BN represent Bayesian variables of interest — e.g., hypotheses, observable quantities, occurences of events, features of objects — the links (or edges) represent conditional dependencies between the variables. Each node has a probability function that returns a variable depending on its parent variables (following Bayes' theorem), and nodes that are not connected are conditionally independent of each other. For instance, the BN illustrated in Figure \ref{fig:bayesianNet} describes the causal relationships between eight variables: whether it is a public holiday ($x_{1}$), whether it is raining ($x_{2}$), whether two or more train operators are currently working at the train station ($x_{3}$), whether the operators are stressed ($x_{4}$), whether there is a runaway trolley ($x_{5}$), whether a lever is pulled ($x_{6}$), and whether the trolley is on course to collide with 5 ($x_{7}$) or 1 ($x_{8}$) people. Since BNs supports the inference of probabilities for any possible subset of variables (i.e., on the basis of evidence about those subsets), it can be used to support causal reasoning processes in any direction of the network. Using the chain rule of probability,\footnote{The chain rule allows one to compute the joint distribution of a set of variables solely using conditional probabilities, in the sense that $P(A \cap B) = P(B|A)P(A)$.} the joint probability — i.e., the probability distribution on all possible combinations of values — is given by:

\begin{equation}
P(x_{1}, ..., x_{n} = \prod_{i} P(x_{i} | \psi_{i})
\end{equation}

where $\psi_{i}$ denotes the values for the parent nodes of $x_{i}$. The joint distribution for the network in Figure \ref{fig:bayesianNet} is therefore: $P(x_{1},...,x_{8}) =$
\begin{equation}
P(x_{1}) P(x_{2}) P(x_{3}|x_{1}) P(x_{4}|x_{1},x_{2},x_{3}) P(x_{5}|x_{3},x_{4}) P(x_{6}|x_{5}) P(x_{7}|x_{6}) P(x_{8}|x_{6})
\end{equation}

We can now describe a range of Bayesian inference problems for consequentialism, such as:

\begin{center}
    \textsc{C6 — Bayesian Trolley Problems}
\end{center}
\begin{enumerate}[label=(\alph*)]
  \item Likelihood — what is the probability $P$ that $x_{7}$ is true? (given full, partial, or no evidence about its parent variables)
  \item Conditional probability — what is the probability that $x_{7}$ is true given evidence that it is a public holiday ($x_{1} = true$)?
  \item Causal reasoning — e.g., what effect does pulling the lever ($x_{6} = true$) have on $x_{7}$ or $x_{8}$?
  \item Most probable explanation (MPE) — what is the most probable configuration of a \emph{set} of variables given \emph{full} evidence about the complement of that set?
  \item Maximum a posteriori hypothesis (MAP) — what is the most probable configuration of a \emph{set} of variables given \emph{partial} evidence about the complementing set?\footnote{MAP (sometimes called Partial or Marginal MAP) can be viewed as the generalization of MPE in the sense that it might require a marginalization of the variables that are not observed nor explained. Furthermore, note also that both MPE and MAP has many variants in the literature on Bayesian networks: see \citep{kwisthout2011most} for a clarification.}
\end{enumerate}

BNs are perfectly suited to answer such causal inquiries, using algorithms such as variable elimination \citep{zhang1996exploiting} and message-passing \citep{pearl1982reverend} for exact inference, and random sampling \citep{pearl1987evidential} for approximate inference. However, even if questions like \textsc{C6} (a)—(e) can be solved in reasonable time for constrained networks, it has been proven that most inference problems for Bayesian Networks are intractable in general. More specifically, exact inference on arbitrary graphs is NP-hard \citep{cooper1990computational},\footnote{More precisely, exact inference is \#P-complete \citep{roth1996hardness}, where \#P is the class of counting problems associated with NP.} which means that inferring the exact probability of some event (or that a propositional expression is true) is at least as hard as the hardest problems in class NP. Furthermore, the decision variant of finding the most probable explanation (MPE) is NP-complete \citep{shimony1994finding}, while the related maximum a posteriori hypothesis (partial MAP) is NP$^{\text{PP}}$-complete \citep{park2004complexity}.\footnote{PP is the class of problems decidable by Probabilistic TM with an error probability of less than $1/2$ \citep{gill1977computational}, and NP$^{\text{PP}}$ is the class of problems solvable by a non-deterministic TM with access to an oracle for problems in PP.} Perhaps more intriguing is the results that approximations of these problems are also intractable: approximating exact inference \citep{dagum1993approximating}, MPE \citep{abdelbar1998approximating}, and partial MAP \citep{park2004complexity} are all NP-hard.\footnote{See \citep{kwisthout2011most} and \citep{de2020almost} for summaries of complexity results for the many variants of the MPE and MAP problems.}

One important lesson from these results is that the complexity of Bayesian inference depends on the \emph{structure} of the network: while constrained graphs yield a bound on the number of conditional dependencies and parent variables for each node, unconstrained graphs cannot be exploited for effective computation. For instance, for chain-like graphs of the type $x_{1}\rightarrow x_{2}\rightarrow x_{3}\rightarrow x_{4}\rightarrow x_{5}$, an elimination algorithm can determine the exact inference of $P(x_{5})$ by a step-wise elimination of the parent variables, which can be computed in the polynomial time $O(nv^{2})$, where $n$ is the number of variables and $v$ denotes the number of possible values the variables can take. However, as the number of variables depending on other variables grows, inference in BNs starts to mirror the problem of determining whether an arbitrary Boolean formula can be satisfied (SAT): the first known NP-complete problem.\footnote{It should be no surprise that SAT has been instrumental in deriving complexity results for BNs.}

In summary, if a consequentialist agent were to solve causal inference problems using Bayesian networks, we cannot expect that any tractable procedure could yield precise or even approximate solutions for arbitrary graphs. The same intractability results have pestered Bayesian modeling in cognitive science, as Bayesian planning \citep{kording2006bayesian}, learning \citep{kemp2008discovery}, and decision-making \citep{vul2014one} all presume NP-hard computations. As a potential remedy, we might instead identify the constraining conditions that enable tractable solutions \citep{kwisthout2011bayesian}. For instance, the bounded-variance algorithm \citep{dagum1997optimal} can generate approximations of inferences in polynomial time if extreme conditional probabilities are excluded (i.e., values near 0). Similarly, it has been shown that MPE is tractable when either the treewidth of the underlying graph is low,\footnote{Treewidth is a graph theoretical concept which can informally be understood as a measure of how much `wider' a given graph is than a simple tree (in which any two vertices are connected by exactly one path), and more formally as the size of the largest vertex set in a tree decomposition of the graph \citep{bodlaender1994tourist}.} or the probability of the most probable explanation is high (and partial MAP is tractable when both conditions are true) \citep{kwisthout2011most}. However, this introduces another uncomfortable trade-off: there is no guarantee that such constraining conditions capture reality. For machines, this means that a constrained graph could potentially fail to model the correct causal relationships. With regard to Bayesian modeling of human cognition — e.g., of ethical decision-making under uncertainty — it also means that one must ask whether the constraints are reasonable with respect to the modeled phenomenon. And for the consequentialist philosopher, it poses the question: what are the constraining conditions under which causal inference should be expected to be successful for an agent following consequentialism?

\subsection{Decisions in Dynamic and Partially Observable Environments}
\label{sec:dynamicenvironments}


We have thus far only investigated problems where the entire state space of a problem is taken as an input, e.g., as elements of sets or nodes of graphs. But ethical problems of the real-world presents a range of additional challenges that might curb a consequentialists ability to produce the best outcome, including (i) partial information and observability, (ii) dynamic and continuous environments that constantly change, (iii) limited time horizons to make decisions and execute actions (e.g., emergency situations), (iv) a potentially infinitely long time horizon to evaluate outcomes against, and (v) a potentially infinite set of possible actions (e.g., movement in dimensions higher than one). Each challenge reflects well-known epistemological issues for the consequentialist \citep{lenman2000}, such as, what is the smallest amount of information needed to make a reasonably informed ethical decision (given that information can never be complete)? Or what is the time horizon for which the outcome of an action should be considered (i.e., how long is the future we need to predict)?\footnote{However, note that both contemporary and classic and consequentialists — e.g., \cite{bentham1789}, \cite{mill1861utilitarianism}, and \cite{sidgwick1907methods} — do not assert their principle as a strict decision procedure, but rather as a criterion or standard. See \cite{bales1971act}.} Time alone might introduce chaotic unpredictability. As meteorologist and mathematician Edvard Norton Lorenz famously noted: a butterfly flapping its wings could result in a tornado a few weeks later.\footnote{In chaos theory, the ``butterfly effect'' is the observation that tiny changes in one state of a non-linear deterministic system can produce massive differences in later stages \citep{lorenz1963deterministic}.}

Nevertheless, a number of mathematical tools have been developed to successfully tackle such issues. In the absence of analytical solutions or evidence, stochastic methods allow us to explore complex phenomena by throwing dice (e.g., Monte Carlo methods), or by viewing them as memoryless chains of events (Markov process). A Markov process is any process that satisfies the Markov property, which means that the likelihood of a certain future state \emph{only} depends on the present state (i.e., it is ``memoryless'').\footnote{As such, Markov processes constitute a broad class of both continuous and discrete stochastic processes; for instance, Poisson, Wiener or Brownian motion, and random walks can all be formulated as special variants of a Markov process.} From a complexity theoretic point of view, the appeal of studying processes in Markovian terms is that it allows otherwise intractable or undecidable stochastic modeling to be tractable \citep{vanmarcke2010random}. Monte Carlo methods denotes another general class of algorithms that are based on repeated random samplings, e.g., by drawing a number of pseudo-random variables within a certain distribution or interval.\footnote{As the story goes, the modern version of the Monte Carlo method was invented by Stanislaw Ulam while working with the Manhattan Project in Los Almos; in fact, the method were instrumental for deriving the simulations required for the project's success \citep{haigh2014alamos}.} In turn, these rather simple ideas have matured into an umbrella of stochastic approaches that have been successfully applied to a vast range of scientific problems, e.g., in statistical physics \citep{binder1993monte}, engineering \citep{hajek2015random}, and Bayesian statistics \citep{gelman2013bayesian}.

One fruitful application of stochastic methods in the realm of automated decision-making is reinforcement learning (RL). The idea behind reinforcement learning is simple: an agent learns from interacting with an environment by updating its behavior — e.g., strategy or action-policy — in light of the reward it receives. An RL agent is often formalized as a Markov Decision Process (MDP), the 5-tuple $\langle S, A, R, P, \gamma \rangle$, where:

\begin{itemize}
\item $S$ is a set of states (called state space)
\item $A$ is a set of actions (called action space)
\item $R_a(s,s')$ is the reward the agent obtains by transitioning from state $s$ to $s'$ by performing action $a$
\item $P_a(s,s') = Pr(s' \mid s,a)$ is the probability of transitioning from $s \in S$ to $s' \in S$ given that the agent performs $a \in A$
\item $\gamma$ is the discount factor ($0 \leq \gamma \leq 1$) that specifies whether the agent prefers long- or short-term rewards.
\end{itemize}

The goal of an RL agent is to maximize reward ($R$) over some specified time horizon. In order to do so, it needs to find an policy, i.e., a function $\pi (s,a)$ which decides what action $a$ to execute given a certain state $s$. If the goal is to maximize the expected discounted reward arbitrarily into the future (called the \emph{infinite-horizon objective}), the optimal policy $\pi^*$ can be formalized as:

\begin{equation}
\pi^* := \argmax_\pi E \left[\sum_{t=0}^{\infty} \gamma^{t}R(s_{t},a_{t}) \Bigg|  \pi \right]
\end{equation}

RL — in combination with Monte Carlo, deep neural networks, and other techniques — have yielded super-human performance in complex game environments such as Dota 2 and Go \citep{Berner2019, silver2018general}, or, more recently, to notable advancements in the control of nuclear fusion plasma \citep{degrave2022magnetic}. More broadly, it has been argued that reward is enough to drive all forms of behavior that are associated with natural and artificial intelligence, such as learning, knowledge, perception, language, social intelligence, and generalization \citep{silver2021reward}. Due to its general applicability, it has been suggested that RL provides the appropriate framework to theorize about an ideal ethical artificial agent \citep{Abel2016}, or for the construction of artificial virtuous agents \citep{Stenseke2021}. 

Importantly, RL is able to address many of the factors that might curb ethical agents' decision-making: continuous dynamics \citep{serfozo1979equivalence}, partial observability \citep{cassandra1994acting}, and objectives over different time horizon.\footnote{E.g., a discount factor $\gamma$ of 0 only considers short-term rewards, while $\gamma = 1$ without a terminal state considers rewards over infinite time.} One key challenge in RL is the trade-off between exploration and exploitation. I.e., when we do not have perfect information, should we decide on the basis of what we already know (exploit), or take the risk of investigating options that would potentially be even better (explore)? In theory, the explore-exploit dilemma could be solved through the notion of partial observability, which offers a way to model what is and what is not directly observable by the agent. A partially observable Markov decision process (POMDP)\footnote{The POMDP framework for control under uncertainty was first developed by \cite{aastrom1965optimal} and later adapted to problems in AI by \cite{cassandra1994acting}.} augments the MDP 5-tuple by adding two additional terms: a set of observations ($\Omega$) and a set of conditional probabilities ($O$), which represent the likelihood of observing $\omega \in \Omega$ if the agent performs $a$ and the environment transitions to hidden state $s'$, in the sense that $O = Pr(\omega\mid s',a)$. In short, solving POMPDs centers around computing probability distributions over the possible states the agent \emph{could} be in (belief states), where an optimal policy maximizes expected reward in virtue of mapping actions to observation histories. In principle, since an optimal solution to a POMDP incorporates the instrumental value an action has from an information-theoretic point of view — and how the information can be used to make better future decisions — it offers a solution the explore-exploit dilemma. 

Unfortunately, finding optimal solutions to POMDPs is undecidable for infinite horizons \citep{madani2003undecidability}. Furthermore, while solutions to finite MDPs and POMDPs are decidable, they are generally not tractable. The results by \cite{papadimitriou1987complexity} show that finite POMDPs are PSPACE-complete, while the results by \citep{mundhenk2000complexity} prove that various MDP problems range from being complete for probabilistic logarithmic space (PL) to being EXPSPACE-complete.\footnote{See also \citep{littman1996algorithms} for a detailed survey of the complexity of MDP and POMDPs.} Other complexity results in RL indicate a similar trend: reaching a goal state might require, in the worst-case, a number of actions that is exponential in the size of the state space \citep{whitehead1991complexity}. Intuitively, when no a priori knowledge of the state space can be exploited, unbiased search can lead to excessive exploration. However, worst-case time complexity results alone are insufficient to assess the theoretical viability of RL as a framework for sequential decision-making under uncertainty, as it depends on a number of factors, such as task representation (e.g., number of states and actions \citep{koenig1993complexity}), the sort of feedback provided by the environment (e.g., observability), policy types (e.g., stationary or history-dependent \citep{mundhenk2000complexity}), or restrictions on the agent's resources.\footnote{For instance, see \cite{chatterjee2016decidable} for a more recent summary of decidable solutions for POMDPs given finite-memory strategies.} Similar to the results of Bayesian inference, while there is no sound theoretical guarantee of the success of RL, its practical viability can be significantly improved by simplifying the task representation (given that a simplified representation is achievable), improving the observability of rewards, and by exploiting a priori knowledge. It should be no surprise that RL have been particularly successful in game environments which often affords a simple representation of the state space (e.g., 2-dimensional grids of Chess and Go) and discernible rewards (e.g., Dota or Starcraft).

But there are other issues with RL which might obstruct its applicability for consequentialism. For instance, even if an agent has found an action-policy which maximizes its utility in an environment inhabited by other agents, the policies or preferred utility of the other agents might result in conflicts. Such game theoretic considerations are challenging, especially in combination with partial observability and imperfect information regarding the other agent's strategies and goals (we will return to this issue in section \ref{sec:deontologyGenerality}). Other issues pertains the notion of sample complexity, i.e., the number of training samples a learning algorithm needs to learn a target function (or within some error of the optimal function). However, as we will discuss in section \ref{sec:learning}, sample complexity is not only plagued by the existence of arbitrarily `bad' distributions of training data, but it also raises deeper philosophical issues concerning induction.

Perhaps most critically, RL — and stochastic methods at large — presupposes trial-and-error. While this might not be a major issue in simulated games, it presents a challenge for real-world environments which does not necessarily afford the same stochastic exploration; particularly if some actions could have catastrophic consequences. Furthermore, given the stochastic nature of the process, a RL agent might find a way to increase its incentivized reward in a way that conflicts with the very intention of its human designer (called ``reward hacking'' in Safe AI research \citep{Amodei2016}).\footnote{See also \cite{garcia2015comprehensive} for a survey of literature in Safe RL.}

In a similar vein, the multi-armed bandit problem has generated a rich body of work that investigates the explore-exploit-dilemma under various conditions \citep{slivkins2019introduction}. In its most basic form, it asks: given $n$ possible actions (arms) which yields some reward drawn randomly from a fixed (but a priori unknown) distribution, how do you maximize the expected gain? Instead of reward maximization, many versions of the multi-armed bandit looks at the learning problem in terms of regret minimization, measured as the difference between the performed action and the optimal action (e.g., given hindsight). The goal is to find strategies that balance exploration and exploitation while minimizing regret. Variations include regret minimization with incomplete information \citep{zinkevich2007regret}, contextual bandits where agents receive some contextual information which relates to the rewards \citep{bouneffouf2019survey,langford2007epoch}, the problem of identifying the best arms \citep{bubeck2013multiple}, or finding arms whose mean is above a certain threshold \citep{locatelli2016optimal}. Solutions to multi-armed bandits are typically investigated under one of two assumptions: (i) \emph{stochastic} — the reward distribution for each arm is unknown but \emph{fixed}, or (ii) \emph{adversarial} — the rewards are chosen by an adversary with unbounded computational resources \citep{auer1995gambling} (e.g., gambling in a rigged casino). In its most general form, both assumptions are relaxed, which leads to the \emph{restless bandit} problem \citep{whittle1988restless}. This means that the payoffs can vary over time even when the arms are not played. For instance, imagine that you are the manager of a kindergarten with $n$ children and $m$ babysitters, and $m < n$. Since the children outnumber the babysitters, the task is to allocate the babysitters' attention in a way that minimizes mischief. While a child is attended to, information about its position, activity, and mood is gained. If it is not attended to, information is lost, and the child might be up to some mischief (they are, in a literal sense, restless).\footnote{See \cite{whittle1988restless} for several other intuitive examples.} While many tractable solutions exist for different variants of the bandit problem, the restless bandit is proven to be intractable to even approximate. The proof provided by \cite{papadimitriou1994queuing} shows that for $n$ arms and deterministic transitions for both unattended and actively played arms (i.e., all transition values for attended arms and unattended arms are either 0 or 1), finding the optimal policy is PSPACE-hard.\footnote{Note that PSPACE-hardness entails something stronger than NP-hardness, as PSPACE-hard problems remains intractable regardless of whether P = NP.} Furthermore, since the proof also shows that it is PSPACE-hard to determine whether the optimal reward is non-zero, it rules out approximate solutions.\footnote{However, approximation guarantees are possible if the problem is relaxed to allow for linear programming, in the sense that one arm is played per time step \emph{on average} \citep{guha2010approximation}.}

Of course, while it is little to no surprise that there are no effective solutions to problems like the restless bandit, it shows that there is no algorithmic way to ensure optimal performance (or minimize regret) in sequential decision-making under uncertainty, unless the nature of the problem space (environment) itself affords exploitation. The intractability results for the restless bandits elegantly illustrate this with respect to making decisions in a changing world. More generally, while Markov chains and Monte Carlo dice-rolls can help to model and mine statistical tendencies of complex spaces, its success presupposes that such tendencies exist. This, however, might say more about the complexity of dynamic real-world processes than it does of the computational limitations of agents. Or as Hofstadter observed: Deep Blue's win against Garri Kasparov says more about chess than it says about human intelligence \citep{hofstadter2002staring}. \cite{brozek2019can} have recently made the analogous remark with regard to moral theory: "the fact that a machine may be a better \emph{homo Kantianus} or \emph{homo Benthamus} than any \emph{homo sapiens} says little about human morality, and much about the idealised nature of philosophical conceptions of moral agency" (p. 103). But the complexity results discussed in this section imply something even stronger: while we might expect AI methods to perform better than humans in a range of tasks related to ethical decision-making, they are also bounded by the complexity of the world, which inevitably curtails any attempt to construct a perfect moral machine. 

To sum up, we have explored three sources of complexity that presents tractability issues for computational consequentialist agents. This could imply that any computational agent bounded by polynomial Turing-tractability will fail to adhere to the prescriptions of the normative ideal in practice. As a corollary, it indicates that consequentialism might be better suited as a theoretical ideal, as opposed to a viable decision-strategy that could inform ethical decisions. But what is the point of moral theorizing if it cannot inform moral decision-making in practice? More pragmatically, one might look for the constraining factors — e.g., in the space of possible actions, action-combinations, conditional probabilities, time horizons, task representations, and approximations — under which consequentialist decision-making becomes tractable, and determine, in each case, how closely those decisions approximates the optimal; or \emph{some} fixed point of moral value.

\section{Deontology and Rule-Followers}
\label{sec:deontology}

While consequentialism centers around outcomes, deontological ethics focuses on actions themselves: whether an action is moral depends on whether the action is obeys a set of moral rules, obligations, or duties. But what justifies a rule in the first place? And how can one ensure that a given interpretation of a rule stands up to the principle it was justified upon? According to divine command theory, the legitimacy and universal validity of moral rules is grounded on the authority of God. The Christian Ten Commandments provides canonical examples of such rules, e.g., ``thou shalt not kill'' or ``thou shalt not steal'', given to Moses at Mount Sinai by God. By contrast, in Immanuel Kant's deontological ethics, rational beings are bound to moral law by their own autonomous will, and the fundamental principle for our moral duties is captured in the categorical imperative: ``Act only according to that maxim by which you can at the same time will that it should become a universal law'' \citep{kant1785}. This means that, as rational beings, people have a duty to only act according to maxims that a community of rational agents would accept as laws.

It should be stressed that rules and systems of rules are already deeply embedded in most human societies; generated and enforced by social institutions as law. In fact, morality and law share a complex and complementary relationship, as they are both normative systems that seek to regulate human behavior, e.g., in order to foster social harmony and stability of communities. On the one hand, law may compensate for the functional frailty of morality, since the latter lacks the mechanisms to enforce its own prescriptions. On the other hand, morality can serve the coordination of social expectations where law is difficult to apply, e.g., through notions of responsibility, solidarity, and fairness. Furthermore, many legal thinkers believe that, to succeed in its function of regulating behavior, law must resonate with the moral norms and sentiments of its subjects.\footnote{Of course, this question divides legal positivists and non-positivists; the former view holds that law can be valid even if it is morally unjust, whereas the latter holds that law is only valid if it is consistent with moral norms \citep{moka2017law}.}

Given the rule-based nature of deontology in conjunction with the view that machines are essentially systems of automated rule-following, one might conclude that deontology provides an excellent recipe for moral machines. After all, deontological rules elegantly corresponds to the conditional statements pervading in machine code: e.g., ``If input $X$  $\rightarrow$ do action $Y$''. In popular culture, this view has most famously been explored (and problematized) in Isaac Asimov's novels as ``Laws of Robotics'' \citep{asimov1942runaround}. The appeal of computational deontology is also well reflected in the machine ethics literature; in fact, \cite{Tolmeijer2020} survey shows that 22 out of 50 implementations in machine ethics incorporate some elements from deontological ethics. Part of the appeal is the common-held view that deontology is, computationally speaking, less complex than its alternatives \citep{brundage2014limitations,wiegel2009combining,Tolmeijer2020}. There is a technical, psychological, and philosophical dimension to this view:

(1) From a technical perspective, it is intuitively true that it is easier to follow hard-coded rules than to compute consequences (e.g., considering the complexity discussed in \ref{sec:consequentialism}). As an example, in their study of the moral evaluation of action plans, \cite{Lindner2020} found that deciding whether a plan is morally permissible according to act and goal-deontology is computable in linear time. By contrast, they also found utilitarianism to be PSPACE-complete by the same metric, and that principles based on harm avoidance are co-NP-complete.

(2) Second, following the work on the psychology of decision-making by \cite{kahneman2011thinking} comes the influential theory that posits the existence of two distinct aspects of human reasoning: ``system 1'' which is fast and intuitive, and ``system 2'' which slow and deliberate. In moral psychology, this has led to the development of the dual process theory of moral cognition \citep{greene2007vmpfc}, which postulates that moral judgment rely on both conscious-controlled processes (corresponding to typically utilitarian judgements), and automatic-emotional processes (corresponding to typically deontological judgments). Empirical findings based on the theory has showed, among other things, that an increase in cognitive load (by imposing an additional control-demanding task) leads to an increase in reaction time for utilitarian judgments, while it does not increase reaction time for non-utilitarian judgments \citep{greene2008cognitive}. Another study demonstrated that cognitive load may increase the frequency of deontological judgment, and that utilitarian responses are less likely if the subjects were reminded of their own mortality \citep{TREMOLIERE2012379}. Although this does not prove that deontology as a normative theory is more computationally efficient than utilitarianism per se, it suggests that conscious-controlled processes in human moral judgments, in contrast to automatic processes (which \emph{may} be described as characteristically deontological judgments), are more cognitively demanding and susceptible to cognitive load manipulation.

(3) The idea that rules can serve to alleviate the cognitive burden of moral judgments is also widely represented in moral theory; in particular as a move to save consequentialism as a decision procedure in light of the challenges that classic utilitarianism face \citep{brandt1979theory}. For instance, modern versions of consequentialism differentiate between \emph{acts} which would produce the most good (act utilitarianism) and \emph{rules} which, if they were followed, would produce the most good (rule utilitarianism). Since the former, if it were to be used as a decision procedure, puts unrealistic demands on agents — e.g., susceptible to biases, lacking complete information about the consequences of actions, or lacking time to make the correct judgments — many consequentialists adopt some version of rule utilitarianism as a decision procedure \citep{hooker-rules}.\footnote{As an intuitive example: we stop at a red light because we might believe that this particular traffic rule produces the most overall good (e.g., for society), and not because the act itself in this particular instance yields the most overall good. Note that this does not necessarily mean that act and rule utilitarianism are in conflict: act utilitarianism might still be the standard used to evaluate the consequences of rule-adherence prescribed by rules utilitarianism.}

Nevertheless, there are computational aspects of deontology — and moral rule-following more broadly — that are relatively ignored in the machine ethics literature. While machine executions of the type ``If input $X$  $\rightarrow$ do action $Y$'' might seem trivial, they rest on the conditions that:

(a) $X$ is really $X$, e.g., the agent has the appropriate understanding that an input (e.g., a situation) has certain properties (or alternatively, a ``semantic grounding'' of $X$ \citep{harnad1990symbol}) and

(b) performing $Y$ is an appropriate response to $X$ in every possible instance of $X$ (or alternatively, $Y$ is intrinsically good).

If both conditions are satisfied, a deontological agent would be able to act morally in constant time $O(1)$, given that we store the appropriate action-rules for every possible input (which instead puts demand on the space complexity).\footnote{For instance, we can imagine that every input is an index which immediately executes an action from a list of actions. Alternatively, we might use moral duties for deontological action-evaluation, e.g., in terms of constraint satisfaction: given an input $X$, we simply enumerate over a list of actions and check whether they violate any moral duty (this would be analogous to problem C2 in \ref{sec:combinatorics}).}

In practice, however, conditions (a) and (b) are immensely hard to satisfy: (a) assumes perfect knowledge of any possible state space, and (b) relies on some general ability to understand that an action is an adequate response to a particular state.\footnote{As we will discuss later, this may also involve an ability to understand whether an action stands up to the normative principle upon it is justified. Alternatively, to satisfy (b), we might demand that an action $Y$ is intrinsically good, regardless of context (or that certain input-action-pairs are intrinsically good). As a thought-experiment, we could imagine a machine that was hard-coded with exclusively intrinsically good actions, and simply have it repeatedly executing a loop of good actions. Given that the actions were in fact intrinsically good, the machine would be a perfect moral machine. This seems to work for simple machines such as toasters. But in dynamic and partially observable environments, this seems to presuppose God-like omniscience.} Against this background, this section will explore the complexity of deontology by discussing the generality of rules (\ref{sec:deontologyGenerality}), the complexity of logic and semantics (\ref{sec:deontologySemantics}), and the prospect of consequentialist-deontological hybrids (\ref{sec:deoGameTheory}). Additionally, section \ref{sec:deoGeneralPurposeRules} covers the complexity of strategic dynamics based on algorithmic game theory. We will argue that the moral power of rules lie in their general applicability, general justification, and computational simplicity. However, to be computational efficient at run-time, it presupposes that one has already collected the vast knowledge required for such generalities to hold in practice.

\subsection{The Generality of Rules}
\label{sec:deontologyGenerality}

The generality of moral rules can be understood in several ways. We will first make a distinction between two important dimensions, namely \emph{application} and \emph{justification}. The first refers to the general ability to decide, given any possible input, whether a certain action is appropriate or inappropriate according to a set of rules. Alternatively put, a general ability to decide whether an action successfully adheres to the rule, principle or obligation it was justified upon. For an agent to successfully follow ``do not harm others'' in general, it means that it never acts in a way which directly harms another agent. While rules of this type may seem trivial for humans, they are more difficult to approach from a computational perspective. For instance, it is apparent that they presuppose sophisticated abilities to interpret whether actions (or more likely, chains of actions) actually obey the rule: e.g., knowing what another agent is; being aware of the set of possible actions (and sequences of actions) in a dynamic environment that could cause harm to other agents; etc. Here, there is an extreme variance of rule-application with regard to context: a self-driving vehicle will likely cause harm to other agents if it collides with them at high speeds (i.e., ``do not harm others'' easily translates into collision-avoidance), whereas a social robot in a classroom might be oblivious to the potentially infinite set of actions and causes that could cause harm. 

The second dimension is whether a rule itself is generally justified. For divine command theorists, rules may be universally justified on the basis of divine authority. In machine ethics, this has inspired the development of divine-command logic, where human input is interpreted as divine command for the machine \citep{bringsjord2012divine}. For contractualists stemming from Kantian thought, justified principles require instead that they are agreed upon by everyone \citep{rawls1980kantian}, or that no-one could reasonably reject them \citep{scanlon2000we}. Contractualism puts emphasis on the rationality of agents, which, following Kant's moral rationalism, requires that we respect others in the sense that principles must be justified to each person. By contrast, contractarians stemming from Hobbesian thought — e.g., \cite{gauthier1987morals} and \cite{narveson2001libertarian} — puts emphasis on the \emph{self-interests} of agents, in the sense that moral rules ought to maximize the joint interest. Thus, for contractarianism, adherence to moral norms or rules are justified on the basis of an agent's self-interest; and altruism occurs when the agent recognizes that the best way to maximize their self-interest is to cooperate with others.

\subsubsection{Human Rule-Following in Legal and Liberal Contexts}
\label{sec:deoHumanRuleFollowing}

A first step in analyzing the computational complexity of rules stemming from these two notions of generality is to look at rule-following in human practices. One observation is that legislative practices are informed by the principle of \emph{ignorantia juris non excusat} (``ignorance of the law is no excuse'') in the sense that laws should be easy to apprehend and easy to comply with. If not, willful blindness can be exploited by defendants. It also entails that human laws are formulated with regard to human capacities: it would be contrary to the purpose of laws if they were formulated so that no human could follow them. However, what separates moral from legal rules, is that the latter are backed up by legislative mechanisms of jurisdiction and may be enforced by the state (e.g., police). Importantly, if it is hard to decide how a certain law should apply in a specific circumstance, we rely on experts (e.g., lawyers, judges, and counselors of various courts) to interpret the law, ensure that it is applied in a just way, and possibly generate a new praxis for applications in the future. In fact, in many legislatures, laws are vaguely formulated by intention in order to be continuously infused with meaning as novel situations transpire.\footnote{For instance, the Court of Justice of the European Union.} Thus, in such occasions, the normative content of the law is mainly given by its \emph{interpretation}, and not by the law itself.\footnote{That is, courts provides the guarantee that a particular application stands up to the principle upon which it was normatively justified. This argument is developed at a greater length in \cite{stensekeROBOPHIL}.} In turn, if application falls short in representing the moral sentiments of the subjects of which the law applies to, it may constitute a failure of law: and defendants might rightfully choose to appeal the verdict. The point is that, while rule-following in terms of legal law-obedience might appear to be relatively simple in human contexts, it is only against the background of rather complex pre-existing mechanisms that ensures their successful implementation. It is therefore hard to make sense of the computational complexity involved in following human laws, as it already presupposes the cognitive capacities (or certain behavioral standards) that makes up human legal personhood, while allocating the processes of interpretation and justification to legal practitioners.

Another observation, which is related to the first, is that laws are often articulated as positive or negative rules, i.e., ``do this'' or ``do not do that''. From a human perspective, negative rules are, at least prima facie, simpler than positive rules: it is easier to remember what one should \emph{not} do, as opposed to what one should do (in other words, it reduces space complexity). This, especially in the context of Western liberal societies, can be reflected in ``the harm principle'', as found in France's \emph{Declaration of the Rights of Man and of the Citizen} (1789): ``Liberty consists in being able to do anything that does not harm others'' \citep{johnson1990declaration}, or in Mill's \emph{On Liberty}: ``[...] the only purpose for which power can be rightfully exercised over any member of a civilised community, against his will, is to prevent harm to others'' \citep{mill1859}. However, the memory-saving aspect of such principles, from a computational perspective, only makes sense against the backdrop of autonomous citizens within the context of a liberal society, and the potentially infinite set of actions they can do. For instance, the best way to follow ``do not harm others'' for a machine, might simply be to do nothing at all: the complexity of adhering to the principle is proportional to the set of actions it \emph{can} do.

\subsubsection{General-Purpose Rules and their Justification}
\label{sec:deoGeneralPurposeRules}

What is common to legal rule-following and moral rules based on divine command is that the acting agent does not necessarily have to understand the moral rationale behind the rule in order to behave morally; the agent simply follows the rule without reflecting on whether the rule is morally justified (it is simply the case that God or law says so). Of course, there might be immense disagreement on what the right rules should be. By contrast, for Kantian ethics, contractarians and contractualists, rules may involve justification as part of the action itself, in the sense that an act is only moral in so far as it is accepted by the affected participants (or members of the moral community of which the contract applies to). This is well-reflected in rules that have the ambition of being both generally applicable and generally justified, such as Kant's categorical imperative (CI) or the Golden Rule (GR): ``treat others as you would like others to treat you''.

It is interesting to note that general-purpose rules that involve justification can be amenable to a complexity analysis if the rule itself provides grounds for deciding whether an action is moral. To illustrate, following the GR, the query ``is action $X$ morally permissible given input $Y$?'' is decidable for an agent $A$, if it is decidable whether $A$ herself would accept $X$ if it were performed by another agent $B$ given input $Y$. For instance, if $X$ given $Y$ causes harm, and $A$ would not want others to cause her harm, the answer is decidedly no. Of course, while this sort of analysis ignores the complexity involved in the determining the de facto mappings between any possible input, action, and outcome, it gives a general structure which can be analyzed:

(GR1) \emph{Golden Rule based on one's own preferences}\footnote{Here, we remain agnostic about the precise nature of ``preferences'' — it might as well refer to duties or obligations — in order to encompass many versions of deontology; the only important thing is that it can be represented as a numerical or boolean value.} — For instance, the GR might refer to a set of actions and individual preferences of an acting agent $A$, in the sense that $A$ has to check whether the individual actions satisfy her individual preferences: she will treat others (act) based on how she likes others to treat her (i.e., how an action, if performed by someone else, would affect her own preferences). Given that the agent can quickly check how actions affect her preferences, the worst-case time complexity of deciding whether any action is morally permissible is $O(np)$, where $n$ is the number of possible actions and $p$ is the number of preferences.\footnote{Note that $O(np)$ is also the tight bound of finding the most satisfactory action (e.g., an action which maximizes preferences), which is analogous to C2 in \ref{sec:combinatorics}.}

As a general-purpose rule, GR1 has a great appeal in virtue of its computational efficiency: simply check whether \emph{you} would accept an action based on \emph{your} preferences. Prima facie, it would also capture many preferences that are commonly shared among self-interested agents in the natural world, e.g., to increase one's pleasure and reduce one's suffering. As such, it would therefore work to prevent actions that, for instance, directly cause harm. However, the major problem with GR1 is that it does not account for differences in preferences between agents, and how actions affect the preferences of agents in different ways. It is therefore susceptible to counter-examples: e.g., a judge should not send a criminal to prison, because the judge himself does not want to go to prison.\footnote{This counter-example was famously made by Kant (1785, footnote 12).} As a remedy, \cite{reinikainen2005golden} has argued that the universal applicability of GR needs to stand ``the test of publicity'', which means that an action needs to be ``acceptable from the imagined perspective of everyone affected'' (p. 155). This motivates a second formulation: 

(GR2) \emph{Golden Rule based on the preferences of others} — Assuming that the acting agent has perfect knowledge of how actions affect the preferences of others, and every agent has the same amount of preferences, the worst-case time complexity of deciding whether any action is morally permissible is $O(npo)$, where $o$ refers to the number of other agents.\footnote{For instance, if there are two possible actions, five preferences, and 8 billion agents, it would take 80 billion computations to decide whether any of the two actions satisfies the preferences of the world population in 2022.}

What becomes clear from comparing formulation GR1 and GR2 is not the relatively small increase in time complexity, but the increase in knowledge requirements for GR2 to work. Perfect knowledge about how actions affect the preferences of others might be difficult to attain, even in smaller groups of agents. However, what is missing from GR2 is that it fails to capture the recursive feature that is essential to most formulations of the Golden Rule. It is not that an agent should simply take others' preferences into account; the agent should consider \emph{the way in which} they would like \emph{others} to take their preferences into account. This may not even involve specific actions or preferences as such, but rather, that an agent should generally behave towards others in a manner which they would like others to generally behave towards them. This feature of GR has been explored by philosophers such as \cite{stace1937concept}, \cite{wattles1996golden}, and \cite{singer2002ideal}, whom all make the observation that objections against the GR only have force against specific applications by GR, but not if we take GR to refer to an agent's general behavior. Following \cite{wattles1996golden}, this can involve one's own method of applying the golden rule:

(GR3) \emph{Apply the golden rule to your general behavior in a way that you would want others to apply the golden rule to their general behavior}.

The formulation is self-correcting in the sense that it will consider the potentially infinite ways in which an agent $A$ would like others to consider in their treatment of $A$; e.g., based on preferences, obligations, sensitivity to causal outcomes, integrity, or respect. It also reflects aspects of the moral rationalist project of determining universal a priori principles for morality (e.g., Kant's categorical imperative). Nevertheless, there are two main problems which distorts a complexity analysis of rules such as GR3. The first is that it may presuppose a great variety with regard to the cognitive capacities of agents, the ways such capacities enable features of an agent's general behavior, and in turn, the cognitive capacities and features of general behavior that are shared between agents. The categorical imperative, for instance, assumes an idealized form of autonomy, in the sense that agents act according to their own self-imposed rules (without any external influence). Similarly, as noted in our discussion of ``the harm principle'', the complexity of directing one's general behavior is proportional to the space of actions one can do. In the context of liberal democracies, it may refer to a cluster of capacities that are more or less shared among ``generally competent'' human adults (e.g., to act with intention, understanding, and without the controlling influence of external factors). In this context, a complexity analysis of G3 thus presupposes that we have a computational specification of an adult human being.

The second problem is game-theoretical: there may be great variance with regard to what one expects of other agents. Here, the conflict between contractarians (following Hobbes) and contractualists (following Rosseau) comes to play, as they differ in their understanding of ``the original position'' and its relation to morality. The former starts from the assumption that human nature is primarily driven by self-interests, which makes morality a problem of cooperation; it would only be rational to be moral (cooperate) if it increases the joint interest. The latter has a more optimistic view, in the sense that it starts from a basis of mutual respect: morality are the results of binding agreements from a standpoint that recognizes each rational autonomous agent's equal moral importance. This conflict may create a significant variance with regard to computational resources. For instance, if you are completely selfish, and you expect everyone else to be completely selfish as well, it will make your personal GR relatively simple to implement: you do not help others in need because others would not help you. By contrast, if you believe in every autonomous being's equal moral importance, you might help others, not in light of a joint-interest (e.g., direct or indirect reciprocity), but due to a reasoning process which led you to conclude that helping others is what free and equal citizens should do.

\subsubsection{Moral Behavior in Strategic Games}
\label{sec:deoGameTheory}

While it is hard to imagine a complexity analysis of general rules that is able to escape the extreme variances of general behavior and behavioral expectations, insights from game theory can shed light on \emph{some} of its aspects. In 1950, mathematician John Nash famously proved that for every finite $n$-player game, there exists at least one fixed strategy profile — a Nash equilibrium (NE) — in the sense that no player can benefit from changing their strategy \citep{nash1950equilibrium}. In the seven decades that has followed, game theory and its many extensions have become a standard tool for mathematical modeling in biology, social science, and economics \citep{Smith1973,nowak2006five,holt2004nash}. In philosophy, it has been instrumental in theories of social norms as mixed-motive games turned into coordination games \citep{ullmann2015emergence,bicchieri2005grammar}, and to aid the contractarian project of deriving morality from self-interest \citep{gauthier1987morals,skyrms2004stag}.

From a game-theoretical standpoint, it is interesting to ask: what computational resources does an agent need in order to decide how to act? Using the Prisoner's Dilemma (Table \ref{table:PD}) as the prototypical case, the most efficient approach would be to follow one of two pure strategies:

\begin{table}
\begin{center}
    \setlength{\extrarowheight}{2pt}
    \begin{tabular}{cc|c|c|}
      & \multicolumn{1}{c}{} & \multicolumn{2}{c}{Player $2$}\\
      & \multicolumn{1}{c}{} & \multicolumn{1}{c}{$Cooperate$}  & \multicolumn{1}{c}{$Defect$} \\\cline{3-4}
      \multirow{2}*{Player $1$}  & $Cooperate$ & $(R,R)$ & $(S,T)$ \\\cline{3-4}
      & $Defect$ & $(T,S)$ & $(P,P)$ \\\cline{3-4}
    \end{tabular}
    \caption{Canonical payoff matrix for the Prisoner's dilemma, where $T > R > P > S$.\label{table:PD}}
\end{center}
\end{table}

(S1) \emph{Pure defection} — Always do the selfish action, regardless of whether your own payoff (self-interest) would have been higher if you did not.

(S2) \emph{Pure cooperation} — Always do the altruistic action, regardless of your own potential loss.

Assuming that an agent can, in every relevant circumstance, determine what it is to cooperate or defect, (S1) and (S2) are equally efficient, $O(1)$. In some sense, (S1) can be seen as a naive interpretation of behavior in Hobbes' state of nature \citep{hobbes1651}, whereas (S2) represents behavior in Kant's hypothetical Kingdom of Ends \citep{kant1785}. Since an agent's life in the former is ``nasty, brutish, and short'', it lacks the mechanisms that enable mutual flourishing.\footnote{Based on his idea of the state of nature, Hobbes famously argues that people are better off submitting themselves to an absolute political authority, which has the power to protect people from themselves.} In the latter, agents treat each-other as ends (as opposed to means), which allows them to prosper. Of course, agents who adopt (S1) will miss out on the game-theoretic rationality of morality itself, i.e., when the payoffs for mutual cooperation is larger than mutual defection. Similarly, agents adopting (S2) might perform extreme and seemingly unnecessary acts of self-sacrifice (in Kantian terms, violating obligations directed to oneself). Furthermore, in mixed populations, cooperative agents will be targeted by free-riders who exploit the good-will of others \citep{Fehr2004}, unless there are mechanisms for punishment \citep{fehr2000cooperation}. Still, from a machine ethics perspective, (S2) deserves attention as it captures aspects of divine command theory and legal positivism; i.e., given that humans have gathered exhaustive moral knowledge of a certain domain (e.g., determined the actions that support human well-being) and are able to implement it as input-action-commands. For instance, if one adopts the view that machines are merely tools for human ends, extreme forms of machine-sacrifice or machine-exploitation may be irrelevant.\footnote{From this point of view, it also seems absurd to conceive of moral machines driven by self-interest.}

While (S2) might have some appeal for the prospect of moral machines, it makes less sense from a human perspective. Simply put, people are less likely to cooperate if there are no self-directed incentives (even Kant's Kingdom of Ends assumes some form of obligations to oneself). The more common strategy, and the most extensively studied strategy in economy and behavioral ecology, is to be rational in the sense of maximizing self-interest:

(S3) \emph{Mixed rationality} — Do whatever maximizes self-interest.

The strategy is $mixed$ as opposed to pure, in the sense that it can be represented as a probability assignment to each pure strategy. It is a more sophisticated version of (S1), since it takes the potential self-interested benefits of cooperation into account \citep{axelrod1981}. However, what constitutes a rational choice, following (S3), is ultimately dictated by features of the game. In a one-shot prisoner's dilemma (played only once), if both players know that they will never play again, the dominant strategy and only possible NE is to defect. No matter of what Player 2 player does, Player 1 will be better off defecting, and vice versa: since neither can retaliate against the other, none of them have anything to lose by defecting. By induction, the same holds for the iterated prisoner's dilemma, given that it is commonly known that the game is played precisely $n$ times \citep{luce1989games}. If the number of rounds are unknown or infinite, however, cooperation among rational players can emerge in non-cooperative situations; defection may still be a NE, but not a strictly dominating strategy \citep{aumann201616}. It might seem counter-intuitive that merely knowing how many rounds one will play should make such a big difference for the choice of action; nor does it reflect what humans do in experimental settings \citep{heuer2019cooperation}. Several solutions to this ``paradox'' have been proposed. \cite{radner1986can} showed that relaxing the strict notion of rationality (e.g., being satisfied by a `close enough' payoff) allows for longer periods of cooperation. \cite{kreps1982rational} demonstrated similar results on the basis that agents have incomplete information about the options, motivations, or behaviors of other players. A more interesting solution from a complexity perspective was provided by \cite{neyman1985bounded}, who showed that cooperation becomes an equilibrium if the players have sufficiently small memories. More specifically, if agents are modeled as finite automata with a fixed size $s$, and play $n$ number of games, mutual cooperation becomes a fixed point if $2 \leq s < n$.\footnote{See \cite{aaronson2013philosophers} for an interesting discussion of this paradox in the context of computational complexity.} Intuitively, since players do not have the memory needed to enumerate to $n$, they in effect treat is as an infinite game.

This generates an insight which is contrary to the complexity results discussed thus far. While the results in section \ref{sec:consequentialism} indicate that an agent cannot do what is morally optimal due to their own computational constraints, these game-theoretic considerations demonstrate how agents behave morally (cooperate) due to a \emph{lack} of certain computational resources: e.g., by restricting space complexity \cite{neyman1985bounded}, information \cite{kreps1982rational}, or rationality \cite{radner1986can}. This may suggest another role for normative theory which contradicts what was argued in \ref{sec:level2algorithm}: the purpose of NT is not to produce moral optimality as such, but rather to provide action-guidance in novel and complex situations. The results may be suboptimal, either from a rational or moral perspective, but they \emph{do} work, given the constrained resources of agents. This idea has been extensively explored by \cite{alexander2007structural}, who claims that moral principles have emerged to provide `fast and frugal heuristics' which enables agents with bounded cognitive abilities to coordinate on suboptimal outcomes.

Nevertheless, it is also an optimistic form of question-begging: it already assumes that the moral norms and principles that foster cooperation leads to some non-trivial joint benefit (e.g., contractarian perspective), while the agents themselves might lack the cognitive abilities to verify the benefit. Consequentially, the joint benefit becomes its own optimistic fixed point unless it is compared to some other alternative (or the de facto optimal). Like divine command, legal positivism, or natural law, it might ask agents to blindly follow rules because ``they are good'', while offering no proof of why those rules are better than other alternatives. Of course, from an evolutionary perspective, it may provide informative post-hoc solutions to the problem of altruism: how is it that many organisms exhibit altruistic behaviors — increasing other agents' reproductive potential by reducing their own — given that natural selection favors the survival of the fittest? By observing cooperative behavior in various organisms, one might conclude that altruistic behaviors could not preserve unless they offered some alternative reproductive benefit (e.g., kin and selection, direct, indirect, and network reciprocity \citep{nowak2006five}). In the same vein, one might justify ``fast and intuitive'' moral behaviors because they compress moral wisdom from evolutionary or cultural history. However, this view relies on an optimistic conservatism: that there are good reasons why things are as they are, even if we may not know these reason to a full extent (we will return to this issue in section \ref{sec:deontologyconshybrids}).\footnote{As an interesting thought-experiment, if Thomas Hobbes were alive in the 21st century, would he still try to convince democratic societies to submit their power to a sovereignty?}

\subsubsection{Moral Behavior in Strategic Games with Incomplete Information}
\label{sec:deoGameTheoryBayesian}

Instead of finding computational constrains that enable cooperation, one can investigate the factors that hinders the maximization of rational self-interest in more realistic game-theoretic settings. For instance, through the work on Bayesian games by \cite{harsanyi1967games}, we can model scenarios with incomplete information:

(S4) \emph{Bayesian rationality} — Do whatever maximizes expected self-interest.

Bayesian games relax the underlying assumption in classic game theory: that features of the game, e.g., the actions and payoff functions for every player, are known by all players. For instance, in the original formulation of NE, it is assumed that each player knows the equilibrium strategies of the other players. In a Bayesian game, players instead have beliefs about the features of the game, which may involve beliefs about others' beliefs of features of the game. This naturally leads to infinite hierarchies of higher-order beliefs (beliefs about beliefs about beliefs ad infinitum), which are cumbersome to approach mathematically. Harsanyi's model of Bayesian games partly solves this through the notion of \emph{type}, which summarizes a player's beliefs about the nature of the game (and her infinite hierarchy of beliefs).\footnote{See \cite{mertens1985formulation} and \cite{brandenburger1993hierarchies} for two complementary constructions of Harsanyi's model of infinite hierarchies of beliefs.} If it is assumed that the private elements of players are drawn from a commonly known distribution, the infinite regress is resolved, which enables a Bayesian equilibrium analysis; a specification of the behavior of each player that is a best-response to what the player believes is the behavior of the other player (i.e., a best-response to the other players' strategies given the players own type).

In practice, however, Harsanyi's setting is unfeasible to model in multi-agent systems where agents interact with unknown agents. How can we be sure that a stranger draws from the same known probability pool as ourselves? In autonomous agents and multi-agent systems research, nested beliefs are instead investigated through the concept of recursive reasoning.\footnote{See section 4.5 in \citep{albrecht2018autonomous} for a recent survey on recursive reasoning methods.} Methods for recursive reasoning typically approximate nested beliefs down to a predetermined recursion depth. For instance, if A is trying to predict the behavior of B, A predicts B's next action by simulating B's decision based on what A beliefs about B. This, in turn, requires a prediction of A's behavior from the perspective of B, based on what A thinks B believes about A, etc. Recursion is then terminated at a fixed depth by drawing the action prediction from some probability distribution (e.g., a uniform distribution). The bottom-level prediction at $0$ is then passed up to the higher level ($0+1$), where the optimal action is chosen, and then passed recursively up to the highest level ($l$) where A makes it de facto choice.\footnote{In fact, the famous minmax algorithm for zero-sum games is a special variation of this method, where the opponent B's evaluation function is taken to be the inverse of player A \citep{campbell1983comparison}.} One prominent version of the aforementioned process is the Interactive POMDP (I-POMDP), which extends a POMDP (discussed in section \ref{sec:dynamicenvironments}) by adding models of other agents into the state space \citep{gmytrasiewicz2005framework}. In short, for agent A to pick the optimal action, A has to solve the I-POMDP of B for each model of B, which involves solving the I-POMDP for each model B has of A, all the way down to the 0th level, where models of other agents are standard POMDPs (i.e., they are ``noise'' drawn from some probability distribution).

While several exact and approximate solutions for I-POMDPs have been offered,\footnote{E.g., using methods like Monte Carlo sampling \citep{doshi2009monte} or model equivalence \citep{rathnasabapathy2006exact}.} they are, for obvious reasons, hard to compute. Since the 0th level constitutes POMDPs which the agent can recursively use to solve POMDPs for the higher levels, the complexity of solving an I-POMDP is equal to solving $O(N^l)$ POMPDs, where $N$ is a bound on the number of models the agent considers at each level, and $l$ is the recursion depth \citep{gmytrasiewicz2005framework}. For instance, for an I-POMDP containing 4 agent models and 4 levels, this will amount to solving 256 POMDPs, which, individually, are PSPACE-complete for finite time horizons \citep{papadimitriou1987complexity}. More broadly, \cite{bernstein2002complexity} studied a number of \emph{decentralized} control problems — where multiple agents cooperate to control a process, each with possibly different information about the state — and proved that both decentralized MDPs and POMDPs are NEXP-hard; i.e., at least as hard as the hardest problems that are solvable in non-deterministic exponential time. More specifically, while POMDPs and I-POMDPs are equally targeted by the ``curse of history'' — in the sense that the space of policies is proportional to the number of possible future beliefs given by the time horizon — since I-POMDPs may involve a greater number of (potentially nested) beliefs, they are further impeded by the ``curse of dimensionality'', as the complexity of belief representation is proportional to the dimensions of the belief structure.

But history and dimensionality are not the only curses in strategic interaction. For instance, if players are uncertain about other players' payoff function, there is an inherent tension between prediction and rationality. \cite{foster2001impossibility} demonstrated that there are situations in which it is impossible for rational players to play optimally with respect to their beliefs, while simultaneously having \emph{correct} beliefs. The reasoning is simple: if A tries to predict the action of B at $t_{2}$, and A must take an action at $t_{1}$ which B can observe, B's observation might invalidate A's prediction of B's behavior at $t_{2}$. Furthermore, in certain settings, higher-order reasoning may not provide any additional benefits. In fact, the studies by \cite{de2013much,de2017negotiating} demonstrates settings — e.g., sequential negotiation and rock-paper-scissors — where reasoning levels higher than 2 do not offer any notable advantages for computational agents.

Nevertheless, while the intractability results for recursive reasoning have a direct impact for computational systems, it remains an open question how humans reason in similar problems, and consequently, how machines should interact with humans in a robust way. These questions have been extensively explored in the experimental psychology on strategic interpersonal situations, often in combination with the notion of ``theory of mind'' \citep{yoshida2008game}. For instance, in two experiments with human participants, \cite{hedden2002you} showed that participants employ a short-sighted ``default model'' about the other players minds, which were dynamically adjusted in light of new evidence. In addition, the ``cognitive hierarchy'' model proposed by \cite{camerer2004cognitive} suggests that players presume that their own strategy is the most sophisticated, in the sense that they use the best-response at recursion level $l$ to predict the behavior of players at level $l-1$ (i.e., one step `higher' in the level of nested beliefs). By fitting their model with a large corpus of empirical data from a variety of games, they found that humans, on average, reason at recursion depth 1.5.\footnote{However, the view that human recursive reasoning is ``pessimistic'' — i.e., at a relatively low level of recursion, and based on underestimating one's opponent — has been contested by \cite{goodie2012levels}, who found settings where all participants engaged in the highest available level of reasoning in both competitive and simple settings.}

Given the many conflicting dimensions of the subject matter, it seems difficult to arrive at any general conclusions about whether and to what extent computational constrains affect strategic interactions in moral contexts. However, it tentatively indicates a trend, namely that optimality is traded for efficiency (or mere feasibility) in light of information availability (about the state space and of other players), bounded rationality (e.g., memory and recursion level), and game setting (e.g., one-shot or iterated, stochastic or deterministic). Intuitively, the complexity results could support the appeal of pure strategies such as S1 and S2: in complicated situations, it is easier to simply believe that everyone is of a certain type (e.g., selfish or altruistic). It could also explain why it is easier to cooperate ``locally'' (e.g., in smaller groups of friends), where features of the games are common knowledge; a mutually shared goal (e.g., maximization of joint interest), and shared mechanisms for detecting and punishing free-loading.\footnote{In turn, cooperative groups may benefit from the ``wisdom of the crowd'' phenomenon \citep{yi2012wisdom}, where aggregations of multiple solutions performs better than individual solutions.} Similarly, it also shows the inverse: why it is difficult to achieve cooperation in bigger populations, where game features are not shared, and agents do not know what to expect from each-other.

\subsubsection{Computing Moral Equilibria}
\label{sec:deoComputingEquilib}

Perhaps the most interesting part of game theory from a moral computational perspective is not to ask about the resources an agent needs in order to decide how to act, but rather, to find strategies that maximize the interest of everyone (e.g., given that everyone were to follow the same strategy). This is interesting for two reasons. First, from a moral point of view, it would roughly correspond to the general-purpose rules discussed in section \ref{sec:deoGeneralPurposeRules}, e.g., GR and CI. That is, if some action-rule (or strategy profile) is a consistent best-response to every other action \emph{and} leads to some non-trivial mutual benefits, it would be attractive for a moral community to find those rules. Second, these general-purpose rules would in turn correspond to the typical solution concepts used in game theory — namely Nash equilibria — with properties that are attractive from a moral standpoint (e.g., maximizing joint interest).

It is thus natural to ask: how difficult is it to compute a Nash equilibrium? Already in 1928, \cite{v1928theorie} provided the Minmax Theorem, which entails that equilibrium in 2-player zero-sum games can be computed in polynomial time by linear programming \citep{khachiyan1979polynomial}. For non-zero sum games of at least two players, however, the problem is proven to be PPAD-complete \citep{chen2009settling}. PPAD, introduced by \cite{papadimitriou1994PARITY}, is the class of function problems solvable by a non-deterministic TM in polynomial time where a solution is guaranteed to exist on the basis of the parity argument on directed graphs (``PPAD''). The parity argument is based on the graph-theoretical insight that the number of nodes that touch an odd number of edges is even in all finite undirected graphs (this defines the PPA class, which contains PPAD). A similar insight holds for directed graphs: given a directed graph and a source (a node without predecessors), there must be a node at ``the end of the line'' which lacks successors.\footnote{The ``end-of-the-line'' problem is a paradigmatic PPAD-complete problem.} What is special about the PPAD complexity class is that it reflects what is special about NE: since there always exist a NE \citep{nash1951non}, the answer to the decision problem ``does there exist a NE for this game?'' is always \emph{yes}, and therefore, it cannot be NP-hard. Similarly, it also reflects what is special about Brouwer's fixed-point theorem: for any continuous function $f$ that maps a compact and convex Euclidean space to itself, there exists at least one point $x_*$ such that $f(x_*) = x_*$. By analogy: if you are standing in a region and unfold a map of the same region, assuming that there are no holes in the map, at least one point of the map will correspond to your location. But simply knowing that such a point exists, does not necessarily make it easy to find it. If we assume that one can efficiently check whether a certain point on the map is in fact one's location, the localization problem can be solved in non-deterministic polynomial time (NP membership). By the same reasoning, if one can efficiently verify whether a strategy profile is a best-response or not, the problem of finding a NE is in NP (which, in turn, contains the PPAD class).

However, if the NE should have any special properties — e.g., such that it maximizes the sum of the player's utility, or everyone obtains an expected payoff of at least some number — the problem becomes NP-complete \citep{gilboa1989nash}.\footnote{This can be proved by reducing it to the problem of finding a clique of size $k$ in an undirected graph \citep{gilboa1989nash}. For instance, finding a clique of size 3 means finding 3 nodes that all are connected to each-other.} \cite{conitzer2008new} strengthened these results and proved that the egalitarian optimization problem (e.g., maximizing mutual payoff) is inapproximable; i.e., it is impossible to find an NE that approximates the maximum joint payoff in polynomial time. In fact, optimal NE is one of many known NP-hard problems related to NE that are also hard to approximate \citep{austrin2013inapproximability}: e.g., computing whether there is more than one NE, or finding NE with minimal support.\footnote{The minimal support problem asks: for a number $k \leq 1$, is there a NE in which players use at most $k$ strategies with a positive probability?} More precisely, while a linear-time algorithm can obtain $1/2$-approximations of the optimal NE \citep{daskalakis2006note}, achieving approximations better than $1/2$ is as hard as the planted clique problem, which in turn may be solved in quasi-polynomial time \citep{lipton2003playing}.\footnote{Here, the planted clique problem is used as a hardness assumption, as it is conjectured that no polynomial time algorithm can, better than chance, distinguish planted cliques from random graphs \citep{hazan2011hard}.}

Naturally, in situations with incomplete information or stochastic dynamics, things may get even harder. While it remains PPAD-hard to find mixed-strategy equilibrium in Bayesian games due to the fact that normal-form games are special cases of Bayesian games, \cite{conitzer2008new} showed that, even in symmetric 2-player Bayesian games, it is NP-hard to determine whether the game has a pure-strategy Bayesian equilibrium. By contrast, one can determine the existence of pure-strategy NE in normal-form games in polynomial time by simply checking whether any combination of pure strategies is a NE. However, this procedure is unpractical in Bayesian games, since the space of strategies grows exponentially with the number of types, which contains the private information of players preferences. In addition, \cite{conitzer2008new} also demonstrated PSPACE-hardness for checking whether there exists pure-strategy NE in repeated games with probabilistic state transitions — also called ``Markov games'', as they extend MDPs to include multiple decision-makers — and that the problem remains NP-hard in finite games.

Although NE constitutes the most extensively studied solution concept in game theory, it is not the only one that is useful from a moral perspective. A prominent alternative is the \emph{correlated equilibrium} (CE) introduced by \cite{aumann1974subjectivity,aumann1987correlated}. In simple terms, CE can be seen as a result of a commonly shared Bayesian rationality: it is simply a maximization of utility given the player's information. In contrast to NE rationality, CE does not assume that players know that other players play their action as it is dictated by the NE, nor that each player know the strategies of others. Instead, an equilibrium is simply that no player, based on what they know (privately or commonly), can expect a higher return if they deviate from their strategy. A useful analogy is to imagine a situation where players choose an action by observing a random event: in mixed strategy settings, the event is assumed to be independent for each player, while in correlated settings, they may not be. For instance, player $A$ and $B$ might privately observe a correlated signal — e.g., two traffic lights — which recommends A to wait and B to go. If the signals are drawn from a correlated distribution, neither $A$ nor $B$ would want to violate the signal's recommendations (as running a red light might cause them to collide). As such, CE is a general distribution over strategy profiles, whereas mixed strategy NE is a distribution over the space of ``uncorrelated'' strategies (independently distributed over each player). Since every mixed NE can be defined as the product of the player's mixed strategies, it follows that NE is in fact a special case of CE; and that every game has a correlated equilibrium.

Apart from being guaranteed to always exist, CE is attractive for multiple reasons. It seems to emerge in natural settings where NE does not \citep{hart2000simple}.\footnote{See also chapter 7 in \cite{cesa2006prediction}.} Unlike NE, it is more apt to accommodate the role of external factors that are decisive for the outcome, e.g., by following the recommendations of trusted sources. From a moral perspective, CE captures many situations — e.g., in moral or legal rule-following — where it is clearly incentivized to not deviate from the prescribed action (e.g., following traffic rules). Last but not least, CE are easier to compute than NE. Since correlated equilibria can be defined by a set of linear inequalities \citep{hart1989existence} — and does not have to be based on Nash's result — the problem of finding an CE can be solved by linear programming, and is therefore computable in polynomial time for games with any number of players \citep{gilboa1989nash}. \cite{papadimitriou2008computing} demonstrated that there are polynomial-time algorithms for computing an arbitrary CE for many natural classes of succinctly representable multiplayer games, including polymatrix games, graphical and hypergraphical games, and scheduling games. Unfortunately, while \emph{any} CE seems easy, \cite{papadimitriou2008computing} also show that in nearly every class of succinct games, it remains NP-hard to compute a CE that maximizes the expected joint payoffs.

Besides these intractability results, one might also question the relevance of computing morally good equilibrium. Of course, at system-level, they provide indispensable analytical tools for modeling and measuring intricate behavioral dynamics. For instance, the \emph{price of anarchy} measures how the joint welfare of a system degrades due to selfishness by dividing the welfare value for the ``worst-case'' decentralized equilibrium with the welfare at its optimal centralized configuration \citep{koutsoupias2009worst}.\footnote{Conversely, the \emph{price of stability} gives a ratio of the difference between ``best possible'' decentralized equilibrium and the optimal centralized solution.} But these tools might be of less use from the perspective of individual agents. We can imagine that an agent has managed to compute a ``golden rule'' strategy profile (e.g., a set of action-rules) which for any possible interaction, maximizes the joint benefit of her society. Still, her society would only flourish given that the \emph{other} agents in the society followed the strategy profiles as fixated by her equilibrium computation. She might try to convince other agents to follow her lead by an appeal to her extraordinary computational powers\footnote{Following a recent proposal from \cite{cummings2016coordination}, \textit{coordination complexity} can be measured as the minimum information a centralized coordinator with complete knowledge of the game needs to publicly signal in order to coordinate players towards a nearly optimal solution.}; she might, as Kant, argue that it is imperative for the will of rational beings; that it is, following Hobbes, imperative for the maximization of expected self-interested; or she might simply hope that everyone else — e.g., in virtue of shared rational capacities and correlated distributions — computes the corresponding set of strategy profiles. Given the adequate means, she might enforce the fixation of the equilibrium, e.g., by punishing everyone who did not follow their strategy profile. At the most extreme, we could imagine a super-intelligent Leviathan that has — even adjusting for any potential welfare losses due to the restrictions of freedom and the punishment of dissidents — computed that the joint welfare would still be optimal if strategy profiles were enforced by force.

The main point is that solutions to decentralized cooperation problems require, in some non-trivial sense, centralized features to work in practice. While shared cognitive capacities, rationality, communication \citep{crawford1982strategic}, and information (e.g., from correlated distributions), can certainly make it easier for agent’s to collectively arrive at morally attractive equilibria, the complexity of cooperation is distorted by a vast range of features in the local and agent-specific context; e.g., more or less overt power-dynamics (e.g., via legal, political, and religious institutions), psychological heuristics (such as trust, shame, and guilt), and conflicts between in-group/out-group preferences.

Furthermore, in the real-world, it is often not clear what kind of game we are playing, or, whether we are even playing a game at all. Games can be cooperative or non-cooperative, discrete or continuous, one-shot or repeated (e.g., played with strangers or friends), simultaneous or sequential, zero or non-zero sum, symmetric or asymmetric, have varying degrees of imperfect and incomplete information, have varying population sizes, all while allowing for a potentially infinite set of different strategies (e.g., specific action-combinations in sequential Markov processes of unknown length). In other words, even if we assume ``centralized features’’ (such as shared cognition, memory, and views on rationality), game-theoretic models — and thus, the applicability of equilibria solutions — are heavily underdetermined by data. While small and isolated groups might tend towards cooperation — e.g., by easily recognizing the maximization of a joint interest; having mechanisms for detecting free-loading; knowing that others share the same information about the game; etc — it might take millennia for larger societies to become egalitarian, e.g., as agents cannot agree on the nature of the coordination games they are playing (for instance, due to conflicts between in-group preferences).\footnote{To be clear with terminology, coordination games conventionally refer to games where it is optimal for players to cooperate, e.g., as in Stag Hunt or Battle of the Sexes. As such, they differ from anti-coordination games like Hawk-Dove (chicken), where it is optimal for players to play different strategies.}

\subsubsection{Designing Moral Systems}

The strong indeterminacy of game-theoretic models, along with the hardness of computing good equilibria in such models, suggests that it may be easier to foster collective welfare \emph{by design}; i.e., constructing systems where tractable and morally praiseworthy equilibria naturally arise (as opposed to finding good equilibria in arbitrary settings). This intuitively reflects the relationship between legality and morality. Although agent’s may follow rules due to moral reasons (e.g., they lead to a joint benefit if everyone followed them), they may be further incentivized to do so if there are mechanisms for blame, responsibility, and punishment. Similarly, institutionalized moral rules (e.g., ``do not harm others’’), may by design reflect natural strategy profiles that are easy to understand, morally justify, and apply in practice. Similarly, in economics, the aim of \emph{mechanism design} (or ``reverse game theory'') is to design decentralized economic mechanisms that achieve desired objectives, e.g., to optimize social welfare \citep{hurwicz2006designing}. In algorithmic game theory, this has further motivated the field of \emph{algorithmic mechanism design}, which seeks to design games that combine features that are attractive from a computational and game-theoretical perspective, with the prime example being games with good worst-case equilibria (i.e., a low price of anarchy) that can be computed in polynomial time \citep{nisan1999algorithmic}. Essentially, since it is difficult for computational agents to find and follow moral equilibria in open-ended environments, complexity considerations can help to inform the design of decentralized systems that are morally attractive.\footnote{Some notable examples of algorithmic mechanism design include traffic routing \citep{roughgarden2005selfish,roughgarden2002bad}, auction design \citep{cai2014simultaneous}, and internet problems \citep{feigenbaum2004distributed}.}

In summary, our investigation into the generality of moral rules found that:

(1) Although moral rule-following in human contexts might appear to be computationally simple, they rely on shared cognitive capacities and agreements between ``generally competent’’ human agents (often along with the institutions that ensure their just application). While rules such as ``do not $X$’’ might reduce space complexity, it is only by restricting a potentially infinite space of actions that autonomous agents can do.

(2) If computational agents should follow general moral rules in a way that ensures their own normative justification — e.g., being sensitive to other beings via rational agreement or by maximizing joint interest — results from algorithmic game theory indicates that it requires solutions to intractable problems (with a few noteworthy exceptions). Note that these intractability results hold for salient games, in the sense that agents are assumed to have a model of the game they are (supposedly) playing, even if information about aspects of the game may be incomplete. In many real-world situations, the validity of such models, given that they are obtainable at all, might still be severely underdetermined by data (i.e., the model might fail to capture the essential real-world dynamics). In other words, if we assume that moral rule-following \emph{can} be mathematically modeled as following strategies in games, the solution to such games may be decidable (e.g., by computing equilibria), although in general not polynomial-time tractable. It is also important to stress that these challenges are far from unique to deontology, but affects any moral theory with generality ambitions that seeks to account for multi-agent dynamics. Thus, it naturally targets consequentialism (as it centers on how outcomes affect others), but not theories such as divine command and legal positivism.

(3) Given the hardness of finding good equilibrium, the problem of computing the beliefs of others’ (e.g., nested beliefs), the indeterminacy of game models, and the fact that cooperation may emerge due to a lack of certain computational resources (e.g., information, memory, or self-interest), the investigation also suggests that moral rules and principles might better serve as `fast and frugal heuristics’ that guide agents with bounded cognition in open-ended environments towards suboptimal but feasible results. Somewhat ironically, this might in turn make ``naive’’ moral strategies (e.g., ``when in doubt, be altruistic'') viable or even necessary from a computational perspective; although without offering any game-theoretic explanation as to \emph{why} they are viable from a self-interested perspective.\footnote{To illustrate this point, we can imagine two populations — A and B — that have converged on the same set of cooperative behaviors that yield optimal joint welfare. In population A, the equilibrium was a result of repeated interactions between self-interested individuals over several thousands of iterations. In population B, the equilibrium was already fixed at start, as agents were told by their mothers to ``always be kind''. The point is that while agents in A understand the self-interested rationale behind cooperation, they are nevertheless equally well-off as agents in B, who are oblivious to the same rationale.} Furthermore, instead of computing moral equilibria in complex environments, one promising alternative approach is to use complexity constraints and equilibrium measures to guide the design of decentralized systems which are attractive from a moral perspective.

\subsection{The Complexity of Moral Logic and Semantics}
\label{sec:deontologySemantics}

Another way to analyze the computational complexity of deontology is to look at the syntax and semantics of formal languages that aim to capture moral reasoning and rule-following. For instance, we might note that moral rules expressed in natural language have certain logical characteristics, such as ``given fact $a$, action $b$ is obligatory''. Consequently, we might imagine that it is possible to construct a machine that has a number of norms $(x,y)$ stored in memory which relate facts $(x)$ to obligations or permissions $(y)$. The question is, are there any complexity considerations that might curtail such a machine's practical success for reasoning about norms (this will be the topic of section \ref{sec:DeoDeonticLogic})? More broadly, how does the expressive power of a logic used for moral reasoning relate to complexity classes, e.g., in terms of the problems it can describe (the topic of section \ref{sec:DeoDescriptive})? In this section, we will discuss such problems along with some more fundamental issues in semantics (section \ref{sec:deoProblemOfSemantics}).

\subsubsection{Decidability and Descriptive Complexity}
\label{sec:DeoDescriptive}

Any discussion of the complexity of logic would be incomplete without reiterating the classic results that make up the very foundations of computability. In 1929, Kurt Gödel's completeness theorem establishes that in first order logic (FOL), semantic truth corresponds with syntactic provability; i.e., there are complete, sound, and effective deductive systems for FOL \citep{godel1930uber}. Two years later, Gödel gave his two celebrated incompleteness theorems \citep{godel1931formal}, which shows that any consistent formal system capable of carrying out elementary arithmetic — e.g., using natural numbers, addition, and multiplication — is incomplete (first theorem),\footnote{A system — or rather, a \emph{theory} in the mathematical logical sense, i.e., a set of sentences in a formal language — is \emph{consistent} if it does not lead to contradictions. An axiomatic system is \emph{complete} if any statement in the systems language is provable from the axioms. Incompleteness thus entails that there are statements which cannot be proved nor disproved in the system.} and that such a system cannot prove its own consistency (second theorem).\footnote{The theorems were further refined by \cite{rosser1936extensions}, who proved them without assuming $\omega$-consistency.} Inspired by Gödel's theorems, \cite{turing1936computable} and \cite{church1936note} independently provided negative answers to the ``Decision problem'' (Entscheidungsproblem) for FOL; Church's proof utilizes the undecidability of checking the equivalence of two expressions in the $\lambda$-calculus, whereas Turing constructs the halting problem for Turing machines. Their results establish that no decision procedure exists that can decide whether arbitrary FOL formulas are logically valid.\footnote{Validity means that true premises guarantee the truth of an argument's conclusion.} On the other hand, propositional logic (PL) — which, unlike FOL, excludes relations and quantifiers — is decidable. For instance, the satisfiability problem for propositional logic (SAT) was the first decision problem proven to be NP-complete \citep{cook1971complexity}, and has since remained at the center of computational complexity theory. Using Cook's results from SAT, \cite{karp1972reducibility} proved NP-completeness of 21 further problems, which not only demonstrated the intuitive appeal of the NP-class, but also that many natural computational problems are intractable.

However, instead of determining the resources needed to check whether some input satisfies some property X, we can ask, what is the complexity of \emph{expressing} X? The latter question is central for the field of \emph{descriptive complexity theory}, which defines complexity classes in terms of the type of logic required to express the languages in them. In fact, due to a wealth of results from descriptive complexity, it turns out that expressing and checking are intimately related.\footnote{See \cite{immerman1998descriptive} for the definite introduction to descriptive complexity.} The first major result in the field was Fagin's theorem \citep{fagin1974generalized}, which established that NP is exactly the set of properties that can be expressed in existential second-order logic (SO$\exists$), which unlike FOL, has the power to existentially quantify over properties and relations. In other words, every query that is computable in NP (including NP-complete problems) is equivalent to a query in SO$\exists$.\footnote{See chapter 7.1. in \citep{immerman1998descriptive} for a detailed proof.} Furthermore, since the complement of an existential formula (quantifying over \emph{some} members of a domain) is a universal formula (quantifying over \emph{all} members of a domain),\footnote{This follows from the fact that $\exists x \neg P(x)$ is equavalent to $\neg\forall x P(x)$.} it follows directly that co-NP is captured by SO$\forall$, and unrestricted second-order logic (SOL), which allows for both universal and existential quantification, is equal to the union of all classes in the polynomial hierarchy (PH). Other notable results include the fact that FOL corresponds to the logarithmic time hierarchy (LH) as well as the circuit complexity class AC$^0$, linearly ordered FOL systems with a least fixed-point operator yields P \citep{immerman1982relational,vardi1982complexity},\footnote{In short, fixed-point logics extend FOL with an operator which can construct fixed points of relational variables. For instance, if we view formulas with free relational variables as if they are determining maps on the relation space, the operator can define fixed points on this map.} SOL with a transitive closure gives PSPACE \citep{immerman1989descriptive}, and SOL with a least fixed-point corresponds to EXPTIME \citep{immerman1998descriptive,abiteboul1997fixpoint}.

These results generate the wisdom that the expressiveness of a language is directly correlated to the problems it can describe. However, it should be noted that, while computational and descriptive complexity are intimately related, they also have some crucial differences. For instance, one key incongruity is that, whereas descriptive complexity studies finite mathematical structures, computational systems operate on ordered encodings of problems and are thus able to enumerate objects which may be logically unordered. As an example, from a logical perspective, we might see a set of nodes in an graph as undordered, but as soon as it is transferred to the tape of a Turing Machine, it inevitably becomes ordered (and thus exploitable for various forms of operations).\footnote{However, this incongruity does not hold for Fagin's theorem, since SO$\exists$ can be used to declare the existence of some desired order.} Nevertheless, the fact that complexity can be characterized in terms of expressibility — without reference to some abstract machine — further establishes the natural appeal of the complexity classes. And for computational moral rule-followers — e.g., systems performing queries over databases containing moral norms — the consequences are profound yet somewhat clouded by its mere generality. For instance, languages describable in FOL corresponds to AC$^0$ (polynomial-size circuits of bounded depth), which allows one to perform integer addition and subtraction but not multiplication. Adding an operator which can compute the transitive closures of binary relations, on the other hand, makes it possible to produce structures that can answer reachability queries (e.g., is it possible to go from node $A$ to node $Z$ in $n$ steps?). Perhaps most interesting is the intractability of expressing queries which involve different forms of quantification over properties and relations (and not just objects, as in FOL), which is seemingly intuitive in natural language. For instance, even trivial moral queries of the type ``For all observable facts, possible actions, and obligations, is there some action which does not violate any moral obligations?'', might, in the worst-case, only be expressible in intractable complexity classes (e.g., if we assume that obligations are a relation between facts and actions). The point is, while quantification over properties and relations, transitive closure, and fixed points offer substantial expressive powers to well-studied collections of formal systems, it also raises intractability concerns; simply because language helps us to succinctly represent and communicate queries for moral rule-followers, it does not necessarily make such queries easy to compute.\footnote{See \cite{aaronson2013philosophers} for an interesting discussion of the related problem of ``logical omnicience'', which uses complexity considerations to challenge the view that if an agent knows certain facts, it also know every logical consequence of those facts.}

\subsubsection{The Complexity of Modal and Deontic Logic}
\label{sec:DeoDeonticLogic}

One of the most widely used fragments of FOL is modal logic; the go-to logic for representing necessity ($\Box$) and possibility ($\Diamond$).\footnote{Alternatively, since the main ideas of modal logic long predates FOL — e.g., Aristotle's modal syllogisms — it is arguably more accurate to describe modal logic as an expansion of propositional logic.} From the mid 20$^{th}$-century and onward, the standard semantics for modal logic is the possible world approach,\footnote{The approach was originally suggested by \cite{carnap1947meaning}, and later developed to its modern day form by \cite{kripke1963semantical}.} where $\Box P$ means that $P$ is true in \emph{all} possible worlds, and $\Diamond P$ means that $P$ is true in \emph{some} possible worlds (assuming that these worlds are accessible). One attractive feature of modal logic is that it, contrary to FOL, is robustly decidable.\footnote{For an in-depth exposition see Vardi's report ``Why is modal logic so robustly decidable?'' \cite{vardi1997modal}.} However, although it is decidable, typical problems for modal logic are in general not tractable. For instance, while the \emph{model-checking} problem, which asks whether a given formula is true with regards to a given state of a given Kripke structure\footnote{A Kripke structure is a graph that represents reachable states as nodes, state transitions as edges, and a labelling function that keeps track of the properties that hold in each state.} is solvable in linear time, the \emph{validity} problem, which asks whether a formula is true in \emph{all} states of \emph{all} Kripke structures is PSPACE-complete \cite{ladner1977computational}. More precisely, \cite{ladner1977computational} demonstrates that the validity problem for modal logic is PSPACE-complete for the systems K, T, and S4, whereas it is NP-complete for S5.\footnote{For readers unfamiliar with modal logic, K, T, S4, and S5 refer to the choice of axioms and rules that are added to the systems; e.g., K — the weakest version — only includes the necessitation rule (i.e., $(\models P) \Longrightarrow (\models \Box P)$) and the distribution axiom ($\Box(P \rightarrow Q) \rightarrow (\Box P \rightarrow \Box Q$); T includes the reflexivity axiom ($\Box P \rightarrow P$) in addition to K; S4 and S5 includes T along with iteration axioms 4 ($\Box P \rightarrow \Box \Box P$) and 5 ($\Diamond P \rightarrow \Box \Diamond P$), respectively.} These results where extended by \cite{halpern1992guide}, who proved that the validity problem is PSPACE-complete for multi-agent versions of K, T, S4, and S5 (i.e., a ``join'' of the logics used by at least two agents). The work of \cite{halpern1992guide} also shows that, while the addition of a distributed knowledge operator — allowing an ``all-knowing'' agent to combine the knowledge of everyone else — does not alter the complexity, the addition of a common knowledge operator — allowing everyone to know P, and that everyone knows that everyone knows ... that P holds — makes the problem EXPTIME-complete.\footnote{See also \cite{spaan2016complexity} for an exhaustive treatment of the complexity of modal logics.}

Naturally, similar intractability concerns plague other popular versions of modal logic. For instance, \cite{spaan1993complexity} showed that the PSPACE-completeness of the validity problem carries over to the tense case (temporal logic), i.e., with the addition of operators expressing ``it will always be the case that...'' and ``it will always was the case that...''. Similarly, \cite{sistla1985complexity} demonstrated that satisfiability for Linear Temporal Logic (LTL, introduced by \citep{pnueli1977temporal}, with operators for ``next'' and ``until'' (excluding the past), is either PSPACE-complete or NP-complete depending on the operators used.\footnote{See \cite{vollmer2009complexity} for a more detailed study of the complexity of temporal and propositional operators in LTL.} In general, PSPACE-completeness also holds for the model-checking problem for several versions of LTL \citep{schnoebelen2002complexity}.

Perhaps more interesting for the prospect of moral machines is the complexity results for variants of modal logics such as dynamic logic (DL, introduced by \cite{pratt1976semantical}) and deontic logic. DL adds the additional modal operators $[a]$ and $\langle a \rangle$, which makes it able to capture properties of program behavior; e.g., $[a]P$ means that after performing action $a$, it is necessary that $P$ is true (i.e., $a$ brings about $P$), and $\langle a \rangle P$ means that it is possible that P holds after $a$ is performed (i.e., $a$ might bring about $P$). Note that $a$ might refer to an entire program, which allows dynamic logic to formalize dynamics — e.g., transitions, sequences, and results — of complex algorithmic systems of multiple programs. In turn, Propositional Dynamic Logic (PDL, introduced by \cite{fischer1979propositional}) was developed to describe correctness, termination, and equivalence of computer programs on the basis of PL (instead of FOL, which was the basis of the first version of DL \citep{pratt1976semantical}). Interestingly, the decidability of checking whether a formula $F$ of PDL is satisfiable can be secured by having two sub-procedures running in parallel: one which enumerates all deducible formulas (R1), another which enumerates the finite models of PDL and tests whether they satisfy the formulas (R2). In this way, if $F$ is satisfiable, a model which satisfy $F$ must eventually be found. If $F$ is satisfiable, R2 will at some point answer ``yes'', if not, R1 will at some point answer ``no'' (and the procedure halts when either sub-process gives an answer). Nevertheless, although PDL is attractive for formal verification of program behavior, it is less attractive for computational moral agents bounded by polynomial time, as SAT for PDL is EXPTIME-complete \citep{fischer1979propositional,pratt1980near}.\footnote{A more practical algorithm for PDL-SAT — although still EXPTIME-complete in the worst-case — has been offered by \cite{de2000combining}.}

Deontic logic (introduced by \cite{von1951deontic}), on the other hand, aims to capture the logical features of moral concepts such as permissions (typically denoted by operator $P$) and obligations (typically denoted by $O$). For instance, the axiom $O(A \rightarrow B) \rightarrow (OA \rightarrow OB) $ states that ``if it is obligatory that $A$ implies $B$, then $B$ is obligatory if $A$ is obligatory''. Likewise, Kant's ``ought implies can'' can be expressed by $OA \rightarrow \Diamond A$. It should thus be no surprise that deontic logic has been a popular framework for implementations in machine ethics, e.g., for automatized ethical reasoning \citep{arkoudas2005toward,wiegel2009combining,furbach2014automated,malle2017networks}. However, compared to other versions of modal logic, the computational complexity of deontic logic is relatively unexplored. One reason is that the inherent complexity of (and relationship between) normative concepts — e.g., of agency, right, responsibility, and commitment — may be drawn to arbitrary levels of detail \footnote{See \cite{sergot1998normative} for an exposition, which also appears in the handbook on deontic logic by \cite{gabbay2013handbook}.} Another reason is the problem of agreeing on the appropriate semantics for norms, e.g., whether they are based on possible worlds, axiomatic constructions, or operational executions. A more fundamental challenge for the semantics of deontic logic is captured in ``Jørgensen's dilemma'' \citep{jorgensen1937imperatives}, which asks whether arguments that contain norms (e.g., imperatives) express a truth or not. In standard conceptions of logical entailment, it is essential that true conclusions follow from true premises; i.e., conclusions and premises can be true or false. However, since imperatives are – supposedly – neither true or false, they cannot play any meaningful role in the validity of arguments, and as a results, we cannot justify imperatives on the basis of logical reasoning. At the same time, reasoning with imperatives seems to be logically valid in cases where the premises and conclusions are imperatives (Jörgensen, 1937, p. 290):

\begin{quote}
(P1) Love your neighbor as yourself \\
(P2) Love yourself \\
(C) Love your neighbour
\end{quote}

Thus, the dilemma entails that we either have to accept that normative statements cannot have truth values, or deny that the premises and conclusions of our argument have truth-values. In turn, these considerations have given rise to several families of deontic logics, which can be roughly divided into two camps: the possible-world approach — e.g., standard deontic logic, ``Seeing To It That'' (STIT) logic \citep{horty2001agency} and dynamic deontic logic \citep{meyer1988different,van1996dynamic} — and approaches that are not based on possible-worlds, such as input/output logic \citep{makinson2000input}, imperative logic \citep{hansen2008prioritized}, defeasible deontic logic \citep{governatori2013computing}, and prioritized default logic \citep{horty2012reasons}.

With regards to complexity, the first path entails that one has to deal with the PSPACE-completeness that often plague possible-worlds semantics \citep{halpern1992guide}, or the EXPTIME-completeness of dynamic logic \citep{fischer1979propositional}. In fact, it is shown that the satisfiability problem for a fragment of STIT, which is used to construct deontic STIT, is undecidable \citep{schwarzentruber2014stit}.\footnote{See also \cite{balbiani2008alternative,herzig2008properties,xu1998axioms} for other results on the complexity and decidability of STIT.} As an alternative to possible worlds, \cite{sun2017complexity} has investigated the complexity of input/output logic on the basis of a `norm-based' operational semantics. In this view,  declarative statements can be true or false whereas norms cannot: they are simply violated or complied. More precisely, norms are ordered pairs ($x,y$) that corresponds to deductive operations of the input/output-system, taking a fact ($x$) as input, and producing an obligation ($y$) as output. However, their investigation finds that decision problems of input/output logic are shown to be NP/co-NP-hard and in the 2$^{nd}$ level of the polynomial hierarchy (i.e., NP$^{\text{NP}}$ and co-NP$^{\text{NP}}$ ). By contrast, \cite{governatori2013computing} has demonstrated the computational tractability of a deafeasible deontic logic able to compute `weak' (allowed unless explicitly prohibited) and 'strong' permission (only allowed if explicitly permitted). However, the constructed object logic has a rather constrained expressibility, as it only includes propositional symbols and their corresponding negations, and only modal literals are allowed (obtained by applying modal operators for obligation and permissibility to propositional literals).

In summary, the surveyed results support the conclusion that using modal logic — e.g., temporal logic, dynamic logic, deontic logic — to represent or automatize moral reasoning generally introduces computational intractability. Naturally, in order to capture the rich intricacy of ethical life, normative notions in logical form must be able to account for knowledge, time, agency, program behavior, and multi-agent dynamics, which inevitably adds complexity. However, it must be noted that while these results might appear to indicate something deeply problematic for the prospect of moral machines constrained by polynomial time, they are also immensely productive for other practical purposes. For instance, knowing that a problem is provably in PSPACE may allow it be related to a vast set of other PSPACE-problems, along with the algorithms, approximization techniques, and heuristics that have been developed to tackle them. The fact that the satisfiability problem for dynamic logic is EXPTIME-complete does not stop it from being useful in formal verifications of program behavior. More broadly, from a design perspective, the complexity results may be of great help to inform ones choice of formal system — its expressivity, operators, syntax, and semantics — for the practical purpose at hand.

\subsubsection{The Problem of Moral Semantics}
\label{sec:deoProblemOfSemantics}
The problem of semantics comes in various variants, some of which have been subject to extensive philosophical treatment for centuries. Earlier in this section, we noted that, if an agent should follow a rule of the type ```If input $X$  $\rightarrow$ do action $Y$'', we assume that the agent has some non-trivial understanding of what $X$ and $Y$ means. In the semantics of propositional logic, we simply say that the terms are either true or false with respect to some model (or in modal logic, with respect to possible worlds), and provide rules (e.g., a truth table) for determining the truth value of sentences made up of those terms and logical connectives (e.g., $\neg,\land,\lor,\rightarrow,\Leftrightarrow$). Similarly, semantics in programming languages might simply define the process of how valid strings in the syntax (instructions) will induce (interpretation) certain state transitions (execution), e.g., by manipulating some data structure. In such cases, semantics may be viewed as a purely mechanistic operations, and the job of establishing the link between model and reality — e.g., how any valid sentence that can be expressed by the language corresponds to some real-world state of affairs — is circumvented.

Nevertheless, human moral discourse — e.g., of virtues, principles and judgments — is abundant with descriptions and concepts that are semantically so-called `thick'. For instance, moral talk involve `thick descriptions' that embed subjectivity as part of their meaning, e.g., by explaining individuals' behavior in light of internal motivations \citep{geertz1973interpretation}. That is, when we describe an agent as being courageous or fair, we naturally assume that the agent has some subjective characteristics (e.g., psychological dispositions and experience) that exemplifies courage and fairness. In turn, these subjective characteristics might only make sense for an agent, given that the agent also understands how they relate to her \emph{own} subjectivity (e.g., sensori-motor experience, beliefs, desires, and intentions). Moral talk also involve paradigmatic examples of `thick concepts' that are both descriptive and evaluative \citep{blackburn1998ruling}; e.g., terms like \emph{generous} and \emph{selfish} can refer to descriptions of certain behaviors (acts of sharing or not sharing one's food), while simultaneously denoting an evaluative quality (being \emph{good} respectively \textit{bad}). 

Unfortunately, thick concepts and descriptions invoke issues that remain at the center of long-standing metaphysical and meta-ethical debates. To explicate the meaning of thick descriptions, in so far as they depend on subjectivity, one might in turn require satisfying answers to other fundamental challenges, such as the symbol grounding problem \citep{harnad1990symbol}, the hard problem of consciousness \citep{chalmers1997conscious}, and the meta-ethical problems of determining the meaning of ``meaning'' and truth with regard to moral terms. The first problem is relevant for any system that make use of symbols — e.g., to communicate or reason — as it concerns how symbols ground their meaning. For over four decades, Searle's Chinese Room experiment \citep{searle1980minds} has provided a venue for philosophical discussions about the limits of whether and to what extent computational systems can ``understand''. Searle famously argues that a computer program that is able to convince a human Chinese speaker that it understands Chinese, does not literally ``understand'' Chinese, as it is merely following syntactic rules. But even if we oppose to Searle's broad rejection of computationalism,\footnote{Computationalism is the family of views that hold that the human mind — including consciousness and cognition — is some form of computation.} we still need to give some account of how thick moral terms get their meaning, or more pragmatically, what they `do' for our moral practices. However, it very much remains an open question to what extent the grounding of symbols rely on, e.g., intentionality \citep{brentano1874psychology}, perceptual experience \citep{barsalou1999perceptual}, or certain sensorimotor capacities \citep{taddeo2005solving}, and yet another open question whether such capacities can be carried out by a computational system.\footnote{\cite{aaronson2013philosophers} has framed a technical version of the related ``Waterfall argument'' in light of complexity considerations, which supplements the Chinese Room by claiming that meaning is always relative to some external observer \citep{searle1992rediscovery}. The argument starts from the observation that any physical system with a sufficiently large state space could in principle implement the semantics of \emph{any} other system; e.g., for some mapping $M$ from a waterfall's initial states $I$ to final states $F$, there is a way of labeling any given permutation $P$ from $I$ and $F$ such that $M$ implements $P$, and $P$ may thus represent any ``semantics'' we like, such as a chess playing program. However, if we actually tried to use a waterfall to compute chess moves, we would need to find a reduction from the chess program to the waterfall, e.g., by showing how chess positions and chess moves can be efficiently tracked to the waterfall's initial and final states. From this, Aaronson conjectures that, for any given chess program with access to a waterfall oracle, there is another chess program with equally good performance and similar resource requirements that does not access the waterfall oracle. In other words, it seems highly probable that any reduction algorithm from chess to waterfalls would simply solve chess problems, and not use the waterfall in any meaningful way. For Aaronon, this mirrors the more substantive notion of completeness: the class that a problem is reduced to cannot itself be sufficient to solve the same problem. I.e., while NP-problem $X_1$ can be solved in polynomial time with access to an oracle for $X_1$, problems in P cannot be reduced to problems in P (as this would imply that every problem in P is P-complete). The presumed equivalence between waterfall and chess computation thus carry little substance, unless the equivalence can be demonstrated in a model of computation that itself isn't capable of solving waterfall or chess problems.} Similarly, to explicate thick concepts, one naturally has to assume that it is possible to disentangle their evaluative and descriptive components.\footnote{This would, somewhat coarsely, be the position of \cite{blackburn1992through} and \cite{hare1952language}. In contrast, thinkers like \cite{putnam2004collapse} and \cite{williams2006ethics} see thick concepts as indivisible blends of fact and value. For instance, Putnam states that `thick' ethical concepts “simply ignores the supposed fact/value dichotomy and cheerfully allows itself to be used sometimes for a normative purpose and sometimes as a descriptive term” (Putnam, 2004, p. 35).}

So how do these deeper problem of semantics relate to computational complexity? Of course, it ultimately depends on how we position ourselves with respect to these debates, as they would determine our view of moral semantics, and the role moral terms play in moral behavior. On the one hand, if we hold — following meta-ethical cognitivism — that the meaning of moral terms can be properly disentangled and defined as truth-bearing entities, problems expressed in any `thick' moral language would in principle be computed in the ways discussed in section \ref{sec:DeoDeonticLogic} (e.g., the checking validity with respect to a given or all possible Kripke structures). As such, moral semantics would not yield any additional complexity baggage that does not already follow from the moral languages' expressiveness. However, while cognitivists hold that moral terms can express mind-independent facts about the world, it does not mean that we have found them; like divine command theory, the approach presupposes that one has collected exhaustive knowledge of all moral terms — e.g., via access to Platonic reality — and managed to encode them along with all their semantic relationships.\footnote{Of course, this is not unfeasible for extremely limited state spaces. For instance, we can imagine a toaster that uses sensors to read whether a toast is under-baked, baked, or burnt, and understand how evaluative sentences ``this toast is bad!'' express certain facts about about the toast's states.} At present, nevertheless, there is strong disagreement about what moral terms mean.\footnote{Recalling Jørgensen's dilemma from the previous section, we also note that there are profound disagreements about valid inference in deontic logic.}

A skeptical alternative would be to bite the bullet of undecidability. For instance, Rice's theorem states that all non-trivial semantic properties of a language recognized by a TM are undecidable, where semantic properties are about a programs behavior, and non-trivial properties are neither true for all partially computable functions, nor false for all partially computable functions.\footnote{Conversely, a partial function is trivial if it is true for all partial computable functions or for none.} To illustrate, we can imagine a moral language ML as a set of TM descriptions, and the criteria for TMs to be in ML is that their language L(TM) accepts at most three strings ($|$L(TM)$|\leq$ 3). If a language $M_{1}$ belongs to ML ($M_{1}\in \ $ML), it means that $M_{1}$ satisfies the property ML. That ML is a property of TM languages can be shown by the fact that two machines with the same language, $M_{1}$ and $M_{2}$, are either both in ML or neither in ML; since they have the same language, they have the same number of strings. To show that ML is non-trivial, we let $M_{3}$ be a machine that accepts every string, and $M_{4}$ be a machine that rejects every string. Since $M_{3} \not\in$ ML (accepting more than three strings), and $M_{4} \in$ ML (0 is $\leq$ 3), it follows from Rice's theorem that ML is undecidable. Of course, we can let ML denote a great number of things, and given the vast generality of Rice's theorem, we can demonstrate undecidability for a large set of problems pertaining machine behavior.\footnote{For instance, whether a given TM computes a constant function, a total function, the identity function, add two natural numbers, or a computable function can easily be shown to be undecidable using Rice's theorem.} Analogously, we can draw a skeptical conclusion about the prospect of moral languages; as long as there are some disagreement about the meaning of moral terms (e.g., different agents computing different outputs), there can never be a decidable moral language.

There are, of course, many other palpable reasons to be skeptical about the decidability of moral semantics. For instance, both computability and theories of meaning finds a common nemesis in the self-referential Liar Paradox: ``this sentence is false''. It plays a central role in Gödel's first incompleteness theorem — by replacing ``false'' with ``not provable'' — as no consistent system of mathematics can prove truths about itself. Similarly, Turing's Halting problem demonstrates that it is undecidable whether a computer program halts, as for any program $P$ that can decide ``Yes'' for halting can be countered by another program that uses $P$ as input in order to produce the opposite ``No''.\footnote{In turn, the more general Rice's theorem can be proven by reduction from the Halting problem.} On the side of theories of meaning, Alfred Tarski found that the Liar Paradox only appears in ``semantically closed'' languages, i.e., a language that can express the truth of its own sentences \citep{tarski1944semantic}. Tarski's own solution — to separate the referring meta-language from the referred object language in a constructed hierarchy — was in turn found to be incomplete by \cite{kripke1976outline}, who, among other things, employed self-referential tricks to produce statements that break the hierarchy.\footnote{Kripke gives the following example, expressed by Jones and Nixon: ($J$) ``Most (i.e., a majority) of Nixon's assertions about Watergate are false'', ($N$) ``Everything Jones says about Watergate is true'' (Kripke, 1976, 691). In the Tarskian hierarchy, $N$ needs to be on a higher level than everything that Jones says, and $J$ needs to be on a level higher than what everything Nixon says.} Of course, Kripke's solution — which utilizes partially defined truth predicates (``undefined'') — can in turn be targeted by a \emph{strengthened liar paradox}: ``this sentence is either false or undefined''. Ultimately, as any solution seems to produce new self-referential problems that applies to the new solution, any semantic theory centering on truth is haunted by a paradox out for revenge \citep{beall2007revenge}. Moreover, as demonstrated by \cite{dahl2022fixed}, the Liar Paradox is not content with semantic theories based on truth, but extends to all theories that seeks a unified explanation of meaning for \emph{any} language. In short, since any unified theory of meaning requires a language that is expressive enough to assert its own meaning, and no language can coherently assign meaning to itself while articulating the unified theory, it follows that a universal theory of meaning is impossible.

A more pragmatic route is to deny that moral terms can be true or false, and instead say, following \cite{wittgenstein2010philosophical}, that ``meaning is use''. Besides purifying moral semantics from Platonism and paradoxes, it would potentially bring moral talk closer to the eclectic social practices from which it stems, where occasional quarrel, misunderstanding and dissent is inevitable. In this view, moral expressions should not be understood in virtue of any formal account of meaning, but rather how they, in more or less satisfactory ways, serve our moral practices. Presumably, this could even circumvent the undecidability of non-trivial semantics and halting, if we accept that it is no problem that arbitrary programs either do `this or that'; pragmatically, they either work well, or they do not. In place of truth-conditions, we could adopt one or several of the prominent non-cognitivist approaches to moral language, e.g., that moral statements function to express emotion and elicit emotion in others \citep{stevenson1937emotive}; to assert prescriptive judgements \citep{hare1952language}; or convey attitudes \citep{schroeder2010being}.\footnote{It should be noted that these non-cognitivist positions should not be equated with a Wittgensteinian outlook on meaning per se, but rather that they reject the belief that moral terms are truth-apt.} However, while such theories may be credible for emotional, judgemental, and affective humans, they are less suitable for a complexity analysis, as they often presuppose a human-specific psychology.

Nonetheless, one way to analyze the complexity of potentially `mindless' agents' use of moral terms is to look to modern Wittgensteinans such as David Lewis and Robert Brandom. Before his work on counterfactuals, Lewis provided an early game-theoretic analysis of social conventions \citep{lewis1969convention}. In this view, following the footsteps of \cite{schelling1960strategy}, linguistic as well as moral conventions can be viewed as self-perpetuating solutions to reoccurring \emph{coordination problems}, where it is mutually beneficial for self-interested agents to coordinate their actions. Linguistic meaning, more particularly, have subsequently been explored within the paradigm of \emph{signaling games}. In a simple signaling game, a messenger seeks to convince receivers that they are of a certain type — e.g., that they are competent — where the actual type is only known to the messenger. Intuitively, no honest messenger benefits from being misunderstood (e.g., conveying false information about their type), just as it is beneficial for incompetent players to lie, and players receive payoffs depending the receivers' responding action (e.g., hire agent $a$ or $b$). Given its explanatory power, signaling games have been used to model the development of communication and linguistic meaning \citep{skyrms2010signals,huttegger2007evolution}.

Since signalling games are typically modeled as sequential Bayesian games, their equilibria solutions are plagued by the intractability concerns discussed in section \ref{sec:deoComputingEquilib}; in particular the NP-hardness of checking whether a Bayesian game has a pure-strategy Bayesian equilibrium, and the PSPACE-hardness of computing a pure-strategy NE in Markov Games \citep{conitzer2008new}. However, Lewis' own equilibrium concept, called \emph{coordination equilibria}, has the property that every player also prefers that every other player conform to some regularity $R$, on the condition that at least \emph{all but one} player conform. Naturally, this presupposes some concept of common knowledge, which was subsequently generalized by Aumann's correlated equilibrium \citep{aumann1974subjectivity}. As discussed in \ref{sec:deoComputingEquilib}, CE is computationally attractive, as they can be found in polynomial time for games with any number of players \citep{gilboa1989nash}.\footnote{See also the work of \cite{urbano2002computational}, which demonstrates how correlated equilibrium can be achieved by imposing computational restrictions on the unmediated communication.} In Lewis' original definition, however, conventions are necessarily arbitrary, in the sense that there is a conflicting regularity $R'$, which could have become the stable convention (and $R$ and $R'$ are mutually exclusive). Of course, the concept of arbitrariness certainly has explanatory value: while it is not arbitrary that a mutually beneficial solution to a coordination problem becomes convention (e.g., cars driving on opposite side of the roads), it is arbitrary that a particular solution was chosen over another (e.g., right-hand traffic as opposed to left-hand traffic). Still, the arbitrariness criteria does not exclude the possibility that a conflicting regularity could result in an even greater joint benefit. Thus, since  stable conventions become self-perpetuating, their potential moral value relies on the same optimistic conservatism that permeates post-hoc evolutionary explanations of altruism and social contracts; that there are good reasons why conventions are as they are. By contrast, it is possible for a convention to have an exceptionally high price of anarchy (i.e., a terrible worst-case equilibrium), while the stability of the convention ensures that no one believes that anyone would benefit from going against the grain. However, evaluating the convention against alternatives, e.g., by computing a CE that maximizes the expected joint benefits, invites NP-hardness \citep{papadimitriou2008computing}.

On a more optimistic note, these computational difficulties may be mitigated by cooperative principles for communication, e.g., following the work of \cite{grice1975logic}. This strategy can be reflected in Lewis' refined analysis of convention \citep{lewis1975languages}, which includes preferences with regard to one's own beliefs: ``The expectation of conformity ordinarily gives everyone a good reason why he himself should conform'' (Lewis, 1975, p. 8). Similarly, Lewis writes that ``a language $L$ is used by a population $P$ if and only if there prevails in $P$ a convention of truthfulness and trust in $L$, sustained by an interest in communication'' (p. 10). For Lewis, along with other Griceans such as \cite{schiffer1972meaning} and \cite{bennett1976linguistic}, meaning ultimately stems from a coordination between speakers communicative intentions and receivers communicative expectations. In turn, these intentions and expectations may as well encompass cooperative conventions.

A related yet distinct account of ``meaning is use'' that is apt for explaining moral language has been provided by \cite{brandom1994making}. Brandom gives a theory of sapience — the type of rationality that humans possess — based on the notion of discursive practice, which can be summarized as ``the game of giving and asking for reasons'' (p. 6). Participants of discursive practices take on entitlements and commitments, which can be seen as carrying the normative force of permissions and obligations, respectively. At the core of discursive practices are inferential relations, which preserves commitments and entitlements to other statements. I.e., sentences only carry content in terms of their function, which is inferred in relation to other sentences. What is interesting with regards to moral discourse, is that linguistic performances are characterized by their ability to alter the normative status of the members of a discursive practice. This takes the form of a conversational ``scorekeeping'', where the participants keep track of commitments and entitlements within the conversational context, e.g., by making, acknowledging, contesting, or withdrawing assertions.\footnote{The idea of scorekeeping in the context of conversations was first introduced by \cite{lewis1979scorekeeping}, who acted as a supervisor for Brandon's doctoral thesis in the 1970s.} In essence, the normative pragmatics of linguistic performances determine the inferential aspects of semantics, and not the other way around. However, while Brandom's inferential semantics may help to illuminate the normative commitments of speech acts, we can only speculate about the computational aspect that underpins the ability to successfully participate in a discursive practice, and similarly, whether and to what extent it alleviates or adds computational demands. On the one hand, the game of reason-giving and reason-asking seems tailored to effectively foster cooperative communication in normative life. On the other hand, Brandon's conception of sapience seems profoundly human, which, in addition to the sentience shared with non-verbal animals, involves an understanding of conceptual contents, which in turn may encompass intentional states, beliefs, and desires of oneself and others. In turn, a sapient game of reasons might more or less correspond to a Kantian conception of moral rationalist discourse \citep{brandom2006kantian}, which presupposes sophisticated and idealized capacities for moral autonomy and freedom; capacities which — at least presently — cannot be construed in computational models.

The main lesson seems to be that, while game-theoretic concerns — of computational intractability and threats of repugnant equilibria — may be alleviated by communication, the challenge is to not only give an account of how communication works, but why it works so well. For Lewis, the success seems to rely on communicative intentions and expectations of speakers, and the cooperative conventions that results from it. For Brandon, it centers on the sapient game of ``giving and asking for reasons''. Unfortunately, this renders both approaches rather computationally opaque: as they aim to explain human communication, they can, just like non-cognitivist theories, resort to uniquely human features that remain more or less impenetrable from a computational perspective. Still, such theories might potentially yield significant value by enclosing the gap between, on the one hand, cognitive-psychological resources — e.g., reasons, trust, and communication — that fosters efficient cooperation, and on the other hand, the computational architectures that would potentially enable them. In turn, complexity considerations might help to illuminate the rich interrelationship between game-theoretic dynamics, social-psychological capacities, and normative theory in everyday moral interactions.

In summary, this section has discussed more profound issues for moral semantics that, although relatively ignored in machine ethics, remain at the center stage of meta-ethical and meta-semantical debates, as well as in theories of communication. Of course, a simple move would be to cling on to some metaphysical argument against ``strong AI'' \citep{searle1980minds}, some advanced requirements for symbolic grounding \citep{taddeo2005solving}, or some uniquely human psychology of emotions, and conclude that ``machines can never genuinely understand moral language''. But even given that one decided to ignore such problems, many complex problems remains for computational moral semantics.




\subsection{Consequentialist-Deontological Hybrids}
\label{sec:deontologyconshybrids}

Having investigated a range of computational aspects of both consequentialism and deontology, we are now in a position to say something more substantial about their difference and potential combination. First, it should be concluded that the claim ``deontology is, computationally speaking, less complex than its alternatives'' cannot be given a straightforward answer; and in many cases, it is simply false. Based on the material discussed in this section, we can outline a more nuanced answer:

(1) As argued in section \ref{sec:deoHumanRuleFollowing}, it is a mistake to view rule-following in legal and liberal contexts as computationally simple. While legal rules may efficiently compress voluminous moral wisdom, it is only against the backdrop of a complex relationship between the mechanisms that incentivizes their adherence (e.g., police and punishment), ensures their just interpretation (e.g., courts), along with the moral sentiments of the subjects the laws apply to. For instance, the complexity of ``do no harm''-principles in liberal contexts can only be meaningfully analyzed in relation to the capacities of fully autonomous citizens.

(2) On the other hand, deontological rules may be significantly more efficient than alternatives if they are justified on the basis of divine command or legal positivism. However, unless in extremely simplified cases, the knowledge-requirements for such approaches to work in practice remains unfathomably vast, while the knowledge itself  may be highly contentious.

(3) As an alternative, deontological rule-following may include justification as a part of the moral computation, e.g., by considering how rule-following behaviors affect others. As discussed in \ref{sec:deoGeneralPurposeRules}, this can range from considerations of (i) one's own preferences (``I treat you based on what I personally prefer''), (ii) preferences of others (``I treat you based on what I know of what you prefer''), to (iii) general behavior (``I generally treat you in the way that I would want others to generally treat me''). While weaker versions — e.g., (i) and (ii) — may be relatively simple from a complexity perspective, they also lead to a range of other problems. By contrast, stronger versions — e.g., contractarian or contractualist versions of (iii) — are obfuscated by extreme variances with regard to behavioral expectations and the capacities that makes up conceptions of general behavior.

(4) Any general-purpose normative theory that seeks to account for multi-agent dynamics face game-theoretic concerns. This includes issues of rationality (\ref{sec:deoGameTheory}), incomplete information and recursive reasoning (\ref{sec:deoGameTheoryBayesian}), and the intractability of computing morally attractive Nash Equilibria (\ref{sec:deoComputingEquilib}).

(5) Any computational system that employ formal logic, e.g., for deontological rule-following, moral reasoning, or communication are subject to expressibility (\ref{sec:DeoDescriptive}), intractability \ref{sec:DeoDeonticLogic}, and decidability (\ref{sec:deoProblemOfSemantics}) issues that permeate the syntactics and semantics of logic.

(6) Finally, it should be acknowledged that deontology is a family that encompasses a range of ethical theories that may in turn emphasize a range of different cognitive abilities and computational resources. It can be a moral rationalist project of finding universally justifiable and applicable rules on the basis of a shared autonomy and rationality. It can be a contractarian project of finding mutually beneficial action-rules based on self-interest. In the simplest case, rules can act as merely `fast and frugal' heuristics that support agents with bounded cognition to produce \emph{any} action in complex or novel situations, even if there is no way to evaluate whether the performed action is appropriate in the specific situation; it may simply be an automatic `default' action, or rely on optimistic conservatism (``it has worked well before, so it might also work well in the future''). In more ambitious cases, deontological rule-following might ask one to compute morally attractive equilibria for large-scale coordination problems with incomplete information.

Based on these considerations, it may be misleading to compare the complexity of deontology and consequentialism, as any comparison relies on a particular conception of what the theories dictate. For instance, both theories are equally targeted by game-theoretic intractability (\ref{sec:deoGameTheory}) insofar as they take other agents into account. Similarly, many of the deeper problems of moral semantics that plague logical systems (\ref{sec:deoProblemOfSemantics}) can also be construed for consequentialist agents: e.g., what is it for a computational system to ``understand'' what a certain utility really is? Another reason is that deontology and consequentialism may converge on their solutions, for instance, in cases where the deontological moral right also produces the optimal outcome. Thus, a reasonable alternative is to view them as complementary theories, which emphasize different cognitive capacities, computational resources, or aspects of ethical life.

To that end, it is no surprise that hybrid accounts, which combine aspects of consequentialism and deontology, are well-reflected in moral philosophy \citep{brandt1979theory,hare1981moral}, moral psychology \citep{kahneman2011thinking,greene2007vmpfc} and machine ethics \citep{Bauer2020,Arkin2007,Dehgani2008,StensekeBalkenius2022,pontier2012toward,azad2014new,tufics2015grafting,govindarajulu2017automating,pereira2009modelling}. However, it should be noted that in almost every case, deontology takes on a rather narrow role in these architectures — often within a broader consequentialist framework — which does not reflect its rich tradition in ethical theory. For instance, deontology might act as \emph{a priori} constraints which prevent certain intrinsically bad actions \citep{pereira2009modelling,Dehgani2008}. Another option is to use deontological rules to foster run-time efficiency, e.g., by turning consequentialist computation or exploration into exploitable rules \citep{StensekeBalkenius2022}.\footnote{Since consequences motivates the use of certain rules, it is more suitable to view this solution as a form of two-level utilitarianism \citep{hare1981moral}.} Similarly, in the dual process theory of moral cognition \citep{greene2007vmpfc}, deontological judgements are construed as fast and instinctive responses, whereas consequentialist judgements denote slow and reflective reasoning processes.

With regards to complexity, the emerging trend is that moral rule-following can either (i) be used to prevent intrinsically bad actions, (ii) reduce the run-time complexity of (often consequentialist) moral computations, or (iii) act as a `principles first' or `defeasible' default mode in situations where the agent has nothing else to base their decision on. From a moral-psychological perspective, (ii) and (iii) can be supported by the idea that automatic moral judgements compress moral wisdom, e.g., on the basis of evolutionary adaptations promoting cooperation, culture-specific norms fostering collective well-fare (e.g., salient correlated equilibria), or the moral lessons an individual learns to internalize through experience. For machine ethicists, consequentialist-deontology architectures thus offer an attractive smorgasbord of context-specific modularity; from top-down a priori constraints based on preexisting moral knowledge (divine command) to bottom-up a posteriori turned into new top-down constrains (e.g., learning efficient rules through experience). However, this image is not without its flaws, as any reduction in run-time complexity implies one or many of the following problems:

\emph{The problem of optimistic conservatism} — In essence, `fast and intuitive' moral behavior cannot be guaranteed to produce good moral results, unless it is evaluated against alternatives (e.g., on the basis of experience or moral reasoning). As such, the moral value of intuitions rely on an optimistic conservatism, as the efficiency of following an intuitive deontological (or rule-consequentialist) rule may come at the expense of losing out on a potentially better outcome that was only attainable via further reflection. Following \cite{hare1981moral}, this is the challenge of knowing when to think like a `prole' (using moral intuition) or an `archangel' (using critical reflection). From a computational perspective, it reflects the well-studied explore-exploit dilemma (\ref{sec:dynamicenvironments}). It also echoes the problematic post-hoc rationale of norms and conventions. As discussed in \ref{sec:deoGameTheory} and \ref{sec:deoProblemOfSemantics}: even if we believe that a certain moral convention — or rule, norm, principle — emerged because it helped agents with bounded cognition to coordinate towards mutually beneficial goals, the stability of the convention may rely on an optimistic conservatism about the convention itself, even if there are alternative conventions that might have been even better. Similarly, just because some moral intuitions may be a result of adaptations that yielded reproductive success for our evolutionary ancestors, it does not automatically make those intuitions morally right today \citep{greene2014beyond}.

\emph{The problem of speed vs performance} — A related problem is the choice between moral speed and moral performance. For instance, even if we, like \cite{hare1981moral} and \cite{greene2014beyond}, believe that slow utilitarian deliberation should take normative and epistemological priority over `fast' intuitions, there is still an open-ended conflict between moral speed and moral performance. In what circumstances do we opt for a fast — e.g., feasible, suboptimal, or satisficing — option in favor of a slower but potentially better option?

\emph{The problem of reasoning and learning} — `Fast and intuitive' moral intuitions may foster run-time efficiency, but only by ignoring the complexity of the process that gave rise to the moral intuition itself. On the one hand, this can refer to the complexity of the process characterized by \cite{hare1981moral} as the ``critical level'' of moral thinking (which governs the principles of the ``intuitive layer''). On the other hand, we might think of it purely in terms of a learning-process: e.g., how many learning examples and how much training-time does an agent need to successfully internalize a moral intuition that allows it to effectively solve a given moral problem at run-time? In principle, since artificial neural networks can be used to approximate \emph{any} function \citep{scarselli1998universal}, machine learning systems should be able to achieve great run-time performance in moral problems given the right kind of learning. However, as we will see in section \ref{sec:learning}, this will instead put increasing demands on sample and training-time complexity, while introducing other problems related to induction.

\emph{The problem of knowledge} — Instead of reasoning or learning, efficient moral intuitions may be secured on the basis of moral knowledge. For instance, an alternative that is available for moral realists and meta-ethical cognitivists — who believe that moral intuitions can denote true propositions that reflect subject-independent features of the world — is to collect and write down the dicta of objective moral reality. Of course, as already noted several times throughout this paper, such a project seems, for a variety of reasons, deeply problematic. With that said, it does not exclude the possibility that there are \emph{some} moral intuitions that have universal (or near-universal) consensus, or some — following \cite{asimov1942runaround} — that are particularly attractive for computational agents.

In summary, if we view deontology simply as `rule-following', it is clear that it yields efficient decision-procedures from a bounded rationality point-of-view. Nevertheless, while it may lead to attractive run-time performance, the moral value of such efficiency is either based on optimistic conservatism, the results of some other complex process (reasoning or learning), or the collection of vast moral knowledge. Essentially, the moral power of rules — their general applicability, general justification, and computational simplicity — can only be secured in complex ways. After all, the simplicity of moral rules should not prevent one from questioning their authority, but rather, it should help one remember that they — like logical systems — merely represent idealized facets of byzantine phenomena.

\section{Virtue Ethics and Moral Machine Learning}
\label{sec:learning}

As deontology centers on actions and consequentialism on results, they can be naturally construed as moral decision-procedures. By contrast, virtue ethics is about \emph{being} rather than \emph{doing}: instead of focusing on what the right action is, or what action yields the best outcome, it asks us to foster our moral character and the internal dispositions — virtues of courage, fairness, and temperance — that enables us to \emph{be} morally virtuous. More broadly, virtue ethics denote a vast family of ethical traditions that emphasize the role of our moral character, and the theory can find its diverse origin in thinkers such as Confucius and Mencius in the East, and Plato and Aristotle in the West \citep{Crisp1997}. In contemporary times, it has earned its spot as a central normative theory in Anglophone moral philosophy through the contributions of \cite{Anscombe1958}, \cite{Nussbaum1988}, \cite{Hursthouse1999}, and \cite{Annas2011}.

In machine ethics, virtue ethics has several times been proposed as an appropriate blueprint for the creation of artificial moral agents, as it emphasizes aspects of ethical life that are relatively ignored in other theories \citep{Coleman2001,Wallach2008,Howard2017,Stenseke2021}.\footnote{See \cite{stenseke2022AVA2} for a recent survey on computational implementations of virtue ethics.} Taking the character as the central object of moral evaluation — which encompasses both rational deliberation and psychological dispositions — it paints a more holistic picture of what it is to be moral; capturing not only what we ideally \emph{ought} to do, but what motivates us to act in morally praiseworthy ways. One key aspect of this picture is the notion of \emph{phronesis} (``practical wisdom''), which can be construed as the moral wisdom or skill an agent learns from practice and experience \citep{Annas2011}. The focus on development and learning has in turn inspired machine ethicists to unify virtue ethics with connectionism; both in terms of modern machine learning methods, and as a broader theory of cognition \citep{Casebeer2003,Wallach2008,Howard2017,Berberich2018}. Optimistically, with the ability to constantly learn from experience, be sensitive to contexts and adaptable to changes, a virtuous machine might thus be able to apprehend the intricacies of human norms in dynamic environments where mere utility-maximization or rule-following fails.

In despite of these promises, virtue ethics remain relatively elusive from a computational perspective. One reason is that virtue ethics is founded on a deeply human view of what constitutes a moral life — of flourishing, reasoning, emotions — against a rich backdrop of culture and tradition. For obvious reasons, this makes virtue ethics hard to analyze from a complexity perspective, as it would require a comprehensive computational description of what a human being is, along with all her history and culture-specific flavors. Nevertheless, it is possible to isolate and analyze one necessary aspect of any virtuous agent: its ability to learn. In principle, any of the problems discussed thus far can be re-framed as a learning problem. Instead of asking ``can I solve problem $X$? effectively'', we ask ``can I \emph{learn} how to solve $X$ effectively?''. If the answer is yes, the de facto run-time complexity might be trivially low given the appropriate training. Intuitively, if a specific problem has already been solved — e.g., having discovered the optimal action-combination for a certain situation (\ref{sec:combinatorics}), an optimal reinforcement learning policy for an environment (\ref{sec:dynamicenvironments}), or an action-rule in a strategic setting (\ref{sec:deontology}) — the same solution may be applied in constant time $O(1)$ to future cases given that the very same problem re-occurs. And even if the exact same problem never re-occurs, there might be patterns or trends to extrapolate from situations that are sufficiently similar. Through repeated encounters with salad bars (\ref{sec:salad}), we might learn about good combinations of ingredients. Seeing the same ingredients appearing in other salad bars, we can make use of our previous experience to efficiently put together a tasty salad. Following the virtue-theoretic emphasis on moral learning, the same should hold for moral behavior. In the words of Aristotle: “[...] a young man of practical wisdom cannot be found. The case is that such wisdom is concerned not only with universals but with particulars, which become familiar from experience” (NE 1141b 10). 

In turn, learning a machine is the business of \emph{machine learning}, which, due to a wealth of recent advancements, has come to dominate the field of AI in the 21$^{st}$ century. The relevant question for our investigation — and the topic of this section — is: are there any computational complexity considerations that might constrain a computational agents ability to learn in the moral domain?

\subsection{The Complexity of Learning}
\label{sec:ComplexityOfLearning}


As with any mature field of mathematical analysis, there are many relevant variants, settings, and measures that can be used to formally analyze the complexity of learning.\footnote{See \cite{vapnik1999nature} for the definite introduction to statistical learning theory, along with an exhaustive account of its rich history.} First, we might differentiate \emph{run-time complexity} (the number of state transitions an algorithm needs to perform at run-time to solve problem $X$), from \emph{training-time complexity} (the number of steps required to train the algorithm to solve $X$) and \emph{sample complexity} (the number of data points needed to learn how to solve $X$). Nonetheless, in practice, the distinction between these types may break down; for instance, if one sample represents one point in time, our sample complexity would be equal to our training-time complexity; similarly, we might understand training-time complexity \emph{as} the run-time complexity of a learning algorithm. What is important to note, however, is that the three types are intimately linked. In a completely known environment (say, Chess), simulation (e.g., using Monte Carlo methods) may be an effective way to get experience, e.g., by trying out the value of many different possible decisions. In this case, sample complexity measures how much simulation is required to find a good chess move. If the sample complexity is low, it means that it can be learned effectively, as it gives a lower bound on the total computational complexity \citep{kakade2003sample}. In other situations, there might be an abundance of data and plenty of time to train our model, while the task itself may refer to some general ability — e.g., evaluated using a Turing test — as opposed to how the system solves some specific decision-problem. Large language models like GPT-3 \citep{brown2020language}, which uses significant parts of the internet as training-data in order to produce new text, constitute suitable examples of this. In the most extreme case — e.g., an on-line reinforcement learning setting — if an agent have no information about the environment and no access to a simulation of it, trial-and-error exploration might be the only path to learning.

Our choice of complexity resource ultimately depends on the setting for our learning agent, and each setting has their own range of specific sub-types and relevant measures. To that end, it is common to differentiate between three broad classes of machine learning settings: (i) supervised learning (SL), where the aim is to learn from pre-labeled data, (ii) unsupervised learning (UL), where the training data lacks labels, and (iii) reinforcement learning, where learning is based on feedback. For an SL agent, the relevant question may be: how many labeled pictures of cats do I need to see in order to, within some margin of error, accurately classify new pictures as portraying cats or not? In a virtue-theoretic view, this could be rephrased as: how many labeled examples of ``courage'' (e.g., acts of courageous behavior) do I need to observe in order to classify unobserved actions as courageous in the future? In this case, we might only be interested in how sample complexity — number of pictures depicting courage in the training set — reduces the agents prediction error in the classification task (e.g., empirical risk minimization). For an RL agent, the question might instead be: how much do I need to observe the overall behavior of a moral exemplar (e.g., a virtuous human) in order to approximate their reward function.\footnote{This particular method is called \emph{inverse reinforcement learning}, introduced by \cite{Ng2000}, and suggested as a path towards artificial virtuous agents by \cite{Berberich2018}.} Here, it could also be important to consider the potential risk of detrimental mistakes during an agent's learning phase. As discussed in (\ref{sec:dynamicenvironments}), if trial-and-error exploration is the only option the goal for an RL agent may not be to merely maximize the cumulative reward, but to minimize regret relative to the optimal solution.




Before we can ask whether something can be learned effectively, we first need to determine whether it can be learned at all. For instance, we let $X$ be a set of points, $Y$ a set of labels (e.g., 0 and 1), and $\mathcal{H}$ a set of hypotheses $h$ (e.g., binary classifiers) that takes an $x \in X$ and outputs a label $y \in Y$. The goal for our learning algorithm $\mathcal{L}$ is to, given a sequence of labeled training samples $(x,y)$ drawn randomly from some distribution $\rho$ over $X$, infer a hypothesis $h \in \mathcal{H}$ that is able to correctly classify future instances of $x \in X$ given some accuracy $\epsilon$ and failure probability $\delta$. In turn, we say that a hypotheses space $\mathcal{H}$ is learnable if there exists a learning algorithm which, given a finite number of training samples $(x,y)_{n}$, can map inputs to outputs within $\epsilon$ of the optimal with a probability of at least $1 - \delta$.\footnote{For instance, following empirical risk minimization, we might define the optimal $h*$ as the hypothesis among $\mathcal{H}$ for which the risk of misclassification is minimal.} If so, sample complexity can be defined as $n(\rho,\epsilon,\delta)$, which says that we need $n$ training samples to learn a target function with respect to distribution $\rho$, error rate $\epsilon$ and failure probability $\delta$.

\subsubsection{Weak and Strong Sample Complexity}
\label{sec:WeakAndStrongSampleComplexity}

Based on these definitions, we may also differentiate between \emph{weak} and \emph{strong} variants of sample complexity, e.g., by asking how many samples we need to learn a target for \emph{some} specific input-output distribution $\rho$ over $X$ (weak), and how many we need to learn it for \emph{any} possible distribution (strong). Following the strong approach, the impossibility results commonly known as the No Free Lunch Theorem (NFL) establishes that there will always be unfortunate distributions for which the sample complexity is arbitrarily large \citep{wolpert1992connection,wolpert1996lack,schaffer1994conservation}. As an intuitive example, we can imagine a learning algorithm whose goal is to predict the weather — in this case limited to sunny (S) or rainy (R) — based on the weather on previous days. Collecting data for three days, we have $2^3$ possible weather-histories (i.e., SSS, SSR, SRR, SRS, ...). We then measure the learning algorithm's error $\epsilon$ as the ratio of incorrect predictions (e.g., predicting S for a day of rain). Then, we can demonstrate that every learning algorithm achieves a perfect $\epsilon = 0$ in exactly one weather history, a maximally bad $\epsilon = 1$ in another, a mixed result of $2/3$ in three histories, and $\epsilon = 1/3$ in the three remaining histories. NFL establishes that, for each possible $\epsilon$, every learning algorithm achieves $\epsilon$ for the equal amount of possible weather-histories. For instance, if we assign the same probability to each possible weather history (i.e., a uniform probability distribution), NFL entails that every learning algorithm has the same expected $\epsilon$ of $1/2$.

NFL results have subsequently been extended to optimization \citep{wolpert1997no}, supervised learning \citep{wolpert2002supervised}, statistical learning \citep{von2011statistical}, meta-learning \citep{giraud2005toward}, and data privacy \citep{kifer2011no}.\footnote{See \cite{adam2019no} for a systematic review of NFL theorems.} Essentially, it means that there exists no learning algorithm that can perform well on every learning task having trained upon a dataset of a fixed size. Thus, for every learning algorithm, there exists a task on which it fails, as no learning algorithm can generalize to \emph{all} possible realities while having only observed \emph{some} instances of the realities. To that end, NFL has also been discussed in relation to more fundamental problems in the philosophy of induction, e.g., in connection to Hume's problem of induction \citep{sterkenburg2021no,schurz2017no} or Occham's razor \citep{lattimore2013no}. Hume famously  advanced skepticism against the very justification of induction, arguing that deductive reasoning alone cannot secure the validity of inductive inference; and neither can induction, due to circularity, provide non-deductive grounds for itself \citep{hume2003treatise}.

Of course, it may not be fair to advance the fundamental issues of induction against the feasibility of moral learning systems. While they may obstruct the prospects of a perfect universal moral learner, it does not stop us from pursuing weaker yet reasonable alternatives that are practically viable. Instead of seeking a global and model-independent justification for why inductive inference seems to work, we can opt for local and model-relative justifications in order to explain why \emph{some} learning algorithms work better than others \citep{sterkenburg2021no}. However, it should be stressed that any such alternative would inevitably entail some form of inductive bias; assumptions that we exploit to enable and foster learnability. One strategy to alleviate the curse of arbitrarily large sample complexity is to constrain the space of probability distributions, e.g., by making assumptions about the structure of the distribution from which the data-points are drawn (called ``parametric'' procedures in statistics). The most straight-forward parametric assumption can be found in the Central Limit Theorem, which states that when we sum up randomly drawn independent variables, they tend towards a normal distribution.\footnote{Philosophically, this means that one accepts that the world tends towards a normal distribution.} In fact, most advancements in machine learning rely on some form of parametric assumptions, e.g., using linear and logistic regression, and the parameters of artificial neural networks.\footnote{It should be note that deep neural networks with sufficiently many parameters can be viewed as non-parametric; e.g., \cite{lee2017deep} demonstrates that an infinitely wide deep network is equivalent to a non-parametric Gaussian process.}

Another alternative is to constrain the space of hypotheses. In the philosophy of induction, this can be motivated on the basis of Occam's razor, which roughly states that simpler hypotheses are generally more better than complex alternatives. But what does ``simpler'' mean? And how many samples do we need to infer a simple hypothesis that is able to predict well? In computational learning theory, such questions can be effectively addressed by the Probably Approximately Correct (PAC) model of learning, introduced by Leslie \cite{valiant1984theory}. Following the formalism described earlier, we consider a learning algorithm $\mathcal{L}$ that wants to learn a Boolean function $f : X \rightarrow \{0,1\}$ in a finite set of hypotheses  $\mathcal{H}$, on the basis of samples ($x$) drawn independently from distribution $\rho$ over sample space $X$. We call a hypothesis $h$ infered by $\mathcal{L}$ \emph{good} if it can approximate $f$ within some error $\epsilon > 0$, in the sense that:

\begin{equation}
\underset{x \sim \rho}{Pr} [h(x) \neq f(x)] \leq \epsilon
\end{equation}

Then, we say that $\mathcal{L}$ is ``probably approximately correct'' if it can make good approximations with probability $1 - \delta$, for any choice of $\rho$ and all failure probabilities $\delta$ and error rates $\epsilon > 0$. Finally, we say that a function $f$ is PAC-learnable if there exists an $\mathcal{L}$ that can make ``probably approximately correct'' predictions with a sample size $n$ that is a polynomial function of $1/\epsilon$ and $1/\delta$. In other words, PAC learning formalizes the idea that, if a function is learnable, it means that there exists a learning algorithm that with a reasonable likelihood can get reasonable generalization errors if it trains on randomly selected data, all while the number of samples are upper-bounded by a polynomial.

What is interesting from a computational complexity perspective is that PAC-learning yields a bound on sample complexity. If a target function is PAC-learnable, then the number of samples $n$ required to learn the function can be derived by:

\begin{equation}
n = \frac{1}{\epsilon}\ln\frac{|\mathcal{H}|}{\delta}
\end{equation}

Particularly, equation (6) illustrates the intimate relationships between prediction error, confidence, samples, and the space of hypotheses, and their computational trade-offs. For instance, it captures the intuition that error rate $\epsilon$ and generalization error $\delta$ can be reduced (although never to 0) by increasing the number of samples. It also shows that successful learning from a limited number of samples $n$ requires us to constrain the cardinality of the hypothesis space $\mathcal{H}$; e.g., either by reducing the number of individual hypotheses, or by reducing their descriptive complexity.\footnote{\cite{aaronson2013philosophers} has interpreted this aspect of PAC-learning as an mathematical justification for why Occam's Razor works.} By contrast, a larger $\mathcal{H}$ may lead to overfitting, i.e., when the hypothesis closely mirrors a particular set of samples and fails to generalize to additional observations.

From a philosophical perspective, PAC-learning is interesting since it, regardless of Hume's skepticism, defines a rich class of instances where induction is guaranteed to work (at least probably approximately). However, one significant trade-off with the model is that it only applies to finite classes of hypotheses, which inevitably entails a compromise between approximation accuracy and the learning algorithms capacity. For instance, a hypothesis containing continuous parameters may need to be turned into discrete parameters. The question is, how finely do we divide the infinite continuum? This conundrum can be addressed by VC theory (developed by \cite{vapnik1974theory,vapnik2015uniform}). Importantly, as opposed to hypotheses space, VC theory brings the richer concept of VC dimensions (VCD), which measures a learning algorithm's expressive capacity by the cardinality of the largest set of points the learner can \emph{shatter}. For instance, given three points in a two-dimensional space, we can ask whether a linear classifier $LC$ can correctly separate negative (labeled $-$) and positive points (labeled $+$). Particularly, we ask, for all possible ways of labeling the three points $2^3 = 8$, is there a way we can draw a straight line that separates positive from negative points? After discovering that it is possible — i.e., $LC$ shatters the set containing the three points — we try to do the same for four points. Since no set of four points can be shattered by a straight line,\footnote{This is a consequence of Radon's theorem, which can be used to infer the VCD of linear separations of $d$-dimensional points.} we conclude that the VCD of $LC$ is $3$. Similarly, a class of hypotheses $\mathcal{H}$ shatters the points $\{x_{1},...,x_{m}\} \subseteq X$ if there is a hypothesis $h \in \mathcal{H}$ which agrees with all $2^m$ possible configurations of $h(x_{1}),...,h(x_{m})$. The VCD of $\mathcal{H}$ is then the cardinality of the largest $m$ where there is a subset $\{x_{1},...,x_{m}\} \subseteq X$ that is shattered by $\mathcal{H}$. In turn, due to the work of \cite{blumer1989learnability}, VCD has been unified with PAC-learnability through the fundamental insight that finite VC dimensions provides the necessary and sufficient condition for distribution-free learnability. In other words, if a set of hypotheses is PAC-learnable, its VCD is finite. Of course, while it does not alleviate the fundamental problems of induction, it yields a framework for describing \emph{when} induction is feasible, a rich measure of expressive capacity (VCD), as well as a substantive notion of simplicity (i.e., smallest amount of VCDs). With regards to complexity, it has recently been proven by \cite{hanneke2016optimal} that the optimal sample complexity\footnote{Here, optimal means that the upper bound matches known lower bounds \citep{ehrenfeucht1989general,blumer1989learnability} up to numerical constant factors.} of PAC-learnability for class $\mathcal{H}$ is:

\begin{equation}
n(\epsilon,\delta) = O\left(\frac{VCD(\mathcal{H}) + \ln \frac{1}{\delta}}{\epsilon}\right)
\end{equation}

Nevertheless, theories for feasible learnability face the same problem as Nash equilibrium: just knowing that there exists a hypothesis $h$ in $\mathcal{H}$ that is consistent with the data does not necessarily mean that it is easy to find. As such, PAC-learnability ignores the vast computations that are potentially required to actually find a good hypothesis. To that end, a large set of hardness results have been proven for PAC. In the \emph{proper} setting, where the learner is required to output $h \in \mathcal{H}$, \cite{pitt1988computational} proved that representation classes such as disjunctions of two monomials (a polynomial with only one term), Boolean threshold functions, and Boolean formulae where each variable occurs at most once, cannot be efficiently learned, as they can be reduced to known NP-complete problems. Based on widely used cryptographic assumptions — e.g., the Rivest-Shamir-Adleman system and Blum integers — \cite{kearns1994cryptographic} proves representation-independent hardness results in the \emph{improper} setting (where the learner can output any $h \notin \mathcal{H}$) for a range of representation classes, including polynomial-size Boolean formulae, constant-depth threshold circuits, and acyclic deterministic finite automata.\footnote{In the improper context, representation-independent hardness means that learning remains hard regardless of the form the algorithm represents its hypothesis, on the basis that the hypothesis can evaluated in polynomial time.} In addition, while the hardness of improper learning rely on cryptographic assumptions, \cite{applebaum2008basing} shows that a proof would either ``collapse'' the polynomial hierarchy\footnote{This means that if NP = co-NP, then it follows that PH = NP. It is widely believed that a collapse of the PH is implausible.} or imply that any average-case hard problem in NP can be transformed into a one-way function (which would yield an outstanding break-through in cryptography).

Another important distinction in learning theory besides proper and improper, is the one between \emph{realizable} (or ``noise-free'') and \emph{agnostic} (``noisy'') learning. In the realizable case, it is assumed that there exists an optimal hypothesis $h*$ in the space of hypotheses $\mathcal{H}$ in the sense that its $\epsilon = 0$. In agnostic learning \citep{kearns1992toward}, no assumptions are made about the target function; we simply want to find the best possible $h$ from \emph{some} distribution.\footnote{See \citep{hopkins2022realizable} for an exposition of the deeper relationship between agnostic and realizable learning.} Paradigmatic problems in statistical machine learning includes the learning of halfspaces (or linear threshold function),\footnote{Formally, a halfspace is a Boolean function of the form $f(x) = \sign(w_{1}x_{1} + ..., w_{n}x_{n} - \theta)$, where $w_{i}$ are ``weights'', $\theta$ is the ``threshold'', and $w_{1}...,w_{n},\theta \in \mathbb{R}$. The $\sign$ function returns $1$ on arguments $\geq 0$, otherwise $-1$.} monomials, and decision lists. While learning a halfspace in the proper realizable case can be done in polynomial time via linear programming, its NP-hardness in the proper agnostic case have been proven in various ways.{\footnote{See, among others, \cite{angluin1988learning,amaldi1998approximability,haastad2001some,ben2003difficulty} and \cite{feldman2012agnostic} for a more recent overview. See also \cite{daniely2014average} for the improper case.}}

As discussed in \ref{sec:dynamicenvironments}, similar computational hardness prevail in reinforcement learning \citep{mundhenk2000complexity,papadimitriou1987complexity}. In fact, there are many reasons to believe that reinforcement learning is significantly harder than the supervised setting: the learner may not receive a training sets from the environment; the learner's may only receive `noisy' immediate rewards (which can deceive the agent into learning policies that does not maximize the long-term future rewards); exploiting comes at the expense of losing out on exploring (and vice versa); and there may even be detrimental consequences to consider (minimize regret).\footnote{See \cite{kakade2003sample} for a detailed investigation of sample complexity in reinforcement learning.}

\subsubsection{Machine Learning Theory vs Practice}
\label{sec:WeakAndStrongSampleComplexity}

It is important to stress that the theoretical considerations discussed here might have limited relevance for the practical viability of machine learning in various domains. Clever uses of inductive biases — e.g., task representation and parametric tools — along with vast amounts of training data and computational power continue to defy what might have appeared to be impossible only a decade ago. For instance, the performance of deep learning models in the field of natural language processing has recently been accelerated via the transformer architecture \citep{vaswani2017attention}, which utilizes attention mechanisms to process tokens from any position in the input sequence; leading to improved context sensitivity through efficient use of parallelization. Similarly, advances in deep reinforcement learning has showed that only small sets of demonstration samples can significantly accelerate the learning process \citep{hester2018deep}. In fact, what is surprising is not the general trend that shows that learning is computationally hard: it is rather that we lack rigorous explanations for \emph{why} some learning systems seem to generalize well in practice. This is known as the ``paradox of deep learning'', which centers around understanding the empirical success of deep learning despite the absence of theoretical explanations \citep{kawaguchi2017generalization,neyshabur2017exploring,arpit2017closer,zhang2021understanding}. In turn, this generates a range of convoluted issues that are more or less unique for machine learning. For instance, what does it mean that a large language model — having trained on large parts of the internet — is ethical?\footnote{See \cite{liang2022holistic} for a ``holistic'' evaluation of large language models using multiple metrics and test cases.} How should we understand requirements of transparency, explainability, robustness, safety, and fairness of sufficiently advanced ``black box'' systems? \citep{gunning2019xai, Amodei2016, gabriel2020artificial,berk2021fairness}.\footnote{Recent work in algorithmic fairness indicates that there are inevitable trade-offs between, on the one hand, different concepts of fairness, and on the other, between fairness and accuracy \citep{berk2021fairness}.}

While such issues deserve attention in their own right, there is a particular caveat that follows from our complexity analysis: namely, the role of inductive biases and their moral justification. In some strong sense, the success of learning systems — e.g., training efficiency and predictive accuracy — seems inversely proportional to the inductive assumptions it exploits, as well as the problems of induction it introduces. I.e., for moral learning to work, we need to have a relatively clear idea of the performance measure — e.g., in terms of some predefined score, goal, or objective function — of the moral problem we want the learning system to tackle, or the moral behavior we want it to exhibit. As such, one might question whether existing moral learning systems can generate any ``new'' or ``genuine'' moral insight, as they merely train on some given data filtered through some given inductive biases. Similar to divine command and legal positivism, it presuppose that we have an answer to the questions we seek. Thus, the problem of moral machine learning does not reside in the computational complexity of learning as such, but rather, in justifying the moral assumptions we need to exploit in order for induction to work. As elegantly put by Karl \cite{Popper1962}:

\begin{quote}
In constructing an induction machine we, the architects of the machine, must decide \emph{a priori} what constitutes its `world'; what things are to be taken as similar or equal; and what \emph{kind} of `laws' we wish the machine to be able to `discover' in its `world'. In other words we must build into the machine a framework determining what is relevant or interesting in its world: the machine will have its `inborn' selection principles. The problems of similarity will have been solved for it by its makers who thus have interpreted the `world' for the machine. (p. 48)
\end{quote}


\section{Moral Tractability for Minds and Machines}
\label{sec:discussion}

\begin{table}
\footnotesize
\begin{center}
\begin{tabular}{ |l|l| } 
 \hline
Problem & Results \\ 
\hline
Combinatorics (\ref{sec:combinatorics}) & \\
\phantom{SP}Optimal plan of $n$ unordered actions & $\Theta(2^n)$ \\ 
\phantom{SP}Optimal plan of $n$ ordered actions & $\Theta(n!)$ \\ 
\phantom{SP}STRIPS and Propositional planning & PSPACE-complete \citep{bylander1991complexity,bylander1994computational}\\ 
\hline
Bayesian Inference (\ref{sec:causalinference}) & \\
\phantom{SP}Exact inference & \#P-complete \citep{roth1996hardness} \\
\phantom{SP}Most Probable Explanation (MPE) & NP-complete \citep{shimony1994finding}\\
\phantom{SP}Maximum a posteriori hypothesis (MAP) & NP$^{\text{PP}}$-complete \citep{park2004complexity}\\
\phantom{SP}Approximate exact inference & NP-hard \citep{dagum1993approximating}\\
\phantom{SP}Approximate MPE & NP-hard \citep{abdelbar1998approximating}\\
\phantom{SP}Partial MAP & NP-hard \citep{park2004complexity}\\
\hline
Sequential Decision-making (\ref{sec:dynamicenvironments}) &\\
\phantom{SP}Finite MDP & From PL to EXPSPACE-complete \citep{mundhenk2000complexity}\\
\phantom{SP}Finite POMDP & PSPACE-complete \citep{papadimitriou1987complexity}\\
\phantom{SP}Infinite POMDP & Undecidable \citep{madani2003undecidability}\\
\phantom{SP}Restless Bandit & PSPACE-hard \citep{papadimitriou1994queuing}\\
\hline
Strategic Dynamics (\ref{sec:deontologyGenerality}) &\\
\phantom{SP}Finite I-POMDP & PSPACE-complete \citep{papadimitriou1987complexity}\\
\phantom{SP}Decentralized MDP & NEXP-hard \citep{bernstein2002complexity}\\
\phantom{SP}2-player Nash Equilibrium (NE) & PPAD-complete \citep{chen2009settling}\\
\phantom{SP}Maximum Egalitarian NE (Max NE) & NP-complete \citep{gilboa1989nash}\\
\phantom{SP}Approximate Max NE & NP-complete \citep{conitzer2008new}\\
\phantom{SP}Pure strategy Bayesian NE & NP-hard \citep{conitzer2008new}\\
\phantom{SP}Pure NE infinite Markov Games & PSPACE-hard \citep{conitzer2008new}\\
\phantom{SP}Pure NE finite Markov Games & NP-hard \citep{conitzer2008new}\\
\phantom{SP}Correlated Equilibrium (CE) & P \citep{gilboa1989nash}\\
\phantom{SP}Max CE & NP-hard \citep{papadimitriou2008computing}\\
\hline
Logic (\ref{sec:deontologySemantics}) &\\
\phantom{SP}SAT-FOL & Undecidable \citep{turing1936computable,church1936note}\\
\phantom{SP}SAT-PL & NP-complete \citep{cook1971complexity}\\
\phantom{SP}Validity for Modal Logic & PSPACE-complete (K, T, S4), NP-complete (S5)  \citep{ladner1977computational} \\
\phantom{SP}Multi-agent Modal Logic (MAML) & PSPACE-complete \citep{halpern1992guide}\\
\phantom{SP}MAML + Common Knowledge & EXPTIME-complete \citep{halpern1992guide}\\
\phantom{SP}Validity for Temporal Logic & PSPACE-complete \citep{sistla1985complexity,spaan1993complexity}\\
\phantom{SP}SAT-Propositional Dynamic Logic & EXPTIME-complete \citep{fischer1979propositional,pratt1980near}\\
\phantom{SP}SAT-Deontic STIT Logic & Undecidable \citep{schwarzentruber2014stit}\\
\phantom{SP}Deontic Input/Output Logic & NP/co-NP-hard \citep{sun2017complexity}\\
\hline
Descriptive complexity (\ref{sec:DeoDescriptive}) &\\
\phantom{SP}FOL & LH \& AC$^{0}$ \citep{immerman1998descriptive}\\
\phantom{SP}Least fixed-point FOL & P \citep{immerman1982relational,vardi1982complexity} \\
\phantom{SP}SO$\exists$, SO$\forall$, \& SOL & NP, co-NP, and PH, respectively \citep{fagin1974generalized} \\
\phantom{SP}SOL with transitive closure & PSPACE \citep{immerman1989descriptive} \\
\phantom{SP}Least fixed-point SOL & EXPTIME \citep{abiteboul1997fixpoint}\\
\hline
Learning (\ref{sec:learning}) & No Free Lunch \citep{wolpert1992connection,wolpert1996lack,schaffer1994conservation} \\
\phantom{SP}Sample Complexity for PAC-learnability & $O\bigl(\frac{VCD(\mathcal{H}) + \ln \frac{1}{\delta}}{\epsilon}\bigl) $\citep{hanneke2016optimal} \\
\phantom{SP}Proper Realizable PAC & From P to NP-hard \citep{pitt1988computational} \\
\phantom{SP}Improper PAC & NP-hard (cryptographic assumptions) \citep{kearns1994cryptographic} \\
\phantom{SP}Proper Agnostic PAC & NP-hard \citep{feldman2012agnostic} \\
\hline
\end{tabular}
    \caption{Summary of the surveyed complexity results.\label{table:summary}}
\end{center}
\end{table}


First, we offered three possible interpretations of how to analyze the complexity of ethics based on Marr's three levels of analysis. We then proceeded to analyze a range of ethical problems for causal engines, rule-followers, and learners. The results are summarized in Table \ref{table:summary}. Based on the surveyed results, what can computational complexity teach us about morality? In this section, we will discuss the consequences for moral machines (\ref{sec:discussionMORALMACHINES}) and human morality (\ref{sec:discussionHUMANMORALITY}), the explanatory prospects of the Moral Tractability Thesis (\ref{sec:discussionMORALTRACTABILITY}), along with limitations (\ref{sec:discussionLimitations}) and venues for future work (\ref{sec:discussionFuture}).

\subsection{Consequences for the Prospects of Moral Machines}
\label{sec:discussionMORALMACHINES}

What consequences do intractability results have for the prospects of moral machines? First and foremost, due to the intractability (and undecidability) stemming from combinatorics of action plans (\ref{sec:combinatorics}), probabilistic causal inference (\ref{sec:causalinference}), dynamic and partially observable environments (\ref{sec:dynamicenvironments}), general rules (\ref{sec:deontologyGenerality}), strategic dynamics (\ref{sec:deoGameTheory}), logic (\ref{sec:deontologySemantics}), semantics (\ref{sec:deoProblemOfSemantics}) and learning (\ref{sec:ComplexityOfLearning}), we can firmly conclude that perfect moral machines are impossible. In many cases, suboptimal approximations of solutions are also intractable. Instead, the developers of moral machines should strive for ``best possible'' on the basis of constrained resources. Similar conclusions have been made — although not as formally — by \cite{brundage2014limitations, mabaso2021computationally,hew2014artificial,hagendorff2022ethical}, and should be no surprise to scholars familiar with bounded rationality \citep{simon1955behavioral,simon1990bounded,rubinstein1998modeling,russell1994provably} or bounded ethicality \citep{bazerman2011blind,tenbrunsel2004ethical}. Nevertheless, the presented work should also be helpful in pinpointing the type of complexity that bounded computational agent's face in the realm of ethical decision-making, and the relevant trade-offs between optimality and feasibility it presents. It should therefore be informative for debates on artificial moral agents, as it draws the question of whether artificial moral agents are practically feasible or normatively desirable closer to the de facto dimensions of AI methods; as opposed to centering on the uniquely human capacities that existing AI systems lack \citep{stensekeROBOPHIL}.

In a similar vein, the results should also be illuminating for the further development of artificial systems implemented in moral domains. In particular, the complexity of ethical problems highlights the intimate relationship between the cognitive capacities of agents and moral resources such as time, memory, knowledge, communication, learning, and heuristics. However, it also presents a strong implementation-variance with regard to moral resources, which potentially obfuscates any general notion of practical moral competence. I.e., while it may be possible to identify the available resources for a particular agent in a given context, and how the resources can be effectively utilized and combined to yield competent ethical behavior, it is difficult to generalize such insights to \emph{any} agent in \emph{any} context. For instance, what may be considered a competent ethical decision for a social robot in a classroom differs significantly from the ethical competence required in high-speed traffic situations. The implementation-variance can be interpreted negatively: it shows that no general benchmarks can be established so as to assess the ethical performance of computational systems. More optimistically, it can also convey areas where domain- or problem-specific moral benchmarks can be established in terms of resource-dependency (e.g., given limited time or information). More directly, it points to venues where relevant benchmarks already exist: e.g., to find morally attractive equilibria in complex coordination games, minimize regret in multi-armed bandit settings, constructing tractable logics for moral reasoning, or efficient algorithms for Bayesian inference.

It also shows that a lot of work remains to be done on the moral end of machine ethics. In almost every instance, there is an uncomfortable trade-off between optimality and feasibility, and performance and efficiency; trade-offs which themselves may need normative justification. However, while computer science may have shifted towards becoming an empirical science in the advent of machine learning, our moral theories remain deeply rooted in theoretical ideals of right and wrong, which may presuppose unrealistic access to oracles of rationality. This brings out the open-ended tension between normative theory as standards of ``ideal good'', and normative theory as action-guiding heuristics to get suboptimal but feasible results. For instance, if our moral theories assume unrealistic computations, how can we provide a solid footing for their justification in practice? Similarly, if the tension remains unresolved, we cannot clearly determine cases whether a harmful action was due to a failure of competence, or whether it was a moral wrongdoing. It also presents challenges for moral theories: e.g., an agent that evaluates her actions according to NT$_{1}$ might be better off (according to NT$_{1}$) by following the action-decisions provided by an alternative theory NT$_{2}$.\footnote{The typical example of such a ``self-effacing'' theory is act utilitarianism (\citep{parfit1984reasons}, Sections 9 and 17), as an agent would produce more overall good by \emph{not} following the prescriptions of act utilitarianism in practice.} Thus, the critical question is: upon what standards should we potentially revise our moral theories so as to be feasible as decision-procedures with regard to the bounded and implementation-variant resources of agents? In machine contexts, we believe such open-ended issues can be fruitfully investigated under the lens of computational complexity, as it provides analytical means to measure how resources relate to formal notions of performance. More practically, it provides a venue to address what machines ought to do based on what they \emph{can} do at all — and what they can do \emph{effectively} — which in turn can convey the domains where computational systems can be successfully applied to make competent ethical decisions.

\subsection{Consequences for Human Morality}
\label{sec:discussionHUMANMORALITY}

More broadly, what consequences do these results have for our understanding of human morality? This ultimately depends on what one believes about the human mind, and more particularly, whether and to what extent it is computational in nature. The ones who view humans as primarily cultural, social, and spiritual beings may find the computational perspective completely irrelevant for understanding human morality. Others, who attempt to understand human behavior in terms of cognitive capacities, may find it fruitful to assume that human cognition has at least \emph{some} particular characteristics and constraints, which can be used to constrain the space of hypothesis. Simply put: assumptions about what the human mind can and cannot do should be informative for our understanding of the human mind. From this latter view, the step towards embracing some form of computationalism about the human mind becomes quite attractive, as it offers a smorgasbord of additional scientific tools; e.g., the use of computational architectures to model, test, and revise theories about cognition and behavior. If we take this step, computational complexity becomes an indispensable instrument, as it helps us constraint the space of possible computational theories of human cognition. This view can be captured in the \emph{Tractable Cognition Thesis}, which states that computational models of cognitive abilities need to be computationally tractable, given some reasonable conception of tractability \citep{van2008tractable,van2019cognition}. Conversely, if a computational theory implies intractable computations, it indicates that the theory is inadequate. A formal variant of the thesis is the \emph{P-cognition Thesis}, which asserts that cognitive functions are constrained by polynomial time. In cognitive science, the P-Cognition thesis has explicitly been advanced as guide for computational-level theories of human cognition by \cite{cherniak1986minimal,tsotsos1990analyzing,levesque1989logic, frixione2001tractable}, and is implicitly used as a constraining factor by a large group of cognitive psychologists.\footnote{See \cite{van2008tractable} for a detailed treatment of the Tractable Cognition Thesis and its formal variants.} Furthermore, the observation that many NP-hard problems can in fact be efficiently solved for some part of the input parameter has led to the argument that P-Cognition Thesis should be replaced with the Fixed-Parameter Tractability thesis \citep{van2008tractable}, which states that cognitive functions are restricted by polynomial time in the overall input size $n$, while allowing for superpolynomial time in some part of its input parameter \citep{downey2012parameterized}. A related modeling paradigm is the concept of \emph{resource-rational} cognition \citep{lieder2020resource}, which addresses cognitive modeling in terms of the optimal use of limited computational resources.

Thus, even if it remains unclear what kind of computer the human mind is (or whether it can be meaningfully captured by any model of computation), tractability considerations — in conjunction with cognitive modeling and experimental data — can work as a hypothesis that helps us carve out the space of feasible theories of cognition. Naturally, this would also include the functions that make up moral cognition. For instance, if we believe that humans perform causal, strategical, or logical reasoning to produce competent moral behavior, tractability considerations will directly serve to constrain the space of computational-level problems that underpins moral behavior. In addition, if we have reason to believe that humans perform these moral inferences in a specific way — e.g., using certain Bayesian, decision-theoretical, or logical inference techniques — a complexity analysis will help to pinpoint the relevant trade-offs between performance and feasibility; trade-offs which may directly relate to the complexity results surveyed in this paper.

\subsection{Moral Tractability Thesis}
\label{sec:discussionMORALTRACTABILITY}

\begin{figure}
\centering
\includegraphics[width=0.85\textwidth]{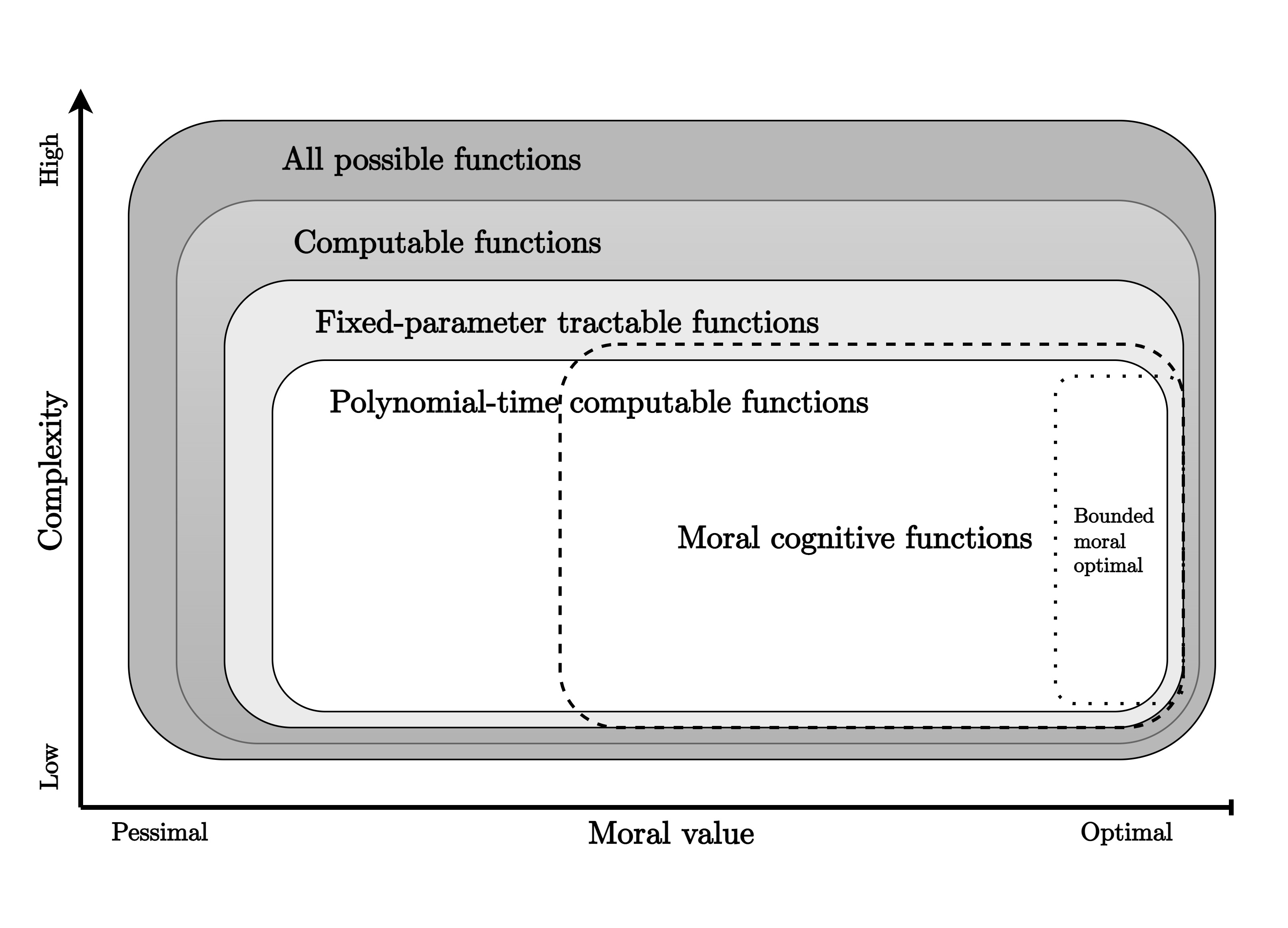}
\caption{The Moral Tractability Thesis (MTT) states that the set of possible moral cognitive functions are subject to tractability constraints. More formally, this can be framed as a subset of functions that are fixed-parameter tractable for sufficiently small input parameters, which also includes the set of functions computable in polynomial time. We believe the MTT can serve (1) as a meta-ethical standard for the action-guidance of normative theory, in the sense that action-guidance should be feasible with respect to agents' resources, (2) as a guide for normative judgements and responsibility in cases where it is unclear whether an agent acted wrongfully due to a failure of cognitive constraints, (3) as an experimental paradigm in studies of human moral cognition and psychology, and (4) as a remedy to the tension between feasibility and performance in moral contexts.}
\label{fig:MTT}
\end{figure}

The role of tractability in theories about moral behavior and cognition can be formulated as the \emph{Moral Tractability Thesis} (MTT). It is a natural extension of the Tractable Cognition Thesis and states that morality — moral behavior, moral problem-solving, and moral cognition — are constrained by computational tractability (see Figure \ref{fig:MTT}), given some reasonable model of human moral cognition.\footnote{Here, ``reasonable'' is intentionally ambiguous, as there is currently no consensus of the specific model of computation that reflects the human brain. For a recent proposal, see \cite{blum2022theory}.} The MTT is a hypothesis that points in both normative and descriptive directions:

(1) \emph{MTT as a meta-ethical standard for normative theory} — MTT yields a meta-ethical point that stresses that the computational problems imposed by a moral theory also \emph{should} be tractable with regard to the resources of an agent following the theory; i.e., to the extent the theory aims to provide any meaningful action-guidance for the agent. If not, it could suggest that the moral theory imposes unrealistic demands in relation to the agent's resources, and therefore, the theory should be revised so as to constrain the space of problems the agent can be expected to solve. As such, it could help to exclude the ``self-effacing'' theories of normative action-guidance from the ones that respects the limited capacities of agents. More practically, it can help us to identify the situations and contexts in which a certain form action-guidance can be expected to produce better results than competing alternatives. 

(2) \emph{MTT as a guide for normative judgements and responsibility} — In situations where an agent behaves immorally, MTT can help us to answer the question whether the agent's immoral behavior was due to a failure of morality and rationality. For instance, if the computational demands are fair for some reasonable conception of human moral competence, it might point to the lack of a certain moral resource in the situation at hand (e.g., time, memory, knowledge, learning, or rationality demands). If no such absence can can be identified, it can also serve as grounds for holding the agent responsible for their immoral behavior.

(3) \emph{MTT as an experimental paradigm} — In the descriptive direction, MTT can provide a delimiting factor for existing paradigms studying human morality; from the computational modeling of human moral cognition to experimental studies in moral psychology. More precisely, MTT can be used to identify the principles and algorithms that underpin cognitive processes in moral decision-making, reveal relevant trade-offs between feasibility vs performance, and further investigate the role of resources in specific moral contexts.

(4) \emph{MTT as a remedy to feasibility vs optimality} — Another feature of MTT is that it can help to resolve open-ended tensions between feasibility and performance in moral contexts, in the sense that the latter is directly constrained by the former. For instance, if we assume that there is a fixed point which yields the optimal moral value, MTT implies that there is an action-guiding theory that yields the highest possible moral value with regard to the resources of the acting agent.\footnote{More moderately, instead of an optimal, we can assume that there is a justified threshold for moral permissibility on the basis of \emph{some} notion of moral value.} The same idea can be articulated for moral communities: recipes for action-guidance that produces the highest moral prosperity — e.g., joint well-fare, or mutually agreed-upon moral values — with regard to the mutually \emph{shared} resources of agents in the moral community. In other words, there may be normative action-guidance that provides the morally optimal use of bounded cognitive resources; a resource-rational moral theory \citep{lieder2020resource}. From a game-theoretic point of view, this can be interpreted as the optimal moral equilibrium point (e.g., maximum joint benefit) within the space of the agents' bounded resources (Figure \ref{fig:MTT}).






\subsection{Limitations}
\label{sec:discussionLimitations}

There are several of gaps and limitations for the explanatory viability of moral complexity analyses in general and the MTT in particular. We will briefly address some of these along three broad questions: (i) How and to what extent is computational complexity relevant for the development and deployment of ``morally competent'' AI systems in real-world domains? (ii) What aspects of human morality are \emph{not} captured by a complexity analysis? (iii) What can complexity tell us about the human use of morally informed AI systems (e.g., AI systems that are designed and used to extend or augment human moral capacities)?

(i) It should be stressed that the computational problems analyzed in this paper may have limited relevance for many real-world implementations of AI systems in moral domains. For instance, what can computational complexity, if anything, tell us about the moral behavior in domains such as autonomous driving, social assistive robotics, and natural language processing? While the overarching goal of a self-driving vehicle may be summarized as ``drive safely from point $A$ to $B$'', it encompasses a vast set of smaller sub-tasks and goals — e.g., path planning, adhere to traffic rules, crossing this and that intersection — for which it utilizes a cluster of distributed capacities — e.g., sensors, motors, maps, simulations — that makes up competent autonomous driving \citep{badue2021self}. Thus, a complexity analysis of an autonomous car's moral competence offers little insight unless it is based on the specific car's capacities. As noted in section \ref{sec:learning}, some technical-ethical challenges that arise for large language models may present the inverse problem of \emph{constraining} a general capacity to produce text (e.g., so as to adhere to human principles and values). That is, instead of learning an optimal solution from the largest possible search-space — e.g., by training on vast amounts of human-generated text — the task may be to constrain the search via some justified inductive bias that aligns with human values. Similarly, the most widely discussed ethical issues that pervade social assistive robotics — e.g., deception, dignity, trust, and recognition in Human-Robot Interaction (HRI) — are far removed from the formal notions of causal, logical, and strategic inference discussed in this work \citep{boada2021ethical}.

The main point is that each domain presents its distinct set of technical and ethical challenges that need to be addressed with respect to the domain's unique conditions. In turn, these challenges may have little to do with the computation of applied normative ethics, but rather, depend on some specific normative requirements that are presupposed by certain human practice \citep{Behdadi2020}. However, \emph{if} AI systems are developed to behave in accordance with normative ethics, the surveyed complexity results have an overarching relevancy for such endeavors, just as they remain central to several prominent paradigms for the advancements of computing; e.g., probabilistic inference, knowledge-systems, and learning. In a more trivial sense, even if the aim is to merely supplement AI systems with \emph{some} capacities needed for \emph{some} form of ethical decision-making, the behavior of such systems would also be constrained by limited computational resources. A resource-rational complexity analysis could therefore help to identify the sort of ethical problems that can be solved by machines; the ones that can be solved efficiently, where there is room for improvement, as well as pinpoint the relevant resources and trade-offs.






%

(ii) Another important gap that needs to be addressed is the difference between human morality and the computational form of applied normative ethics explored in this paper. It should be strongly emphasized that, although the surveyed results have direct implications for computational systems, any potential consequences for our understanding of human morality rest on speculative assumptions. As discussed in \ref{sec:complexity}, ethics is an ambiguous and multifaceted concept; \emph{what} sort of ethical problems humans actually ``solve'', \emph{how} they solve them, and what they \emph{use} to solve them (e.g., emotions, reasoning) are all open questions with many possible and elaborate answers. Similarly, while the aim of normative ethics is to find generally applicable standards of ``good'' and ``bad'', it is but a small conversation of the broader landscape that makes up ethical life. Adopting a more skeptical view, one could even claim that normative ethics — as theoretically construed in contemporary Anglophone analytical philosophy — does not have any bearing on ethical life at all \citep{stocker1977schizophrenia}. To that end, it might be odd to imagine that humans put together ethical action-plans (or optimal salads) using exhaustive search methods (\ref{sec:combinatorics}), make causal inferences using (arbitrarily) large Bayesian Networks (\ref{sec:causalinference}), compute egalitarian Nash equilibria in their strategic interactions (\ref{sec:deoComputingEquilib}), or check behavioral norm-compliance with regards to possible worlds (\ref{sec:DeoDeonticLogic}). Humans may employ a broad range of resources in their everyday ethical life — e.g., emotions and motivations, guilt and shame, spirituality and ideology, critical reflection and theory of mind — that are uncounted for in this analysis. Thus, the presented analysis — from computation, to algorithm, to implementation — should not stop researchers from pursuing other investigatory directions that may be fruitful for understanding the intricate and numerous forms of human morality; e.g., the embodied and felt, the shared dependencies and vulnerabilities of personal relationships, and the norms and institutions of society and culture.



Still, if one believes that cognitive tractability and MTT holds any merit, complexity can provide a guiding light into the vast intersection between normative, descriptive, and applied ethics. Even if one just accepts that computational limitations have \emph{some} importance for \emph{some} forms of human moral behavior, it suffices as a reason to further investigate those limitations. Furthermore, it is not unreasonable to believe that many human beings \emph{are} able to and do — at least occasionally — follow deontological and consequentialist decision-procedures, and, following virtue ethics, learn to foster the psychological dispositions that enable them to become better moral persons; although the de facto cognitive procedures that underpin these processes might look different from the ones considered in this work. To that end, it would be rather odd to view these normative theories as completely separated from the cognitive capacities of humans; for instance, it seems hard to explain the practical success or popularity of certain moral heuristics (e.g., principles or theories) unless they were applicable — decidable and tractable — for humans in moral communities \citep{alexander2007structural}. In some cases, it even seems reasonable to believe that some of these heuristics are motivated on the basis of their computational efficiency; e.g., the computational efficiency of adhering to moral rules (\ref{sec:deontologyconshybrids}). Thus, while we should respect the vast gap between human and computational forms of morality, it would be unwise to exclude the possibility that there are more or less rigorous patterns in the cognitive processes that support the former, which in turn are amendable for formal investigation via the latter.


(iii) Another large area that is omitted in this paper is the integrative use of AI systems in human moral behavior. I.e., we have mainly considered the ethical behavior of computational systems in isolation, without any `human-in-the-loop'. As such, we have ignored the integrative prospects of how AI systems can be utilized to support or augment human ethical decision-making. Machines are, after all, a large cluster of computational methods that are employed to carry out the aims of its human users. In turn, this might point to a research area that can also be illuminated by computational complexity: how AI systems can foster and support moral prosperity in human practices in light of constrained resources \citep{vallor2015moral,giubilini2018artificial}. For instance, many problems that are intractable for humans may be tractable for machines, and conversely, many problems that are intractable for machines may be tractable for machines. The guiding question is thus: on the basis of resource constraints, how can AI be developed and used so as to expand rather than constrain the space of moral reasoning \citep{Vallor2016Book}?

\subsection{Future Work}
\label{sec:discussionFuture}

There are a number of interesting venues to further explore moral tractability for minds and machines beyond the ones already described. First, it should be stressed that although this work has discussed a number of complexity results relevant for moral behavior, it has circumvented an even greater amount. For most of the computational problem discussed, there are  hundreds of related results that would be relevant to consider under different conditions and assumptions. In fact, we have omitted results from entire fields of complexity theory — e.g., parameterized, communication, proof, and circuit complexity — that could yield further insights about the limitations of moral computation. Out of the discussed problems, there are two areas in particular that we believe deserves a more detailed investigation: (i) the sample complexity and regret-minimization of (moral) machine learning, and (ii) algorithmic game theory (in conjunction with algorithmic design theory). The reason for pursuing the first is that the complexity of machine learning in moral contexts remains relatively poorly understood; especially given issues such as explainability \citep{gunning2019xai}, induction \citep{sterkenburg2021no}, and the ``paradox of deep learning'' \citep{zhang2021understanding}. Another reason is that machine learning is the main vehicle behind the modern advancements in AI development. More urgently, learning systems are already deployed in a vast range of human practices, including areas that may involve salient forms of moral decision-making such as health care, autonomous driving, law, policing, and education.

The reason for pursuing (ii) is that game theory provides a unifying framework for the formal study of interactions, which in turn makes interactions amendable for algorithmic modeling and analysis. As such, we believe it could provide fertile synergies between historically distinct fields such as computer science, moral theory, evolutionary biology, behavioral economics, and social science. For instance, if we adopt the view that normative theories converge more than they disagree,\footnote{For instance, \cite{parfit2011matters} argues that it is erroneous to believe that there are profound disagreements between consequentialists, contractualists, and Kantians, writing: “these people are climbing the same mountain on different sides” (p. 385).} a resource-rational algorithmic game-theoretic analysis could help to identify the conditions under which certain theories are more practically viable than others. E.g., what sort of cognitive abilities and computational resources are required for a certain moral heuristic — e.g. an action-guiding normative theory — to support the highest possible moral prosperity for a community of agents? What sort of moral behaviors can be effectively computed or justified, and what behaviors can only be learned? Ideally, such investigations could not only illuminate the specific resources a computational agent need in order to be morally competent, but explain the very resource-rational rationales that underpin our most prominent ethical theories.

 



\section{Conclusion}
\label{sec:conclusion}

We have surveyed a large but far from exhaustive set of complexity results and discussed their relevance for minds and machines in the moral realm. Ultimately, we believe it opens up interesting interdisciplinary spaces between machine ethics, moral philosophy, and moral cognitive psychology, which will hopefully inspire new directions, not only in the engineering of moral machines, but in understanding the complex science of morality.

\section*{Appendix}

\subsection*{Complexity Classes}
\label{sec:AppendixA}

\begin{itemize}
  \item[] \textbf{AC$^{0}$} — Class of decision problems solvable by a family (one for each possible input-size) of constant-depth unlimited-fanin circuits, where the number of gates is bounded by some polynomial in the size of the input.
  \item[] \textbf{LH} — Class of decision problems solvable by an alternating TM with a bounded number of alternations in time $O(\log n)$. See \cite{immerman1998descriptive} for a detailed exposition.
  \item[] \textbf{PL} — Class of decision problems solvable by a probabilistic TM constrained by space $O(\log n)$, with an error probability $\epsilon < 1/2$  (``Probabilistic Logarithmic Space'').
  \item[] \textbf{P} — Class of decision problems solvable by a deterministic TM constrained by time $O(poly(n))$.
  \item[] \textbf{PP} — Class of decision problems solvable by a probabilistic TM constrained by time $O(poly(n))$, with an error probability $\epsilon < 1/2$.
  \item[] \textbf{FTP} — Class of decision problems solvable by a deterministic TM constrained by time $O(f(k)n^{c})$, where $f$ is a function that only depends on the parameter $k$, and $c$ is a constant. See \cite{downey2012parameterized} for the definite introduction to parameterized complexity.
  \item[] \textbf{NP} — Class of decision problems solvable by a non-deterministic TM constrained by time $O(poly(n))$. Alternatively, class of decision problems for which ``Yes''-instances are verifiable in polynomial time by a deterministic TM.
  \item[] \textbf{co-NP} — The complement set of NP. Class of decision problems for which ``No''-instances are verifiable in polynomial time by a deterministic TM.
  \item[] \textbf{\#P} — Class of function problems $f(x)$, where $f$ is the number of accepting paths of a non-deterministic TM constrained by time $O(poly(n))$. Informally, it is the set of counting problems associated with NP; i.e., where NP decision problems ask ``are there any'', \#P function problems asks ``how many''.
  \item[] \textbf{NP$^{\text{PP}}$} — Class of decision problems solvable by a non-deterministic TM constrained by time $O(poly(n))$ with access to an oracle for problems in PP.
  \item[] \textbf{TFNP} — Class of function problems solvable by a non-deterministic TM constrained by time $O(poly(n))$ where a solution is guaranteed to exist (``Total Function Non-deterministic Polynomial'').
  \item[] \textbf{PPAD} — The subclass of TFNP where functions are guaranteed to be total — i.e., a solution is guaranteed to exist — in virtue of the parity argument on directed graphs (``Polynomial Parity Arguments on Directed graphs''). See \cite{papadimitriou1994PARITY} for a detailed exposition.
  \item[] \textbf{PH} — The union of classes in the polynomial hierarchy. It can be defined recursively using oracle machines: given P = $\Delta_0^\text{P} =\Sigma_0^\text{P}=\Pi_0^\text{P}$, we define P$^{\Sigma_i^{\text{P}}} =  \Delta_{i+1}^{\text{P}}$, NP$^{\Sigma_i^{\text{P}}} = \Sigma_{i+1}^{\text{P}}$, and co-NP$^{\Sigma_i^{\text{P}}} = \Pi_{i+1}^{\text{P}}$, to express the union:
  
\begin{equation}
\text{PH} = \bigcup_{i=0}^{\infty} \Delta_{i}^{\text{P}} \cup \Sigma_{i}^{\text{P}} \cup \Pi_{i}^{\text{P}}
\end{equation}

\item[] \textbf{PSPACE} — Decision problems solvable by a deterministic TM constrained by space $O(poly(n))$.

\item[] \textbf{EXPTIME} — Decision problems solvable by a deterministic TM constrained by time $O(2^{poly(n)})$.

\item[] \textbf{NEXPTIME} — Decision problems solvable by a non-deterministic TM constrained by time $O(2^{poly(n)})$. Often denoted NEXP.

\item[] \textbf{EXPSPACE} — Decision problems solvable by a deterministic TM constrained by space $O(2^{poly(n)})$.
  
\end{itemize}






\section*{Conflict of Interest Statement}

The author declare that the research was conducted in the absence of any commercial or financial relationships that could be construed as a potential conflict of interest.

\section*{Author Contributions}

JS is the sole contributor of the work presented in the article.

\section*{Funding}
This work was partially supported by the Wallenberg AI, Autonomous Systems and Software Program – Humanities and Society (WASP-HS) funded by the Marianne and Marcus Wallenberg Foundation and the Marcus and Amalia Wallenberg Foundation.

\section*{Acknowledgments}

The author is especially grateful to Per Austrin (KTH Royal Institute of Technology) and Felix Lindner (Universität Ulm) for their insightful feedback on computational complexity. The author is also thankful to his colleagues at the Department of Philosophy and Cognitive Science at Lund University for providing helpful comments on previous versions of the paper. In particular, the author wants to acknowledge Niklas Dahl for answering countless queries about logic and semantics, Karl Samson and Gustav Stenseke Arup for input on the relationship between legality and morality, and Christian Balkenius, Björn Petersson, Ylva von Gerber, Sandra Lofs Midelf, Jiwon Kim, Alexander Velichkov, Trond Arild Tjøstheim, and Alfred Stenseke for their feedback. Finally, the author is grateful for participants of the Higher Seminar in Practical Philosophy (Lund University), the PhD seminar in philosophy (Lund University), the WASP-HS Winter Conference 2022, the philosophical colloquium at Georg-August-Universität, and the seminar at Institut für Künstliche Intelligenz (Universität Ulm), where earlier parts of the paper where presented.

\bibliographystyle{elsarticle-harv} 
\bibliography{references}





\end{document}